\documentclass[12pt]{article}
\usepackage{epsfig,amsfonts,amsbsy}
\newtheorem{theorem}{Theorem}[section]
\newtheorem{lemma}[theorem]{Lemma}

\newtheorem{definition}[theorem]{Definition}

\makeatletter
\@addtoreset{equation}{section}
\makeatother

\newcommand{\proof}{\par\bigskip\noindent{\em Proof:\ }}
\newcommand{\proofof}[1]{\par\bigskip\noindent{\em #1:\ }}
\newcommand{\qed}{~\rule{2mm}{2mm}\par\bigskip}
\newcommand{\ret}{\nonumber\\}

\newcommand{\abs}[1]{\left|#1\right|}
\newcommand{\norm}[1]{\left\Vert#1\right\Vert}
\newcommand{\rbk}[1]{\left(#1\right)}
\newcommand{\sqbk}[1]{\left[#1\right]}
\newcommand{\cbk}[1]{\left\{#1\right\}}
\newcommand{\bkt}[1]{\left\langle#1\right\rangle}
\newcommand{\set}[2]{\left\{#1\,\Bigl|\,#2\right\}}
\newcommand{\sumtwo}[2]%
{\mathop{\sum_{#1}}_{#2}}
\newcommand{\sumthree}[3]%
{\mathop{\mathop{\sum_{#1}}_{#2}}_{#3}}
\newcommand{\sumfour}[4]%
{\mathop{\mathop{\mathop{\sum_{#1}}_{#2}}_{#3}}_{#4}} 
\newcommand{\suptwo}[2]%
{\mathop{\sup_{#1}}_{#2}}
\newcommand{\supthree}[3]%
{\mathop{\mathop{\sup_{#1}}_{#2}}_{#3}}
\newcommand{\supfour}[4]%
{\mathop{\mathop{\mathop{\sup_{#1}}_{#2}}_{#3}}_{#4}} 
\newcommand{\inftwo}[2]%
{\mathop{\inf_{#1}}_{#2}}
\newcommand{\infthree}[3]%
{\mathop{\mathop{\inf_{#1}}_{#2}}_{#3}}
\newcommand{\inffour}[4]%
{\mathop{\mathop{\mathop{\inf_{#1}}_{#2}}_{#3}}_{#4}} 

\newcommand{\calE}{{\cal E}}
\newcommand{\calF}{{\cal F}}

\newcommand{\calH}{{\cal H}}
\newcommand{\calI}{{\cal I}}

\newcommand{\calK}{{\cal K}}

\newcommand{\calS}{{\cal S}}

\newcommand{\calU}{{\cal U}}

\newcommand{\La}{\Lambda}
\newcommand{\up}{\uparrow}
\newcommand{\dn}{\downarrow}
\newcommand{\bs}{\backslash}

\newcommand{\cxu}{c^\dagger_{x,\up}}
\newcommand{\cxd}{c^\dagger_{x,\dn}}
\newcommand{\axu}{c_{x,\up}}
\newcommand{\axd}{c_{x,\dn}}
\newcommand{\cxs}{c^\dagger_{x,\sigma}}
\newcommand{\axs}{c_{x,\sigma}}

\newcommand{\ays}{c_{y,\sigma}}
\newcommand{\cyt}{c^\dagger_{y,\tau}}
\newcommand{\ayt}{c_{y,\tau}}
\newcommand{\cd}{c^\dagger}
\newcommand{\nxs}{n_{x,\sigma}}
\newcommand{\nxu}{n_{x,\up}}
\newcommand{\nxd}{n_{x,\dn}}

\newcommand{\adjs}{a^{\dagger}_{j,\sigma}}
\newcommand{\ad}{a^{\dagger}}
\newcommand{\Cds}{C^{\dagger}_{\sigma}}
\newcommand{\Cs}{C_{\sigma}}
\newcommand{\bys}{b^{\dagger}_{y,\sigma}}
\newcommand{\byu}{b^{\dagger}_{y,\up}}
\newcommand{\byd}{b^{\dagger}_{y,\dn}}

\newcommand{\Hhop}{H_{\rm hop}}
\newcommand{\Hint}{H_{\rm int}}
\newcommand{\Stot}{S_{\rm tot}}
\newcommand{\Sztot}{S_{\rm tot}^{(3)}}
\newcommand{\Smax}{S_{\rm max}}
\newcommand{\vac}{\Phi_{\rm vac}}
\newcommand{\GS}{\Phi_{\rm GS}}
\newcommand{\UP}{\Phi_\uparrow}
\newcommand{\xiL}{x\in\Lambda}
\newcommand{\Sch}{Schr\"{o}dinger}

\newcommand{\Ne}{N_{\rm e}}
\newcommand{\Ns}{N_{\rm s}}
\newcommand{\Sop}{\hat{\bf S}}
\newcommand{\Sopt}{\hat{\bf S}_{\rm tot}}
\newcommand{\Sopc}[2]{\hat{S}^{(#1)}_{#2}}
\newcommand{\Sopct}[1]{\hat{S}^{(#1)}_{\rm tot}}
\newcommand{\Hub}{Hubbard model}
\renewcommand{\phi}{\varphi}
\newcommand{\ep}{\varepsilon}

\newcommand{\tH}{\widetilde{H}}
\newcommand{\tPsi}{\widetilde{\Psi}}

\newcommand{\tc}{\tilde{c}}
\newcommand{\Sup}{S_{\up}}
\newcommand{\Sdn}{S_{\dn}}
\newcommand{\Lup}{L_{\up}}
\newcommand{\Ldn}{L_{\dn}}
\newcommand{\hnu}{\hat{\nu}}
\newcommand{\mT}{{\sf T}}
\newcommand{\mA}{{\sf A}}
\newcommand{\mB}{{\sf B}}
\newcommand{\tLa}{\tilde{\Lambda}}
\newcommand{\Pas}{P_{-}}
\newcommand{\LBCO}{La422}
\newcommand{\hilb}{\frak{h}}
\newcommand{\comp}{\Bbb{C}}
\newcommand{\real}{\Bbb{R}}
\newcommand{\inte}{\Bbb{Z}}
\newcommand{\vphi}{\boldsymbol{\phi}}
\newcommand{\vpsi}{\boldsymbol{\psi}}
\newcommand{\vlambda}{\boldsymbol{\lambda}}
\newcommand{\veta}{\boldsymbol{\eta}}
\newcommand{\tsigma}{\boldsymbol{\sigma}}
\newcommand{\ttau}{\boldsymbol{\tau}}
\begin{document}
\begin{center}
{\bf
From Nagaoka's ferromagnetism to flat-band ferromagnetism 
and beyond\\
---An introduction to ferromagnetism in the Hubbard model\footnote{
Invited paper; Prog. Thoer. Phys. {\bf 99} (1998).
Archived as cond-mat/9712219.
}}
\par\vfil
Hal Tasaki\footnote{hal.tasaki@gakushuin.ac.jp,
http://www.gakushuin.ac.jp/\( \tilde{\ } \)881791/
}
\par\bigskip
{\footnotesize\sl Department of Physics, Gakushuin University,
Tokyo 171-8588, JAPAN}
\end{center}
\par\vfil
\begin{abstract}
It is believed that strong ferromagnetic interactions in some solids 
are generated by subtle interplay between quantum many-body effects 
and spin-independent Coulomb interactions between electrons.
It is a challenging problem to verify this scenario in 
the Hubbard model, which is 
an idealized model for strongly interacting electrons in a solid.

Nagaoka's ferromagnetism is a well-known rigorous example of 
ferromagnetism in the Hubbard model.
It deals with the limiting situation in which there is one fewer 
electron than in the 
half-filling and the on-site Coulomb interaction is infinitely large.
There are relatively new rigorous examples of ferromagnetism in 
Hubbard models called flat-band ferromagnetism.
Flat-band ferromagnetism 
takes place in carefully prepared models in which the lowest bands 
(in the single-electron spectra) are ``flat.''
Usually, these two approaches are regarded as
two complimentary routes to ferromagnetism in the Hubbard model.

In the present paper we describe 
Nagaoka's ferromagnetism and flat-band ferromagnetism
in detail,
giving all the necessary background as well as complete 
(but elementary) mathematical proofs.
By studying an intermediate model called the long-range hopping model,
we also demonstrate that there is indeed a deep relation 
between these two seemingly different approaches to ferromagnetism.

We further discuss some attempts to go beyond these approaches.
We briefly discuss recent rigorous example of ferromagnetism in the 
Hubbard model which has neither infinitely large parameters nor 
completely flat bands.
We  give preliminary discussion regarding possible experimental 
realizations of the (nearly-)flat-band ferromagnetism.
Finally, we focus on some theoretical attempts to understand 
metallic ferromagnetism.
We discuss three artificial 
one-dimensional models in which the existence of metallic 
ferromagnetism can be easily proved.

We have tried to make the present paper as 
self-contained as possible, keeping in mind readers who are new to 
the field.
Although the present paper is written as a review, it contains
some material which appears for the first time.
\end{abstract}
\newpage
\tableofcontents
\newpage
\section{Introduction}
\label{s:itro}
\subsection{Hubbard model and the origin of ferromagnetism}
\label{s:HubF}
The origin of strong ferromagnetic ordering observed in some
materials has been a mystery in physical science for a
long time. 
Since quantum mechanical many-electron systems without interactions 
universally exhibit
paramagnetism, the origin of ferromagnetism
should be sought in electron-electron interactions.
In most solids, however,
the dominant part of the interaction between electrons
is the Coulomb interaction, which is perfectly spin-independent.
We are thus faced with a very interesting and 
fundamental problem in theoretical physics 
to determine {\em whether spin-independent
interactions in an itinerant 
electron system can be the origin of ferromagnetic 
ordering.}
This problem is important not only because ferromagnetism is 
a very common (and useful) phenomenon, but because 
it focuses on
the fundamental role of nonlinear interactions in many-body quantum
mechanical systems.

It was Heisenberg
\cite{Heisenberg28} who first realized that ferromagnetism is an
intrinsically quantum mechanical phenomenon.
In Heisenberg's approach to ferromagnetism, one starts
from the picture that each electron (relevant to magnetism)
is almost localized in an atomic orbit.
By treating the effect of the Coulomb interaction and
overlap between nearby atomic orbits in a perturbative manner
as in the Heitler-London theory,
Heisenberg concluded that there appears
an ``exchange interaction'' between nearby electronic spins
which determines the magnetic properties of the system.

In a different approach to the problem of ferromagnetism,
which was originated by Bloch \cite{Bloch29}, one starts from
the quantum mechanical free electron gas, in which
electrons are in plane-wave like states.
One then treats the effect of the Coulomb interaction perturbatively,
and tries to find instabilities against certain
magnetic ordering.
When combined with the Hartree-Fock approximation (or a mean-field
theory), this approach leads to the picture that there is an instability
with respect to 
ferromagnetism when the density of states at the fermi energy
and the Coulomb interaction are sufficiently large.

In spite of a considerable number of  attempts to improve these ideas,
neither of these two approaches has yet produced a truly convincing 
explanation about the origin of ferromagnetism.

A modern version of the problem of the origin 
of ferromagnetism was formulated
by Kanamori \cite{Kanamori63}, Gutzwiller
\cite{Gutzwiller63}, and Hubbard \cite{Hubbard63} in the 
1960's\footnote{
A similar formulation was given earlier, for example, 
in \cite{Slater53}.
}.
They
studied simple tight-binding models of electrons with
on-site Coulomb interaction.
This model is usually called the
`Hubbard model.'
When there is no
electron-electron interaction,
the model 
exhibits paramagnetism as an inevitable consequence
of the Pauli exclusion principle. 
Among other things,
Kanamori, Gutzwiller, and Hubbard asked 
{\em whether the paramagnetism found for a non-interacting system can
be converted into ferromagnetism when there is a sufficiently
large Coulomb interaction.}
This is a concrete formulation of the fundamental problem
that we alluded to in the opening of the previous subsection.

It is worth noting that the on-site Coulomb interaction
itself is completely independent of electronic spins, and it does
not favor any magnetic ordering. 
One does not find any terms in the Hubbard Hamiltonian which 
explicitly favor ferromagnetism (or any other 
ordering).
Our theoretical goal is to show that magnetic ordering arises as a 
consequence of the subtle interplay between 
the kinetic motion of electrons and 
the short-ranged Coulomb interaction. 
It is interesting to compare this situation with that in spin systems, 
where one is often given a Hamiltonian which favors some kind of magnetic
ordering, and the major task is to investigate if such ordering really 
takes place.
We can say that the Hubbard model formulation goes deeper 
into fundamental mechanisms of 
magnetism than that of spin systems.
It offers a
challenging problem
to theoretical physicists to derive magnetic interaction from 
models which do not explicitly contain such interactions.

From a more global point of view, the importance of the Hubbard model
may be understood 
from the philosophy of ``universality'', which, in our 
opinion, is at the 
very heart of contemporary physics.
We believe that nontrivial physical phenomena or mechanisms found 
in a suitable 
idealized model can also be found in other systems in the same 
``universality class'' as the idealized model.
We expect that the universality class is 
often large and rich enough to contain 
various realistic systems with complicated 
details which are ignored in the idealized model.
As for strongly interacting electron systems, the Hubbard 
model is regarded as one of the most promising candidates for an 
idealized model to be used in our search 
of possible universality classes.

Perhaps the best justification of the Hubbard model as a standard model 
of itinerant electron systems comes from such theoretical 
considerations, rather than its accuracy in modeling narrow band electron 
systems. 

\subsection{Rigorous results}
\label{s:rigorous}
The problem of ferromagnetism in the Hubbard model
was studied extensively using various heuristic methods.
The Hartree-Fock approximation discussed above leads one to the
so called Stoner criterion.
It states that the Hubbard model exhibits ferromagnetism when the
condition \( UD_{\rm F}>1 \) is satisfied, where \( U \) is the 
the strength of the on-site Coulomb interaction and \( D_{\rm F} \) 
is the density of 
states of the corresponding single-electron problem measured
at the fermi level (of the corresponding non-interacting system).
Although the criterion cannot be trusted literally\footnote{
There are many systems which fully satisfy the criterion but do 
not exhibit ferromagnetism.
Flat-band Hubbard models with low electron densities \cite{92e,93d} are 
typical examples.
}, it guides us
to look for ferromagnetism in models in which \( U \) is 
large and/or
the density of states is large.

The first rigorous result about ferromagnetism in the Hubbard model,
which is one of the main topics of the present paper,
was provided by Nagaoka\footnote{
Thouless \cite{Thouless65} reached a similar but slightly weaker 
conclusion.
} \cite{Nagaoka66}.
It was proved that certain Hubbard models have ground states with
saturated ferromagnetism when there is exactly one hole and the
Coulomb repulsion $U$ is infinite.

In 1989, Lieb proved an important general theorem for the Hubbard model
on a bipartite lattice at half-filling \cite{Lieb89}.
As a corollary of this theorem, Lieb showed that a rather general class of 
Hubbard model exhibits ferrimagnetism\footnote{
Ferrimagnetism is a kind of antiferromagnetism on a bipartite lattice 
such that the numbers of sites in two sublattices are different.
}.
See also \cite{ShenQiu94}.

In 1991, Mielke \cite{Mielke91a,Mielke91b} 
found a new class of rigorous 
examples of ferromagnetism in the Hubbard model.
He showed that the Hubbard models on a general class of line 
graphs have ferromagnetic ground states.
A special feature of Mielke's models is that the corresponding 
single-electron Schr\"{o}dinger equation
has highly degenerate ground states. 
In other words, Mielke's models have flat (or dispersionless) bands.
Mielke' original results were for the case in which the number
of electrons 
corresponds to the half-filling of the lowest flat band, but later it was 
extended to different electron densities in two-dimensional 
models \cite{Mielke92}.

A similar but different class of examples of ferromagnetism in  
Hubbard models, which we 
will discuss in detail in the present paper, 
were proposed in \cite{92e,93d}.
These models are defined on a class of decorated lattices
with ``cell structures'', and are also 
characterized by flat bands at the bottom of the single-electron spectrum.
In a class of models in two and higher dimensions, it was proved that the 
ferromagnetism is stable against fluctuations in the electron number
\cite{92e,93d}.
The ferromagnetism in Mielke's models and those in \cite{92e,93d}
is now called ``flat-band ferromagnetism'', and regarded as one of 
reliable
starting points for the problem of ferromagnetism.

There are also rigorous results for ferromagnetism in
`nearly-flat-band models' obtained by perturbing the flat-band models
of \cite{92e,93d}.
In a general situation, local stability of ferromagnetism is known
\cite{94c,95b},
and for a special class of models, global stability of ferromagnetism 
has been established \cite{95c,97e}.
We shall also briefly discuss the latter results in the present paper.

Nagaoka's ferromagnetism takes place in models with \( U\to\infty \),
while (nearly-)flat-band ferromagnetism takes place in models 
characterized by 
large (or infinite) density of states \( D_{\rm F} \).
One might say that these two rigorous results realize the Stoner criterion
\( UD_{\rm F}>1 \) through complementary paths.

\subsection{About the present paper}
\label{s:about}
One of the main aims of the present paper is to give complete 
descriptions of Nagaoka's ferromagnetism and flat-band ferromagnetism
of \cite{92e,93d}.
In particular, the proof of the generalized version of
Nagaoka's theorem is 
described in complete detail for the first time.
Another important aim is to show that there is a close relation 
between these two approaches which are usually regarded as 
complementary.
We demonstrate this 
fact by studying an artificial model which we call the `long 
range hopping model.'
Our hope is that 
a clarification of the relation between the two approaches
will lead us to a more global view of 
ferromagnetism in the Hubbard model, and that this might lead us 
in the long run to a better 
understanding of the essence of the fascinating phenomenon of 
ferromagnetism.

We also discuss three topics related to these results.
First, we briefly discuss important attempts to go beyond flat-band 
ferromagnetism by treating non-singular Hubbard models.
One of the main achievements in this direction is a proof of the 
existence of ferromagnetism in a Hubbard model which has a finite 
Coulomb repulsion \( U \) and a finite density of states \( D_{\rm F} \).
Next, we give a preliminary discussion of 
some experimental results which may be relevant to flat-band 
ferromagnetism.
The ferromagnetism seen in La$_{4}$Ba$_{2}$Cu$_{2}$O$_{10}$,
 which was discovered back in 1990, shows 
some striking similarities with flat-band ferromagnetism.
Finally we describe some attempts to obtain theoretical examples of 
metallic ferromagnetism.
There are strong indications that some one-dimensional 
Hubbard models exhibit 
metallic ferromagnetism, but there are no rigorous results yet.
We discuss three artificial models 
in one-dimension which are easily shown to exhibit 
metallic ferromagnetism.
The results for the third model, the limiting \( U \)-\( V \) model, 
are perhaps new.

We have tried to make the present paper as self-contained as 
possible\footnote{
For related reviews of mathematically rigorous results 
in the Hubbard model, see \cite{Lieb95,97d}.
Unfortunately there are many interesting related 
topics that we do not even 
mention in the present paper.
For recent reviews of related topics from complementary points of 
view, see 
\cite{Fazekas96,Vollhardt97,HanischUhrigMuellerHartmann97,Shen97}. 
}.
We carefully start from basic assumptions and definition of the 
Hubbard model, keeping in mind readers who are new to the field.
We give complete proofs to all the 
theorems which are directly related to the main subjects of the paper.
The proofs, however, 
may not be optimally organized from a mathematical point of 
view.
Instead, we have tried to present ``readable'' proofs from which the 
readers can learn physical mechanisms underlying the theorems.
(This comment applies even to the proof of purely mathematical 
theorems such as Theorem~\ref{t:PF}.)

The present paper is organized as follows.
In Section~\ref{s:defHub}, we give a complete definition of the 
Hubbard model.
We start from the description of the single-electron  problem, and 
proceed by defining fermion operators, many-body Hilbert space, and 
the Hubbard Hamiltonian.
An expert can safely skip this entire section, provided that he/she 
takes a brief look at Sections~\ref{s:tb} and \ref{s:free} to note our 
notation in the coordinate-free formalism of fermion operators.

Section~\ref{s:mag} is also introductory and standard.
We define the spin angular momenta of the model, and give a precise 
definition (Definition~\ref{d:ferro}) of what we mean by 
ferromagnetism.
We also present some results which rule out the possibility of 
ferromagnetism under several conditions.

In Section~\ref{s:Nagaoka}, we present a complete description
and a proof of 
Nagaoka's ferromagnetism in its most generalized form.

In Section~\ref{s:lrh}, we introduce and discuss 
an artificial model that we call the `long-range 
hopping model.'
Ferromagnetism in this model is first regarded as a special case 
of Nagaoka's ferromagnetism, but a different point of view is presented.
This new picture naturally leads us to flat-band 
ferromagnetism.

In Section~\ref{s:flat}, we introduce flat-band ferromagnetism as a 
natural generalization of the long-range hopping model.
We also discuss some results for the nearly-flat-band models obtained by 
adding a perturbation to the flat-band models.

In Section~\ref{s:exp}, we focus on experimental results which may be 
relevant to flat-band ferromagnetism.

In Section~\ref{s:metallic}, we present some results about
the possibility 
of metallic ferromagnetism in the Hubbard model.
Preliminary rigorous results for related models in
one-dimension are discussed.

Technical material is contained in the Appendices.
In Appendices~\ref{s:gauge} and \ref{s:HP},
we discuss the gauge transformation and the
hole-particle transformation, 
respectively.
In Appendix~\ref{s:positivity}, we summarize some useful properties of 
positive semidefinite operators.
In Appendix~\ref{s:Hilb}, we present (mainly for mathematicians)
an explicit construction of the 
Hilbert space and fermion operators.
In Appendix~\ref{s:band}, we explain (for readers without 
background in condensed matter physics) the elementary
notion of band structures 
for a single-electron in a tight-binding description.
In Appendix~\ref{s:detail}, we present technical calculations required 
in the proof of Theorem~\ref{t:var2}.

\section{Definition of the Hubbard model}
\label{s:defHub}
\subsection{Tight-binding description of a single electron}
\label{s:tb}
\begin{figure}
\centerline{\epsfig{file=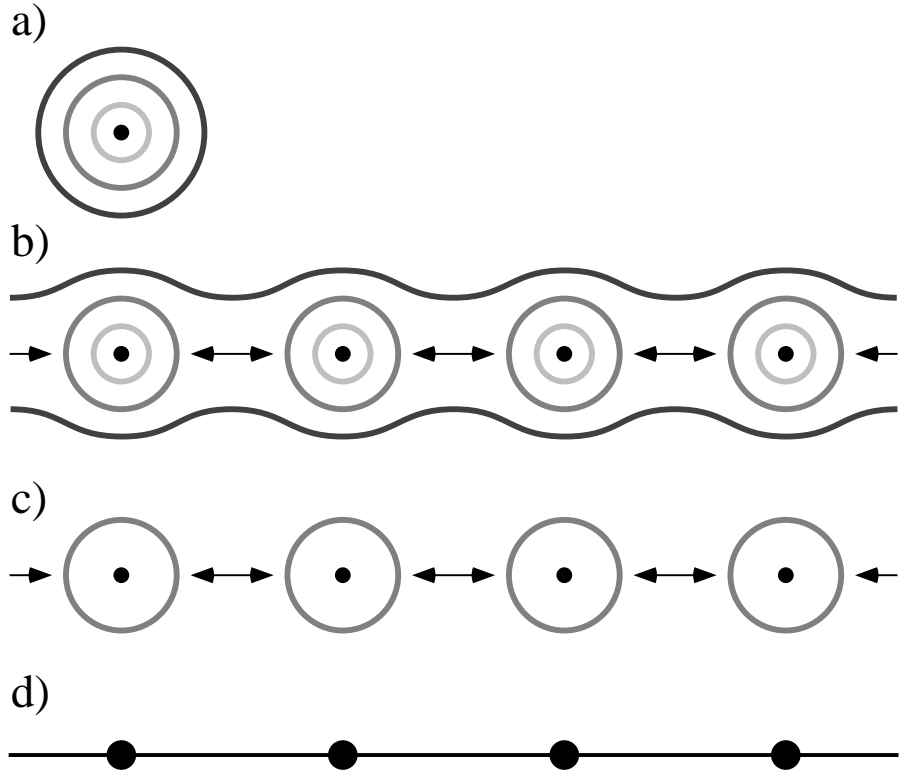,width=12cm}}
\caption[dummy]{
Highly schematic figures which explain the philosophy of tight-binding 
descriptions.
a)~A single atom which has multiple electrons in different orbits.
b)~When atoms come together to form a solid, electrons in the black 
orbits become itinerant, while those in the light gray orbits are 
still localized at the original atomic sites.
Electrons in the gray orbits are mostly localized around the atomic 
sites, but tunnel to nearby gray orbits with non-negligible 
probabilities.
c)~We only consider the electrons in the gray orbits, which are 
expected to play essential roles in determining various aspects of
low-energy physics of the system.
d)~If the gray orbit is non-degenerate, we get a lattice model in 
which electrons live on lattice sites and hop from one site to another.
}
\label{f:tb}
\end{figure}

Before introducing the Hubbard model, we describe the corresponding 
single-electron problem.
Let lattice \( \Lambda \) be a collection of \( \Ns \) sites.
Lattice sites \( x,y,\ldots\in\Lambda \) represent atomic sites in a solid.
In the tight-binding description, which is a kind of low-energy 
effective theory, we declare that {\em electrons can live only on 
lattice sites}.
In the {\em single-orbital model} that we study here, we further assume 
that each atomic site carries a single non-degenerate\footnote{
In some solids, the degeneracy in the original atomic orbit is lifted 
by crystalline anisotropy.
} orbital state.
Of course, actual atoms can have more than one 
orbit (or band).
The philosophy behind the model building is that those electrons in other 
states do not play significant roles in determining 
the low-energy physics in which we are 
interested, and can be ``forgotten'' for the moment.
See Figure~\ref{f:tb}.

Then, a quantum mechanical state of a single electron (with a fixed 
spin) is described by a vector state 
\( \vphi=(\phi_{x})_{x\in\Lambda} \) with 
\( \phi_{x}\in{\comp} \).
We denote by \( \hilb\cong\comp^{\Ns}(\cong\ell^{2}(\Lambda;\comp)) \) 
the \( \Ns \)-dimensional Hilbert space 
formed by such \( \vphi \).
The inner product of two states 
\( \vphi\) and \( \vpsi=(\psi_{x})_{x\in\Lambda}\in\hilb \) is
\begin{equation}
	\bkt{\vphi,\vpsi}=\sum_{x\in\Lambda}(\phi_{x})^{*}\psi_{x}.
	\label{phipsi}
\end{equation}

For \( x,y\in\Lambda \) with \( x\ne y \), we denote by 
\( t_{x,y} \) the quantum 
mechanical amplitude that an electron hops (i.e., tunnels) from 
the site $y$ to $x$.
We assume $t_{x,y}$ is real\footnote{
In a system under magnetic field, $t_{x,y}$ is  generally complex 
and satisfies $(t_{x,y})^{*}=t_{y,x}$.
} and symmetric as $t_{x,y}=t_{y,x}$.
For $x\in\Lambda$, we denote by $t_{x,x}$ the 
(real valued) potential energy\footnote{
We do not want to refer to \( V_{x}(=t_{x,x}) \) as 
``site-dependent chemical potential.''
The notion of the chemical potential makes sense in the 
contexts of thermodynamics or statistical physics.
This notion,
in our opinion, should be 
distinguished from the purely quantum mechanical notion of the potential 
(although there is a  relation between the two).
} (which we usually write \( V_{x} \))
for the electron at site $x$.
Then the \Sch\ equation for stationary states becomes
\begin{equation}
	\ep\phi_{x}=\sum_{y\in\Lambda}t_{x,y}\,\phi_{y},\quad
	\mbox{for any \( x\in\Lambda \)},
	\label{sSch2}
\end{equation}
where \(\ep\) is the energy eigenvalue\footnote{
It is also common to put a minus sign in front of \( t_{x,y} \) in 
(\ref{sSch2}).
We note that the hopping amplitude \( t_{x,y} \) 
depends on the delicate overlap between orbital wave functions, and 
there are no simple principles with which we can determine their 
magnitudes or signs.
Also see Appendix~\ref{s:gauge} for discussion of 
the arbitrariness of the sign of
\( t_{x,y} \).
}.
By introducing the hopping matrix
\( \mT=(t_{x,y})_{x,y\in\Lambda} \), which is a matrix on \( \hilb \),
this can be written in a coordinate free vector form as
\begin{equation}
	\ep\vphi=\mT\vphi.
	\label{Schv}
\end{equation}

Since \( \mT \) is real symmetric, the eigenvalue equation 
(\ref{Schv}) has \( \Ns \) eigenvalues \( \ep_{1},\ldots,\ep_{\Ns} \).
The corresponding eigenstate\footnote{
In notation like \( \psi^{(j)}_{x} \), the superscript \( j \) is the 
name of the state and the subscript \( x \) is the index for its 
component.
} 
\( \vpsi^{(j)}=(\psi^{(j)}_{x})_{x\in\Lambda} \) satisfies
\begin{equation}
	\ep_{j}\vpsi^{(j)}=\mT\vpsi^{(j)},
	\label{seig}
\end{equation}
for each 
\( j=1,\ldots,\Ns \).
We can assume that the energy eigenvalues are ordered as
\( \ep_{j}\le\ep_{j+1} \),
and eigenstates are orthonormal in the sense that
\( \bkt{\vpsi^{(j)},\vpsi^{(j')}}=\delta_{j,j'} \).

Let us consider a standard example.
We take the one-dimensional lattice 
\( \Lambda=\cbk{1,2,\ldots,\Ns} \),
and set
\( t_{x,x+1}=t_{x+1,x}=-t \)
for all \( x\in\Lambda \), and  
\( t_{x,y}=0 \) otherwise.
We impose a periodic boundary condition and identify \( \Ns+1 \) with 
\( 1 \).
The eigenvalues and eigenstates can be most naturally indexed 
as \( \veta^{(k)}=(\eta^{(k)}_{x})_{\xiL} \) and
\( \ep(k) \), respectively, by the 
wave number
\( k=2\pi n/\Ns \) where 
\( n=0,\pm1,\pm2,\ldots,\pm\cbk{(\Ns/2)-1},\Ns/2 \)
(assuming \( \Ns \) is even).
Then we have
\begin{equation}
	\eta^{(k)}_{x}=\frac{1}{\sqrt{\Ns}}e^{ikx},
	\label{veta}
\end{equation}
and
\( \ep(k)=-2t\cos k \).
Note that the eigenstates are described by plane waves.

\subsection{Fermion operators and Hilbert space}
\label{s:fo}
We define the Hilbert space for many-electron 
problems using the fermion operator formalism\footnote{
The formalism is also called ``second quantization.''
One should recall, however, that we are working with a many-body 
problem which is ``quantized'' only once.
The explicit relation to the ``first quantization'' formalism can be 
found in Appendix~\ref{s:Hilb}.
Readers with background in fields like 
functional analysis are suggested to 
take a look at this appendix.
}.
For each lattice site \( x\in\Lambda \) and spin index 
\( \sigma=\up,\dn \), 
we associate a fermion operator 
\( \axs \).
One can freely take the conjugate\footnote{
We denote the conjugate of an operator \( A \)
by \( A^{\dagger} \).
We write \( \cxs \) instead of \( (\axs)^{\dagger} \).
}, products, and linear combinations (with complex coefficients) of 
these operators (and the identity operator)
to get new operators.
We require that these operators satisfy  
the anticommutation relations\footnote{
The right-hand side of (\ref{ac1}) means 
\( \delta_{x,y}\delta_{\sigma,\tau} \) times the identity operator.
Throughout the 
present paper, we refrain from writing the identity operator explicitly.
}
\begin{equation}
\cbk{\cxs,\ayt}=\delta_{x,y}\delta_{\sigma,\tau},
\label{ac1}
\end{equation}
and
\begin{equation}
\cbk{\cxs,\cyt}=\cbk{\axs,\ayt}=0,
\label{ac2}
\end{equation}
for any \( x,y\in\Lambda \) and \( \sigma,\tau=\up,\dn \),
where $\cbk{A,B}=AB+BA$.
Note that (\ref{ac2}) implies \( (\axs)^{2}=(\cxs)^{2}=0 \).

Physically, \( \axs \) and \( \cxs \) are interpreted as the 
operators which respectively annihilate and create an electron at 
site \( x \) with spin \( \sigma \).
The corresponding number operator is defined as
\begin{equation}
	\nxs=\cxs\axs.
	\label{nxs}
\end{equation}
From (\ref{ac2}), we find that number operators
with different indices commute with 
each other.
From (\ref{ac1}) and (\ref{ac2}), we see that
\( (\nxs)^{2}=\cxs\axs\cxs\axs=
\cxs(1-\cxs\axs)\axs=\cxs\axs=\nxs \).
Thus we obtain
\( \nxs(1-\nxs)=0 \),
which means that 
\( \nxs \)
can only have eigenvalues 0 and 1.
This is a mathematical realization of the Pauli exclusion principle.
The number operator for site \( x \) is defined as
\( n_{x}=\nxu+\nxd \).

We can now construct the Hilbert space for many-electron problems.
We start from a single (vector) state \( \vac \),
which physically corresponds to a (fictitious) state with no 
electrons in the lattice.
This property is mathematically represented as
\begin{equation}
	\axs\vac=0
	\quad
	\mbox{for any \( x\in\Lambda \) and \( \sigma=\up,\dn \)}.
	\label{cvac=0}
\end{equation}

\begin{figure}
\centerline{\epsfig{file=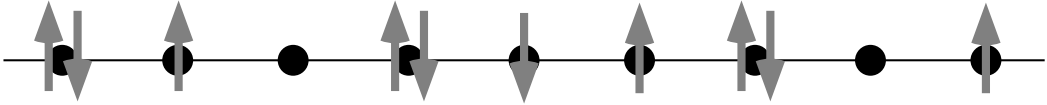,width=14cm}}
\caption[dummy]{
An allowed configuration in a tight-binding model with a single orbit 
per site.
Each site in the lattice can be either empty, singly 
occupied by an electron with up or down spin, or doubly occupied by 
electrons with opposite spins.
If we consider the Hubbard type interaction, we get an extra 
energy \( U>0 \)
whenever two electrons occupy a single site.
(See Section~\ref{s:int}.)
The interaction energy for the above configuration is therefore
\( 3U \).
}
\label{f:config}
\end{figure}

We fix a positive integer \( \Ne \)
such that
\( 1\le\Ne\le2\Ns \), which is the total number of 
electrons in the system.
It is useful to express the electron number in terms of the filling 
factor \( \nu=\Ne/(2\Ns) \), which takes a value in the range
\( 0<\nu\le1 \).
Take \( \Ne \) arbitrary sites 
\( x_{1},x_{2},\ldots,x_{\Ne}\in\Lambda \) 
(with possible overlaps) and
spin indices
\( \sigma_{1},\sigma_{2},\ldots,\sigma_{\Ne}=\up,\dn \),
and define the state
\begin{equation}
	\Phi_{x_{1},\sigma_{1};x_{2},\sigma_{2};\ldots,
	;x_{\Ne},\sigma_{\Ne}}
	=
	\cd_{x_{1},\sigma_{1}}\cd_{x_{2},\sigma_{2}}
	\cdots
	\cd_{x_{\Ne},\sigma_{\Ne}}
	\vac.
	\label{basis1}
\end{equation}
We interpret (\ref{basis1}) as the state in which there is an electron 
at site \( x_{i} \) with spin \( \sigma_{i} \) for 
\( i=1,\ldots,\Ne \).
We allow all the possible states of the form (\ref{basis1}), but make 
identifications according to the anticommutation relations (\ref{ac2}).
In particular, the state (\ref{basis1}) is vanishing whenever
\( (x_{i},\sigma_{i})=(x_{j},\sigma_{j}) \)
for some \( i\ne j \).
This is nothing but the Pauli exclusion principle\footnote{
If we set \( \Ne>2\Ns \), then the state (\ref{basis1}) always vanishes.
}.
In other words, each site in the lattice can be either empty, singly 
occupied by an electron with up or down spin, or doubly occupied by 
electrons with opposite spins (Figure~\ref{f:config}).
Also note that the anticommutation relation (\ref{ac2}) leads to  
relations like
\begin{equation}
	\Phi_{x_{1},\sigma_{1};x_{2},\sigma_{2}}
	=
	-\Phi_{x_{2},\sigma_{2};x_{1},\sigma_{1}},
	\label{antisym}
\end{equation}
which is the well-known rule that a state changes its sign when one 
exchanges the {\em names} of two fermions.

The Hilbert space \( \calH_{\Ne} \) for the given electron number 
\( \Ne \) is generated by all the basis states of the form 
(\ref{basis1}). 
The inner product\footnote{
We do not use Dirac notation since we encounter many non-hermitian 
operators.
Try rewriting the equation
\( \bkt{A\Phi,B\Psi}=\bkt{\Phi,A^{\dagger}B\Psi} \) 
in the Dirac notation.
}, which we again denote as \( \bkt{\cdot,\cdot} \),
is defined by \( \bkt{\vac,\vac}=1 \) and the anticommutation relations.
We denote the norm of a state \( \Phi\in\calH_{\Ne} \) as
\( \norm{\Phi}=\bkt{\Phi,\Phi}^{1/2} \).

We remark that there can be no {\em physical} situations which are 
represented by linear combinations of states with different electron 
numbers.
To see this, we take two states 
\( \Phi\in\calH_{\Ne} \) and 
\( \Phi'\in\calH_{\Ne'} \)
with \( \Ne\ne\Ne' \).
Since the total charge in the universe is conserved (as far as we know), 
the only (physically) 
possible way to linearly combine \( \Phi \) and
\( \Phi'\) is to form a state like
\( \Xi=\Phi\otimes\Gamma_{\Ne}
+ \Phi'\otimes\Gamma_{\Ne'} \).
Here 
\( \Gamma_{\Ne} \) and \( \Gamma_{\Ne'} \) 
are some states of the whole outside world with  total charge 
\( \Ne \) and \( \Ne' \), respectively.
Note that the states \( \Gamma_{\Ne} \) and \( \Gamma_{\Ne'} \) 
are inevitably orthogonal.
Therefore if \( A \) is any operator of the electron system,
its expectation value in \( \Xi \) becomes
\begin{eqnarray}
	\bkt{A}_{\Xi} & = & 
	\bkt{\Xi,(A\otimes{\bf 1}),\Xi}
	\ret
	&=&
	\bkt{\Phi,A\Phi}\bkt{\Gamma_{\Ne},\Gamma_{\Ne}}
	+\bkt{\Phi',A\Phi'}\bkt{\Gamma_{\Ne'},\Gamma_{\Ne'}}
	\ret
	&=&
	\omega\bkt{\Phi,A\Phi}+\omega'\bkt{\Phi',A\Phi'},
	\label{mixed}
\end{eqnarray}
where \( \bf 1  \) is the identity operator for the outside world.
This means that the state \( \Xi \) must be represented as a 
mixed state\footnote{
Note that the expectation value in the (unphysical) pure state 
\( \Xi'=\alpha\Phi+\alpha'\Phi' \) becomes
\( \bkt{A}_{\Xi'}=
|\alpha|^{2}\bkt{\Phi,A\Phi}+|\alpha'|^{2}\bkt{\Phi',A\Phi'}
+\{\alpha^{*}\alpha'\bkt{\Phi,A\Phi'}+{\rm (h.c.)}\} \),
which has extra cross terms.
} if we restrict ourselves to the electron 
system\footnote{
Although this remark might sound entirely trivial (and it indeed is), 
many beginners (and even some experts) are confused when they 
encounter BCS-type states which are linear combinations of states 
with different \( \Ne \).
One should always keep in mind that such linear combinations are 
introduced for purely mathematical
(or theoretical) reasons (which are indeed deep and 
beautiful) and have nothing to do with realistic particle number 
fluctuations.
When one has such a linear combination state, one {\em always} gets 
physically meaningful states with fixed electron numbers by taking  
suitable projections.
}.
\subsection{Coordinate free notation}
\label{s:free}
We introduce the standard coordinate-free notation for 
fermion operators.
The systematic use of this notation simplifies some 
routine calculations.
For a single-electron state 
\( \vphi=(\phi_{x})_{\xiL}\in\hilb \),
we define
\begin{equation}
	\Cds(\vphi)=\sum_{\xiL}\phi_{x}\,\cxs,
	\quad
	\Cs(\vphi)=\sum_{\xiL}(\phi_{x})^{*}\,\axs.
	\label{C}
\end{equation}
These operators satisfy the generalized canonical anticommutation 
relations
\begin{equation}
	\cbk{\Cs(\vphi),C^{\dagger}_{\sigma'}(\vpsi)}=
	\delta_{\sigma,\sigma'}\bkt{\vphi,\vpsi},
	\label{gac1}
\end{equation}
and
\begin{equation}
	\cbk{\Cs(\vphi),C_{\sigma'}(\vpsi)}=
	\cbk{\Cds(\vphi),C^{\dagger}_{\sigma'}(\vpsi)}=0,
	\label{gac2}
\end{equation}
for any \( \vphi,\vpsi\in\hilb \) and \( \sigma,\sigma'=\up,\dn \).

Again (\ref{gac2}) implies 
\( (\Cds(\vphi))^{2}=(\Cs(\vphi))^{2}=0 \).
We also have the following property.

\begin{lemma}[Linear independence and the product of 
creation operators]
	\label{l:CCC}
	Let\par\noindent
	\( \vphi^{(1)},\ldots,\vphi^{(n)} \) be \( n \) arbitrary states 
	in \( \hilb \).
	The state
	\begin{equation}
		\Cds(\vphi^{(1)})\cdots\Cds(\vphi^{(n)})\vac,
		\label{Opro}
	\end{equation}
	which is usually called the Slater determinant state
	(see Appendix~\ref{s:Hilb}),
	is nonvanishing if and only if 
	\( \vphi^{(1)},\ldots,\vphi^{(n)} \)
	are linearly independent.
\end{lemma}
\proof
Let \( O=\Cds(\vphi^{(1)})\cdots\Cds(\vphi^{(n)}) \).
Assume that the \( n \) states are linearly dependent.
Then we can write 
\( \vphi^{(1)}=\sum_{j=2}^{n}\alpha_{j}\vphi^{(j)} \)
with some coefficients \( \alpha_{j}\in\comp \), and hence
\( \Cds(\vphi^{(1)})=\sum_{j=2}^{n}\alpha_{j}\Cds(\vphi^{(j)}) \).
The product \( O \) vanishes because
\( (\Cds(\vphi^{(j)}))^{2}=0 \).

Next we 
assume that the \( n \) states are linearly independent and show 
that \( O\vac \) is nonvanishing.
By repeatedly using (\ref{gac1}), (\ref{gac2}) and \( \Cs(\vphi)\vac=0 \),
we get
\begin{eqnarray}
	\norm{O\vac}^{2} 
	& = & 
	\bkt{\vac,O^{\dagger}O\vac}
	\ret
	& = & 
	\bkt{\vac,
	\Cs(\vphi^{(n)})\cdots\Cs(\vphi^{(1)})
	\Cds(\vphi^{(1)})\cdots\Cds(\vphi^{(n)})
	\vac}
	\ret
	& = & 
	\sum_{p:(1,\ldots,n)\rightarrow(p(1),\ldots,p(n))}
	(-1)^{p}
	\bkt{\vphi^{(1)},\vphi^{(p(1))}}
	\cdots
	\bkt{\vphi^{(n)},\vphi^{(p(n))}},
	\label{OO1}
\end{eqnarray}
where \( p \) is summed over the \( n! \) permutations of 
\( (1,2,\ldots,n) \), and \( (-1)^{p} \) denotes the parity of \( p \).
The Gramm matrix \( G \) is an 
\( n\times n \) matrix defined by
\( (G)_{i,j}=\bkt{\vphi^{(i)},\vphi^{(j)}} \).
Then (\ref{OO1}) implies
\begin{equation}
	\norm{O\vac}^{2}=\det(G).
	\label{detG}
\end{equation}
It is a well-known fact in linear algebra that \( G \) is regular
(and hence \( \det(G)\ne0 \)) if 
and only if \( \vphi^{(1)},\ldots,\vphi^{(n)} \) are linearly 
independent.\qed

The following lemma gives a basic characterization of the Slater 
determinant states (\ref{Opro}).
Although we do not use the lemma in the present paper, we state it 
here since it is enlightening.
We encourage  interested readers to prove the lemma.

\begin{lemma}[Slater determinant depends on a subspace]
	\label{l:Slater}
	Suppose that two sets\par\noindent
	\( \{\vphi^{(1)},\ldots,\vphi^{(n)}\} \)
	and
	\( \{\vpsi^{(1)},\ldots,\vpsi^{(n)}\} \)
	of states in \( \hilb \) span the same \( n \)-dimensional subspace of 
	\( \hilb \).
	Then there is a nonvanishing constant \( c\in\comp \), and we have
	\begin{equation}
		\Cds(\vphi^{(1)})\cdots\Cds(\vphi^{(n)})\vac
		=c\,
		\Cds(\vpsi^{(1)})\cdots\Cds(\vpsi^{(n)})\vac.
		\label{Slater}
	\end{equation}
\end{lemma}

The following lemma is elementary, but  of fundamental importance.

\begin{lemma}[General basis of \( \calH_{\Ne} \)]
	\label{l:basis}
	Let \( \vphi^{(1)},\ldots,\vphi^{(\Ns)} \) be \( \Ns \) arbitrary 
	states in 
	\( \hilb \) which are linearly independent (but not necessarily 
	orthogonal with each other).
	Then the states
\begin{equation}
	\Gamma_{\Sup,\Sdn}
	=
	\rbk{\prod_{j\in\Sup}C^{\dagger}_{\up}(\vphi^{(j)})}
	\rbk{\prod_{j\in\Sdn}C^{\dagger}_{\dn}(\vphi^{(j)})}
	\vac	
	\label{Gamma}
\end{equation}	
	for arbitrary subsets \( \Sup,\Sdn \) of \( \{1,2,\ldots,\Ns\} \)
	such that\footnote{
	Throughout the present paper, we denote by \( |S| \) 
	the number of 
	elements in the set \( S \).
	} \( |\Sup|+|\Sdn|=\Ne \) span the Hilbert space 
	\( \calH_{\Ne} \).
\end{lemma}

\proof
Since the \( \Ns \) vectors are linearly independent, we have
\( \delta_{x,y}=\sum_{j=1}^{\Ns}a_{x,j}\phi^{(j)}_{y} \)
with some regular matrix 
\( (a_{x,j})_{\xiL,j=1,\ldots,\Ns} \).
Thus we have
\begin{equation}
	\cxs=\sum_{j=1}^{\Ns}a_{x,j}\Cds(\phi^{(j)}),
	\label{inv}
\end{equation}
which means that the basis state (\ref{basis1}) can be written as a 
linear combination of the states (\ref{Gamma}).\qed

Let \( \mA=(a_{x,y})_{x,y\in\Lambda} \) be a matrix on \( \hilb \).
We define the corresponding bilinear form of fermion operators
\begin{equation}
	B(\mA)=\sumtwo{x,y\in\Lambda}{\sigma=\up,\dn}
	\cxs\,a_{x,y}\,\ays,
	\label{B(A)}
\end{equation}
which is usually called the ``second quantization'' of \( \mA \).
The easily verified commutation relation
\begin{equation}
	\sqbk{B(\mA),\Cds(\vphi)}=\Cds(\mA\vphi),
	\label{cAcC}
\end{equation}
where \( [A,B]=AB-BA \),
sheds light on the relation between the single-particle quantum mechanics 
and the operator formalism.
This will be useful later.
A similar commutation relation for two bilinear forms
\begin{equation}
	\sqbk{B(\mA),B(\mB)}=B([\mA,\mB])
	\label{BABB1}
\end{equation}
is also easy to prove.

We have so far treated the spin index \( \sigma \) separately from 
the coordinate \( x \).
It is of course possible to treat the combination \( (x,\sigma) \) as
the coordinates of the system and consider the corresponding 
\( 2\Ns \)-dimensional single-electron Hilbert space\footnote{
We shall use such a formalism in Appendix~\ref{s:Hilb}.
}.
We can then restate all of the above results in the new language
with suitable modifications.
Let us only mention the relation corresponding to (\ref{BABB1}), 
because it will be useful later.
Let \( \tilde{\mA} \) be a \( 2\Ns\times 2\Ns \) 
 matrix indexed by \( (x,\sigma) \) with \( x\in\La \) and 
\( \sigma=\up,\dn \), and define
\( \tilde{B}(\tilde{\mA})=\sum_{x,y\in\La}\sum_{\sigma,\tau=\up,\dn}
\cxs(\tilde{\mA})_{(x,\sigma),(y,\tau)}\,c_{y,\tau} \).
Then for two such matrices \( \tilde{\mA} \) and 
\( \tilde{\mB} \), we have
\begin{equation}
	[\tilde{B}(\tilde{\mA}),\tilde{B}(\tilde{\mB})]
	=
	\tilde{B}([\tilde{\mA},\tilde{\mB}]),
	\label{BABB}
\end{equation}
where \( [\tilde{\mA},\tilde{\mB}] \) is the commutator as 
\( 2\Ns\times 2\Ns \) matrices.

\subsection{Non-interacting system}
\label{s:nonint}
Now that the Hilbert space has been prepared, 
we can introduce Hamiltonians.
The Hamiltonian which describes the quantum mechanical hopping with 
the hopping matrix \( \mT=(t_{x,y})_{x,y\in\La} \) is 
\begin{equation}
	\Hhop=
	B(\mT)=
	\sumtwo{x,y\in\Lambda}{\sigma=\up,\dn}
	t_{x,y}\,\cxs\ays.
	\label{Hhop}
\end{equation}
To see that (\ref{Hhop}) describes the desired hopping, 
define the single-electron state 
\( \Phi_{\vphi}=\Cds(\vphi)\vac \)
for \( \vphi\in\hilb \).
Then from the commutation relation (\ref{cAcC}), we
immediately find that the \Sch\ equation
\( \ep\Phi_{\vphi}=\Hhop\Phi_{\vphi} \)
is equivalent to the single-electron \Sch\ equation (\ref{sSch2})
or (\ref{Schv}).

When one has diagonalized the single-electron \Sch\ equation 
(\ref{sSch2}), it is easy to also diagonalize the many-body 
Hamiltonian (\ref{Hhop}).
As in Section~\ref{s:tb}, we let 
\( \vpsi^{(j)} \) 
be the eigenstate satisfying (\ref{seig}) with energy 
\( \ep_{j} \), for
\( j=1,2,\ldots,\Ns \).
Define the new fermion operator by
\begin{equation}
	\adjs=\Cds(\vpsi^{(j)}),
	\label{aj}
\end{equation}
for \( j=1,2,\ldots,\Ns \) and \( \sigma=\up,\dn \).

Let \( \Sup \) and \( \Sdn \) be arbitrary subsets of 
\( \cbk{1,2,\ldots,\Ns} \)
such that
\( |\Sup|+|\Sdn|=\Ne \), and define
\begin{equation}
	\Psi_{\Sup,\Sdn}
	=
	\rbk{\prod_{j\in\Sup}\ad_{j,\up}}
	\rbk{\prod_{j\in\Sdn}\ad_{j,\dn}}
	\vac.
	\label{PsiSS}
\end{equation}
Lemma~\ref{l:basis} shows that these states form a basis of 
\( \calH_{\Ne} \).
Note that the commutation relation (\ref{cAcC}) and
the eigenstate equation (\ref{seig}) imply
\( [\Hhop,\adjs]=[B(\mT),\Cds(\vpsi^{(j)})]=
\Cds[\mT\vpsi^{(j)}]=\ep_{j}\adjs \).
This and \( \Hhop\vac=0 \) immediately imply
\begin{equation}
	\Hhop\Psi_{\Sup,\Sdn}
	=\rbk{\sum_{j\in\Sup}\ep_{j}+\sum_{j\in\Sdn}\ep_{j}}
	\Psi_{\Sup,\Sdn}.
	\label{Hhope}
\end{equation}
We have thus diagonalized \( \Hhop \).

\begin{figure}
\centerline{\epsfig{file=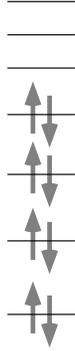,width=1cm}}
\caption[dummy]{
Schematic picture of the ground state of a non-interacting 
many-electron system.
The lowest \( \Ne/2 \) single-electron energy levels are ``filled'' 
by both up spin and down spin electrons.
The state naturally exhibits paramagnetism known as `Pauli 
paramagnetism.'
}
\label{f:pauli}
\end{figure}

By choosing subsets \( \Sup,\Sdn \) which minimize 
the energy eigenvalue 
\( \sum_{j\in\Sup}\ep_{j}+\sum_{j\in\Sdn}\ep_{j} \), 
we get ground state(s) 
of the present non-interacting model.
In particular, if the corresponding single-electron energy 
eigenvalues are 
nondegenerate, i.e., $\ep_{j}<\ep_{j+1}$, and 
if $\Ne$ is even, the ground 
state of $\Hhop$ is unique and written as
\begin{equation}
\GS=\rbk{\prod_{j=1}^{\Ne/2}a^\dagger_{j,\up}a^\dagger_{j,\dn}}\vac.
\label{GS1}
\end{equation}
This is nothing but the state obtained by ``filling up'' the low 
energy levels with up and down spin electrons, as one learns in 
elementary quantum mechanics (Figure~\ref{f:pauli}).

In a single-electron eigenstate of the 
one-dimensional example in Section~\ref{s:tb}, the 
electron is in a plane wave state (\ref{veta}) 
with a definite wave number  $k$. 
The same is true for any translation invariant
model, as we see in Appendix~\ref{s:band}.
The fact that the Hamiltonian $\Hhop$ is diagonalized in the basis
(\ref{PsiSS})  
implies that the electrons behave as ``waves'' in this non-interacting 
(Hubbard) model.

\subsection{Interaction}
\label{s:int}
It is believed that interactions between electrons in a solid mainly 
come from the electrostatic Coulomb force.
Although the Coulomb force in vacuum is long-ranged, we consider 
{\em an extremely short-range interaction which acts only when two 
electrons occupy the same site (i.e., the same atomic orbit)}.
A crude justification of such a short-range (Coulomb) interaction 
comes from the observation that the Coulomb force should be most 
dominant when two electrons approach within the minimum possible distance.
In a slightly more sophisticated justification, one argues that 
the long range Coulomb force is screened by 
electrons in different orbital states which we have decided to forget.
Our point of view, however, is that models with artificial 
short-range interactions are worth studying because they are
among  {\em the
minimum models which can be studied to elicit universal 
properties of strongly interacting electron systems}.

The Hamiltonian for the short-range Coulomb repulsion is 
\begin{equation}
	\Hint=U\sum_{x\in\Lambda}\nxu\nxd,
	\label{Hint}
\end{equation}
where \( U\ge0 \) is the energy from the repulsive interaction.
See Figure~\ref{f:config}.

The expression (\ref{Hint}) is already diagonalized, and it 
is trivial to write down eigenstates of \( \Hint \).
Let \( L_{\up},L_{\dn} \) be arbitrary subsets of the lattice 
\( \Lambda \) such that
\( |L_{\up}|+|L_{\dn}|=\Ne \), and define
\begin{equation}
	\Psi_{L_{\up},L_{\dn}}
	=
	\rbk{\prod_{x\in L_{\up}}\cxu}
	\rbk{\prod_{x\in L_{\dn}}\cxd}
	\vac.
	\label{PsiLL}
\end{equation}
Then by using the commutation relation
\( [\nxs,\cd_{x',\sigma'}]
=\delta_{x,x'}\delta_{\sigma,\sigma'}\cxs \),
we find
\begin{equation}
	\Hint\Psi_{L_{\up},L_{\dn}}
	=
	U|L_{\up}\cap L_{\dn}|\Psi_{L_{\up},L_{\dn}}.
	\label{Hinte}
\end{equation}
The ground state for a given electron number $\Ne$ can be constructed by 
choosing subsets 
\( L_{\up},L_{\dn} \) that minimize the energy
eigenvalue
\( U|L_{\up}\cap L_{\dn}| \).
When $\Ne\le\Ns$, one can always choose  \( L_{\up} \)
and \( L_{\dn} \) so that 
\( L_{\up}\cap L_{\dn}=\emptyset \) holds.
In this case, the ground states are highly degenerate and have
energy equal to  $0$.

It is clear (from the beginning)
that the interaction Hamiltonian \( \Hint \) is most 
naturally treated if we regard electrons as 
(classical) ``particles'' which live on lattice sites.

\subsection{Hubbard model}
\label{s:Hub}
The Hubbard model describes a tight-binding electron model in which 
{\em electrons hop around the lattice and 
interact with each other through short-range repulsive interactions}.
The full Hamiltonian of the single-orbital Hubbard model\footnote{
This model is often called the single-band Hubbard model.
We find this terminology confusing since a single-orbital model can 
have multiple bands depending on the structure of the hopping matrix 
\( \mT \).
See Appendix~\ref{s:band}.
} is simply
\begin{equation}
	H=\Hhop+\Hint.
	\label{H}
\end{equation}

We have already seen 
that both $\Hhop$ and  $\Hint$ can be easily diagonalized.
We have observed, however, that electrons behave as ``waves'' in $\Hhop$, 
while they behave as ``particles'' in $\Hint$.
How do they behave in a system whose Hamiltonian is a sum of 
these totally different Hamiltonians?
This is indeed a fascinating problem which is deeply related to the 
wave-particle dualism in quantum physics.
We might say that many of the important models in many-body problems, 
including the $\phi^4$ quantum field theory and the Kondo problem, 
are minimum models which take 
into account both the wave-like nature and the particle-like 
nature (through point-like nonlinear interactions) of matter.

 From a technical point of view, the wave-particle dualism implies that 
the Hamiltonians $\Hint$ are $\Hhop$ do not commute with each other.
Even when each Hamiltonian is diagonalized, it is still highly nontrivial 
(or impossible) to find the properties of their sum.
Of course, mathematical difficulty does not automatically guarantee that the 
model is worth studying.
A truly exciting characteristic of the Hubbard model is that, though the 
Hamiltonians $\Hhop$ and $\Hint$ do not favor any nontrivial order, their 
sum  $H=\Hhop+\Hint$ is believed to generate various types of 
nontrivial order 
including antiferromagnetism, ferromagnetism, and superconductivity.
When we sum up the two innocent Hamiltonians $\Hhop$ and $\Hint$, 
competition between their wave-like and particle-like characters (or 
between linearity and nonlinearity) takes place, and one gets various 
interesting ``physics''.
To confirm this fascinating scenario is a challenging problem in 
theoretical and mathematical physics.

\section{Magnetism in the Hubbard model}
\label{s:mag}
\subsection{Spin operators and $SU(2)$ invariance}
\label{s:su2}
We introduce the total spin operators\footnote{
We use the index \( \alpha=1,2,3 \) 
to indicate the three axes, because the 
symbols \( x,y,z \) are already used as lattice sites.
} 
\( \Sopt=(\Sopct{1},\Sopct{2},\Sopct{3}) \)
of the system by
\begin{equation}
	\Sopct{\alpha}=
	\frac{1}{2}\sum_{\xiL}\ \sum_{\sigma,\tau=\up,\dn}
	\cxs(p^{(\alpha)})_{\sigma,\tau}\,\axs,
	\label{Stot}
\end{equation}
for  $\alpha=1,2$, and $3$, where $p^{(\alpha)}$ are the Pauli matrices.
It is clear from the general commutation relation 
(\ref{BABB}) that\footnote{
Note that
\( \Sopct{\alpha}=\tilde{B}(\tilde{p}^{(\alpha)}) \)
with
\( (\tilde{p}^{(\alpha)})_{(x,\sigma),(y,\tau)}
=\delta_{x,y}(p^{(\alpha)})_{\sigma,\tau} \).
} 
\( \Sopct{1},\Sopct{2},\Sopct{3} \)
satisfy the standard commutation relation for 
quantum mechanical angular momentum 
operators.
We define the raising and lowering operators by
\( S^{\pm}_{\rm tot}=\Sopct{1}\pm i\Sopct{2} \).

The operators \( \Sopt \) are the generators of the global \( SU(2) \)
rotations in the spin space.
We say an operator \( A \) is \( SU(2) \) invariant if it
commutes with \( \Sopct{\alpha} \) for 
\( \alpha=1,2 \), and 3.
Intuitively speaking, an \( SU(2) \) invariant operator does not 
change if we change the ``quantization axis'' of spins in an arbitrary 
manner.

A typical \( SU(2) \) invariant operator is the number operator
\( n_{x}=\nxu+\nxd \).
The hopping Hamiltonian \( \Hhop \) of (\ref{Hhop}) 
is also \( SU(2) \)
invariant. These facts can be checked easily by using
the general commutation relations
(\ref{BABB}). The interaction Hamiltonian \( \Hint \) of
(\ref{Hint}) is also
\( SU(2) \)  invariant. 
This fact becomes evident if we note
\( \nxu\nxd=\{(n_{x})^{2}-n_{x}\}/2 \).

Knowing that the Hamiltonian
\( H=\Hhop+\Hint \)
is \( SU(2) \) invariant, we can make use of the standard 
representation theory of quantum mechanical angular momenta.
We note that the three operators \( H \), \( \Sopct{3} \),
and
\( (\Sopt)^{2}=\sum_{\alpha=1,2,3}(\Sopct{\alpha})^{2} \)
commute with each other.
We will always consider simultaneous 
eigenstates of these operators.
We denote by \( \Sztot \) the eigenvalue of \( \Sopct{3} \),
and by \( \Stot(\Stot+1) \) the eigenvalue of 
\( (\Sopt)^{2} \).
We call \( \Stot \) the total spin of the state.
Let 
\begin{equation}
	\Smax=\cases{
	\Ne/2&if \( \Ne\le\Ns \);\cr
	(2\Ns-\Ne)/2&if \( \Ne\ge\Ns \).}
	\label{Smax}
\end{equation}
Then the allowed values of \( \Stot \) are
\( \Stot=0,1,\ldots,\Smax  \) if \( \Ne \) is even, and
\( \Stot=1/2,3/2,\ldots,\Smax \) if \( \Ne \) is odd.

\subsection{Ferromagnetism}
\label{s:ferro}
Ferromagnetism is probably the most intuitive among 
various magnetic phenomena exhibited by solids.
Here we shall be rather restrictive and only consider the strongest 
form of ferromagnetism, namely, saturated ferromagnetism in ground 
states.

\begin{definition}
\label{d:ferro}
A Hubbard model is said to exhibit ferromagnetism if any ground 
state of \( H \) has the total spin \( \Stot=\Smax \).
\end{definition}

Recall that any state with \( \Stot=\Smax \) has its copies in the 
subspaces with \( \Sztot=-\Smax,-\Smax+1,\ldots,\Smax \).
Therefore the ground states are at least 
\( (2\Smax+1) \)-fold degenerate when there is ferromagnetism (in 
the above strong sense).
Let \( \GS \) be a ground state with \( \Sztot=\Smax \).
Since \( \GS \) contains only up spin electrons, we must have
\( \Hint\GS=0 \).
This means that \( \GS \) is the lowest energy state of \( \Hhop \) 
within the subspace of \( \Ne \) up spin electrons.
Recalling the  
general 
eigenstates (\ref{PsiSS}) of \( \Hhop \), 
we find that the desired ferromagnetic 
ground state is
\begin{equation}
	\GS
	=
	\rbk{\prod_{j=1}^{\Ne}\ad_{j,\up}}\vac,
	\label{ferroGS}
\end{equation}
with \( \ad_{j,\up} \) defined in (\ref{aj}), which has
the ground state energy
\begin{equation}
	E_{\rm ferro}=\sum_{j=1}^{\Ne}\ep_{j}.
	\label{Eferro}
\end{equation}
When we have \( \ep_{\Ne}<\ep_{\Ne+1} \), (\ref{ferroGS}) is 
the unique ground state\footnote{
When \( \ep_{\Ne} \) is degenerate with 
\( \ep_{n},\ldots,\ep_{m} \)
(where \( n\le\Ne<m \)), we can replace the states 
\( j=n,\ldots,\Ne \) with arbitrary states from 
\( j=n\ldots,m \) to get other ground states.
The ground states are degenerate even in the subspace with 
\( \Sztot=\Smax \).
} in the subspace with 
\( \Sztot=\Smax \).
The ground state \( \Phi_{M} \) with
\( \Sztot=M \) can be obtained from this ground 
state by the standard relation
\begin{equation}
	\Phi_{M}
	=
	\frac{
	(S^{-}_{\rm tot})^{\Smax-M}\GS
	}{
	\norm{(S^{-}_{\rm tot})^{\Smax-M}\GS}.
	}
	\label{PhiM}
\end{equation}

The above elementary construction of the ground states (provided that 
the model exhibits ferromagnetism) is based on the fact that any 
state with \( \Stot=\Smax \) {\em does not feel} the on-site Coulomb 
repulsion.
This reduces the problem to that of non-interacting spinless fermions.
We stress that this is a very special feature of the Hubbard model, 
which has only on-site interactions.

\subsection{Instability of ferromagnetism}
\label{s:noferro}
To see that ferromagnetism is indeed a delicate phenomenon, we  
discuss some results which show that the Hubbard model with certain 
conditions does {\em not} exhibit ferromagnetism.

We first look at the non-interacting model 
with the Hamiltonian \( H=\Hhop \) that we studied in
  Section~\ref{s:nonint}.
Assuming that the single-electron energy eigenvalues are 
non-degenerate, we found that the ground state is uniquely given by 
(\ref{GS1}).
By noting that the operator \( \ad_{j,\up}\ad_{j,\dn} \) is
\( SU(2) \) invariant and \( \Sopct{\alpha}\vac=0 \), we immediately 
find 
\( \Sopct{\alpha}\GS=0 \).
Thus the ground state has the total spin\footnote{
That \( \Stot=0 \) also follows from the uniqueness of the ground state.
} \( \Stot=0 \), and there is no ferromagnetism.
This is nothing but the well-known Pauli paramagnetism.
(See Figure~\ref{f:pauli}.)

It is also trivial that the non-hopping model with 
\( H=\Hint \) we studied in Section~\ref{s:int} does not exhibit 
ferromagnetism.
In this case any state (\ref{PsiLL}) for \( \Lup,\Ldn \) with minimum 
\( |\Lup\cap\Ldn| \) is a ground state.
Thus the 
ground states are highly degenerate and exhibit a kind of 
paramagnetism.

Therefore neither \( \Hhop \) nor \( \Hint \) alone favor 
ferromagnetism.
As we have stressed in Section~\ref{s:Hub}, our hope is that, when 
these two Hamiltonians are added into a single Hubbard Hamiltonian 
\( H=\Hhop+\Hint \), their ``competition'' generates totally new 
phenomena, including ferromagnetism.

Next we consider the situation in which the Coulomb interaction \( U \) 
is finite but small.
We find here that there cannot be (saturated) ferromagnetism.

\begin{theorem}[Impossibility of ferromagnetism for small \( U \)]
\label{t:var1}
Suppose
\( 0\le U<\ep_{\Ne}-\ep_{1} \).
Then the ground state of the Hubbard model does not have
\( \Stot=\Smax \), i.e., the model does not exhibit ferromagnetism.
Note that the fermi energy \( \ep_{\Ne}-\ep_{1} \) is usually 
independent of the system size for a fixed filling factor \( \nu \).
\end{theorem}
\proof
One of the lowest energy states with \( \Stot=\Smax \) is given by 
(\ref{ferroGS}) and has the energy (\ref{Eferro}).
Consider a normalized trial state
\begin{equation}
	\Psi
	=
	\ad_{1,\dn}\rbk{\prod_{j=1}^{\Ne-1}\ad_{j,\up}}\vac,
	\label{var1}
\end{equation}
which is obtained from (\ref{ferroGS}) 
by removing the up  spin electron with the highest energy 
and then adding a down spin electron with the lowest energy.
Noting the \( SU(2) \) invariance of  \( \ad_{1,\dn}\ad_{1,\up} \),
one finds that \( \Psi \) has \( \Stot=\Smax-1 \).
We want to evaluate the energy expectation value 
\( \bkt{\Psi,H\Psi} \).
For the kinetic energy, we have
\( \bkt{\Psi,\Hhop\Psi}=\ep_{1}+\sum_{j=1}^{\Ne-1}\ep_{j} \).
As for \( \Hint \), we note that the inequality\footnote{
See Definition~\ref{d:pos} for the definition of inequalities for 
operators.
} \( n_{x,\up}\le1 \) implies 
\( \Hint\le U\sum_{x\in\Lambda}n_{x,\dn} \) to get
\( \bkt{\Psi,\Hint\Psi}\le
\bkt{\Psi,U\sum_{x\in\Lambda}n_{x,\dn}\Psi}=U \).
Therefore we have
\begin{equation}
	\bkt{\Psi,H\Psi}-E_{\rm ferro}
	\le
	\ep_{1}-\ep_{\Ne}+U
	<0,
\end{equation}
where \( E_{\rm ferro} \) is defined in (\ref{Eferro}), and
the final bound follows from the condition of the theorem.
From the variational principle, we see that \( E_{\rm ferro} \) is not 
the ground state energy of \( H \).\qed

Although the above theorem ensures that the ground state cannot be 
ferromagnetic, it does not provide any information about the nature 
of the true ground state of the model.
To study the latter 
explicitly is in general a very difficult problem which (for 
the moment) is possible only in the simplest one-dimensional model.

Finally, we discuss the situation in which the interaction may be large 
but the density of electrons is very low.
It is expected that the chance of electrons to collide with each other 
in this case becomes very small.
It is likely that the model is close to an ideal gas, and there is no 
ferromagnetism.

This naive guess is easily justified for ``healthy'' models in 
dimensions three (or higher).
The dimensionality of the lattice is taken into account by assuming 
that there are positive constants\footnote{
\( n_{0} \) and \( d \) are the degeneracy of the single-electron 
ground states and the dimension of the system, respectively. 
} \( c \), 
\( n_{0} \), \( \rho_{0} \), and 
\( d \), and the single electron energy eigenvalues satisfy
\begin{equation}
	\ep_{n}-\ep_{1}\ge c\rbk{\frac{n-n_{0}}{\Ns}}^{2/d},
	\label{healthy}
\end{equation}
for any \( n \) such that 
\( n/\Ns\le\rho_{0} \).
Note that the right-hand side represents the \( n \) dependence of 
energy levels in a usual \( d \)-dimensional quantum mechanical 
system.
Then we have the following theorem due to
Pieri, Daul, Baeriswyl, Dzierzawa, and Fazekas \cite{Pieri97}.

\begin{theorem}[Impossibility of ferromagnetism at low densities]
	\label{t:var2}
	Take a \( \Hhop \) which has translation invariance
	(as in Appendix~\ref{s:band}) 
	and satisfies (\ref{healthy}) with positive 
	\( c \), \( n_{0} \), \( \rho_{0} \), and \( d>2 \).
	Then there exists a constant \( \rho_{1}>0 \), and the corresponding 
	Hubbard model does not exhibit ferromagnetism for any \( U\ge0 \)
	if \( \Ne/\Ns\le\rho_{1} \) holds.
\end{theorem}

\proofof{Outline of proof}
The naive trial state (\ref{var1}) does not work for large \( U \).
We follow \cite{Shastry90}, and consider the Roth state \cite{Roth67}
\begin{equation}
	\tPsi
	=
	P_{0}\Psi,
	\label{var2}
\end{equation}
where \( \Psi \) is defined in (\ref{var1}), and
\begin{equation}
	P_{0}
	=
	\prod_{x\in\Lambda}(1-\nxu\nxd)
	\label{PG}
\end{equation}
is the orthogonal projection (called the `Gutzwiller projection') onto 
the space with no doubly occupied sites.
Because of the projection, the state (\ref{var2}) minimizes the Coulomb 
interaction as
\( \Hint\Psi=0 \).
Thus we only need to evaluate the expectation value of \( \Hhop \).
After a tedious but straightforward calculation
whose details can be found in Appendix~\ref{s:detail}, we find
\begin{equation}
	\frac{\bkt{\tPsi,H\tPsi}}{\bkt{\tPsi,\tPsi}}
	-E_{\rm ferro}
	\le
	\ep_{1}-\ep_{\Ne}+c'\rho,
	\label{varesti}
\end{equation}
where \( \rho=\Ne/\Ns \) is the electron density and 
\( c'>0 \) is a constant.
From the assumption (\ref{healthy}), we find that the right-hand side 
becomes strictly negative for sufficiently small \( \rho \) provided 
that \( d>2 \).\qed

That we have a restriction on dimensionality in Theorem~\ref{t:var2} 
is not merely technical.
In a one-dimensional system, moving electrons must eventually collide 
with each other for an obvious geometric reason.
Thus a one-dimensional model cannot be regarded as close to ideal no 
matter how low the electron density is.
We do not know whether the inapplicability of the theorem to
two-dimensional systems is physically meaningful or not.

\subsection{Two more theorems for the absence of ferromagnetism}
\label{s:theorems}

We briefly discuss (without proofs) 
two strong theorems which also rule out 
ferromagnetism.

The classical Lieb-Mattis theorem \cite{LiebMattis62} states (among other 
things) that one can never have ferromagnetism in the one-dimensional 
Hubbard model
with only nearest neighbor hoppings\footnote{
The present theorem appears in the Appendix of \cite{LiebMattis62}.
The main body of \cite{LiebMattis62} treats interacting electron 
systems in continuous spaces.
}.
One-dimensional Hubbard models with next nearest neighbor hoppings may 
exhibit ferromagnetism as we explicitly see in Sections~\ref{s:flat} and 
\ref{s:metallic}.

\begin{theorem}[Lieb-Mattis theorem]
\label{t:LiebMattis}
Consider a Hubbard model with even \( \Ne \)
on a one-dimensional lattice $\Lambda=\cbk{1,2,\ldots,\Ns}$ with 
open boundary conditions.
We assume that the hopping matrix elements satisfy 
$|t_{x,y}|<\infty$ when 
$x=y$,  $0<|t_{x,y}|<\infty$ when $|x-y|=1$, and are vanishing 
otherwise.
Then for any real \( U \), the ground state of the model is unique 
and has \( \Stot=0 \).
\end{theorem}

The next important theorem is due to Lieb \cite{Lieb89}.

\begin{theorem}[Lieb's theorem, special case]
\label{t:Lieb}
Suppose that the lattice \( \Lambda \) is decomposed into two 
sublattices as \( \Lambda=A\cup B \) with \( |A|=|B| \), and we have
\( t_{x,y}=0 \) when \( x,y\in A \) or \( x,y\in B \).
We also assume that the entire lattice is connected by 
nonvanishing \( t_{x,y} \).
When \( \Ne=\Ns \), the ground state of the model is unique 
and has \( \Stot=0 \) for any \( U\ge0 \). 
\end{theorem}

The electron number \( \Ne=\Ns \) is usually referred to as 
``half-filling'', since \( 2\Ns \) is the maximum possible number 
for \( \Ne \).
The low energy properties of a Hubbard model at half-filling is believed 
to be described by the antiferromagnetic Heisenberg model.
Lieb's theorem stated above gives a partial justification to this belief.
This theorem applies to  models on lattices with 
\( |A|\ne|B| \) as well.
In this case, it implies the existence of {\em ferrimagnetism},
as is explicitly stated in \cite{ShenQiu94}.

\section{Nagaoka's ferromagnetism}
\label{s:Nagaoka}
\subsection{Weak version of Nagaoka's theorem}
\label{s:weakN}
We are now ready to discuss Nagaoka's ferromagnetism\footnote{
As for the presentation of the results and the proofs, 
we follow \cite{89c},
where a generalization of Nagaoka's theorem was discussed.
The proof briefly described in \cite{89c} is now presented in its 
full detail.
} \cite{Nagaoka66}.
Briefly speaking, Nagaoka's theorem establishes that some Hubbard models 
exhibit saturated ferromagnetism when the number of electrons is one 
less than the
half-filling (i.e., \( \Ne=\Ns-1 \)) and the Coulomb repulsion 
\( U \) is infinitely large.
Given the general fact that a half-filled system never exhibits 
ferromagnetism, this is a rather striking result,
which demonstrates that strongly interacting electron systems can 
produce very rich and sometimes surprising phenomena.

\begin{figure}
\centerline{\epsfig{file=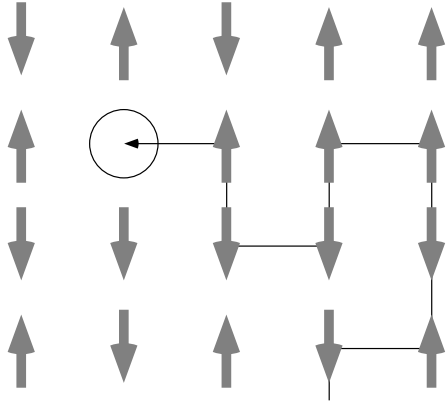,width=6cm}}
\caption[dummy]{
Schematic picture of the origin of Nagaoka's ferromagnetism.
When the hole hops around the lattice, the spin configuration is 
changed.
For a model with \( t_{x,y}\ge0 \), the hole motion produces a precise 
linear combination of various spin configurations which leads to a
ferromagnetic state.
}
\label{f:nagaoka}
\end{figure}

When \( U=\infty \) and \( \Ne=\Ns-1 \), 
states with finite energies have no 
doubly occupied sites, and there is 
exactly one empty site which we call ``hole.''
The basic mechanism of Nagaoka's ferromagnetism is 
that the hole hops around the lattice and generates a suitable linear 
combination of the basis states, in such a way that the resulting 
state exhibits ferromagnetism.
See Figure~\ref{f:nagaoka}.
Thouless \cite{Thouless65} also discussed the same mechanism 
of ferromagnetism in 
slightly more restricted situations.
Also see \cite{Lieb71}.

We start from the following weaker result which is easy to state 
and prove.

\begin{theorem}[Weak version of Nagaoka's theorem]
	\label{t:N1}
	Consider an arbitrary Hubbard model with 
	\( t_{x,y}\ge0 \) for any \( x,y\in\La \),
	\( \Ne=\Ns-1 \), and \( U=\infty \).
	Then among the ground states there are 
	\( (2\Stot+1) \) states with total spin
	\( \Stot=\Smax(=\Ne/2) \).
\end{theorem}

Note that the theorem does not establish the existence of 
ferromagnetism since it does not state ferromagnetic states are the 
only ground states\footnote{
\label{fn:gs}
Note that the statement of the theorem is valid in the trivial model 
with \( t_{x,y}=0 \) for all \( x,y \), where all the possible spin 
states are degenerate.
We must remark that some of the published ``proofs'' of Nagaoka's 
theorem and its extensions only prove such weak statements.
See also footnote~\ref{fn:stab}.
}.
Stronger statement will be proved in the next section as 
Theorem~\ref{t:N2}.

Let us prove Theorem~\ref{t:N1}.
We first consider the treatment of the \( U\to\infty \) limit.
Let us decompose the Hilbert space \( \calH_{\Ne} \) as
\begin{equation}
	\calH_{\Ne}=\calH_{\Ne}^{(0)}\oplus\calH_{\Ne}',
	\label{Hnedecomp}
\end{equation}
where \( \calH_{\Ne}^{(0)} \) consists of the states\footnote{
We are assuming \( \Ne\le\Ns \).
If \( \Ne>\Ns \), the relevant condition is replaced by
\( \Hint\Phi=U(\Ne-\Ns)\Phi \), and Lemma~\ref{l:U} is still valid.
} satisfying \( \Hint\Phi=0 \).
We denote by \( P_{0} \) (see (\ref{PG})) the orthogonal projection onto
 \( \calH_{\Ne}^{(0)} \).
Then we have the following.

\begin{lemma}[Characterization of the \( U\to\infty \) limit]
	\label{l:U}
	The \( U\to\infty \) limit of the Hubbard model is equivalent to 
	studying the Hamiltonian\footnote{
	In the present case, we will eventually find that the ground states 
	in the \( U\to\infty \) limit are eigenstates of \( \Hhop \).
	This is of course an accidental situation for the ferromagnetic states. 
	Examples of exact ground states in the \( U\to\infty \) limit which 
	are not eigenstates of \( \Hhop \) can be found in 
	\cite{BrandtGiesekus92,94a}.
	} \( \tH=P_{0}\Hhop \) on the Hilbert space
	\( \calH_{\Ne}^{(0)} \).
\end{lemma}

\proof
By continuity, the eigenstates of \( H=\Hhop+\Hint \) can be 
classified into those with energies diverging as \( U\to\infty \) 
and those with finite energies in the \( U\to\infty \) limit.
We are only interested in the latter.
Let \( \Phi \) be an eigenstate with energy \( E \)
which has a finite limit. 
We can assume both \( \Phi \) and \( E \) are continuously 
parameterized by \( U \).
Note that we have \( P_{0}\Phi\to\Phi \) in the \( U\to\infty \)
limit, 
because otherwise \( E \) diverges in this limit.
By applying the projection \( P_{0} \) onto the \Sch{} equation
\( E\Phi=(\Hhop+\Hint)\Phi \), we get 
\( EP_{0}\Phi=P_{0}\Hhop\Phi \) because
\( P_{0}\Hint\Phi=0 \).
In the \( U\to\infty \) limit, this becomes
\( E\Phi=\tH\Phi \) for \( \Phi\in \calH_{\Ne}^{(0)} \).\qed

We now prepare a basis for the Hilbert space \( \calH_{\Ne}^{(0)} \).
We use the basis states 
characterized by the position \( x \) of the hole and the spin 
configuration\footnote{
By \( \La\bs x \) we mean the lattice obtained by removing \( x \) 
from \( \La \).
}
 \( \tsigma=(\sigma_{y})_{y\in\La\bs x}\in\calS_{\La\bs x} \), 
defined as
\begin{equation}
	\Phi_{x,\tsigma}
	=
	\axu\rbk{\prod_{y\in\La}\cd_{y,\sigma'_{y}}}\vac
	=
	\axd\rbk{\prod_{y\in\La}\cd_{y,\sigma''_{y}}}\vac.
	\label{Nbasis1}
\end{equation}
Here the product is taken over all the sites in \( \La \) with an 
arbitrary but fixed order.
We gave two equivalent expressions for \( \Phi_{x,\tsigma} \).
The spin configurations \( (\sigma'_{y})_{y\in\La} \)
and \( (\sigma''_{y})_{y\in\La} \) are essentially the same as 
\( \tsigma=(\sigma_{y})_{y\in\La\bs x} \), and defined by
\( \sigma'_{y}=\sigma''_{y}=\sigma_{y} \) for all
\( y\in\La\bs x \).  
As for the missing site \( x \), we set \( \sigma'_{x}=\up \)
and \( \sigma''_{x}=\dn \).
In the definition (\ref{Nbasis1}), we are simply supplying an electron 
at the hole site \( x \), and then annihilating the electron by 
\( \axu \) or \( \axd \).
This may sound meaningless, but we get fermion signs appropriate for 
our purpose in this way.

We wish to examine the action of \( \tH=P_{0}\Hhop \) on the states 
(\ref{Nbasis1}). 
In order to get a finite contribution, an electron must hop 
into the hole site \( x \) from other sites.
So we examine the action of 
\( \sum_{\sigma=\up,\dn}\cxs c_{z,\sigma} \) 
on
\( \Phi_{x,\tsigma} \).
Using the two equivalent expressions (\ref{Nbasis1}) for 
\( \Phi_{x,\tsigma} \), we get
\begin{eqnarray}
	\sum_{\sigma=\up,\dn}\cxs c_{z,\sigma}\Phi_{x,\tsigma} 
	& = &
	-c_{z,\up}n_{x,\up}
	\rbk{\prod_{y\in\La}\cd_{y,\sigma'_{y}}}\vac
	-c_{z,\dn}n_{x,\dn}
	\rbk{\prod_{y\in\La}\cd_{y,\sigma''_{y}}}\vac
	\ret
	& = & 
	-c_{z,\up}
	\rbk{\prod_{y\in\La}\cd_{y,\sigma'_{y}}}\vac
	-c_{z,\dn}
	\rbk{\prod_{y\in\La}\cd_{y,\sigma''_{y}}}\vac
	\ret
	& = & 
	-\Phi_{z,\tsigma_{z\to x}},
	\label{action}
\end{eqnarray}
where \( \tsigma_{z\to x}\in\calS_{\La\bs z} \) is the new spin 
configuration on \( \La\bs z \) obtained from \( \tsigma \) by moving 
\( \sigma_{z} \) to \( x \).
Note that in the second line of (\ref{action}), only one of the two 
terms survive depending on the value of \( \sigma_{z} \).
Therefore the matrix elements of the effective Hamiltonian 
\( \tH \) are given by
\begin{equation}
	\bkt{\Phi_{y,\ttau},\tH\Phi_{x,\tsigma}}
	=\cases{
	-t_{x,y}&if \( \ttau=\tsigma_{y\to x} \);\cr
	0&otherwise.
	}
	\label{ytHxs}
\end{equation}

Let \( \GS \) be a ground state of \( H \).
Since it has a finite energy, it is expanded as
\begin{equation}
	\GS=
	\sum_{\xiL}\ \sum_{\tsigma\in\calS_{\La\bs x}}
	\phi(x,\tsigma)\Phi_{x,\tsigma}.
	\label{expPhi}
\end{equation}
Since the matrix elements (\ref{ytHxs}) of \( \tH \)  
are real, we can assume that the coefficients \( \phi(x,\tsigma) \) 
are real\footnote{
If this is not the case, we redefine 
\( \{\phi(x,\tsigma)+\phi(x,\tsigma)^{*}\}/2 \) as
\( \phi(x,\tsigma) \).
The corresponding \( \GS \) is also a ground state.
}.
We define
\begin{equation}
	\xi_{x}=
	\rbk{\sum_{\tsigma\in\calS_{\La\bs x}}
	(\phi(x,\tsigma))^{2}}^{1/2},
	\label{xi}
\end{equation}
and a ferromagnetic state
\begin{equation}
	\UP=\sum_{\xiL}\xi_{x}\Phi_{x,(\up)},
	\label{UP}
\end{equation}
where \( (\up) \) denotes the spin configuration with all spins up.

By using (\ref{expPhi}) and (\ref{ytHxs}), we find
\begin{eqnarray}
	\bkt{\GS,\tH\GS} 
	& = & 
	\sum_{x,y\in\La}
	\sumtwo{\tsigma\in\calS_{\La\bs x}}{\ttau\in\calS_{\La\bs y}}
	\phi(y,\ttau)\phi(x,\tsigma)
	\bkt{\Phi_{y,\ttau},\tH\Phi_{x,\tsigma}}
	\ret
	& = & 
	-\sum_{x,y\in\La}t_{x,y}\sum_{\tsigma\in\calS_{\La\bs x}}
	\phi(y,\tsigma_{y\to x})\phi(x,\tsigma)
	\ret
	& \ge & 
	-\sum_{x,y\in\La}t_{x,y}
	\rbk{\sum_{\tsigma\in\calS_{\La\bs x}}
	\{\phi(y,\tsigma_{y\to x})\}^{2}}^{1/2}
	\rbk{\sum_{\tsigma\in\calS_{\La\bs x}}
	\{\phi(x,\tsigma)\}^{2}}^{1/2}
	\ret
	& = &
	-\sum_{x,y\in\La}t_{x,y}\xi_{y}\xi_{x}
	\ret
	& = &
	\bkt{\UP,H\UP},
	\label{main}
\end{eqnarray}
where we have used the Schwartz inequality 
(with the assumption \( t_{x,y}\ge0 \))
to get the third line.
This bound shows that \( \UP \) is also a ground state.
This completes the proof of Theorem~\ref{t:N1}.

\subsection{Nagaoka's theorem}
\label{s:fullN}
We now state a stronger and the most general version of Nagaoka's 
theorem.
Under an additional condition (which can be easily verified in some 
typical cases), 
we will prove that the ferromagnetic states are the only possible 
ground states.
To determine the additional conditions, it is better to start 
from mathematics.

Let us recall the Perron-Frobenius theorem, which is standard in linear 
algebra.
The following is the simplest version of the theorem\footnote{
The full version of the theorem applies to non-symmetric matrices
as well.
See, for example, page 130 of \cite{Simon93}.
}.

\begin{theorem}[Perron-Frobenius theorem for a real symmetric matrix]
	\label{t:PF}
	Let \( M=(m_{i,j})_{i,j=1,\ldots,N} \) be an 
	\( N\times N \) real symmetric matrix 
	(i.e., \( m_{i,j}=m_{j,i}\in\real \) )
	with the properties that
	\par\noindent
	i) \( m_{i,j}\le0 \) for any \( i\ne j \).
	\par\noindent
	ii) All \( i\ne j \) are connected via nonvanishing matrix elements 
	of \( M \).
	More precisely\footnote{
	Another way of stating the condition is that,
	for any \( i\ne j \), there is \( K \) such that 
	\( (M^K)_{i,j}\ne0 \).
	}, for any \( i\ne j \), we can take a sequence
	\( (i_{1},\ldots,i_{K}) \) such that
	\( i_{1}=i \), \( i_{K}=j \), and
	\( m_{i_{k},i_{k+1}}\ne0 \) for all \( k<K \).
	
	Then the lowest eigenvalue of \( M \) is nondegenerate and the 
	corresponding eigenvector 
	\( {\bf v}=(v_{i})_{i=1,\ldots,N} \)
	can be taken to satisfy 
	\( v_{i}>0 \) for all \( i \).
\end{theorem}

\proof
Let us present a standard elementary proof based on a 
variational argument.
The essence of the argument is that a state without ``nodes'' has low 
energy.
This idea is familiar in quantum mechanics\footnote{
See, for example, Section 20 of \cite{LandauQM}.
}.

1)~We first prove that if an eigenvector 
\( {\bf u}=(u_{i})_{i=1,\ldots,N} \) of \( M \) satisfies
\( u_{i}\ge0 \) for any \( i \), then it inevitably satisfies
\( u_{i}>0 \) for any \( i \).
To do this, we assume the converse, i.e., \( u_{i}\ge0 \) 
for all \( i \), and 
\( u_{j}=0 \) for some \( j \).
Then from the eigenvalue equation, we see that
\( \sum_{i}m_{j,i}u_{i}=\mu u_{j}=0 \).
Since \( m_{j,i}u_{i}\le0 \), this means
\( u_{i}=0 \) for all \( i \) with 
\( m_{j,i}\ne0 \).
Because we have the connectivity 
 ii), we can repeat this argument until we see
\( u_{i}=0 \) for all \( i \), which is a contradiction.

2)~We then prove that a normalized eigenvector 
\( {\bf v}=(v_{i})_{i=1,\ldots,N} \)
for the lowest eigenvalue \( \mu_{0} \) can be chosen to satisfy
\( v_{i}>0 \) for all \( i \).
Since all  \( m_{i,j} \) and \( \mu_{0} \) are real, 
we can assume that the \( v_{i} \) are all real.
With 1) in mind, we assume the converse, i.e., 
\( v_{j}>0 \) and \( v_{k}<0 \) hold for some \( j \) and \( k \).
Define 
\( {\bf u}=(u_{i})_{i=1,\ldots,N} \)
by
\( u_{i}=|v_{i}| \).
Then from i), we see
\begin{equation}
	\mu_{0}
	=\sum_{i,j=1}^{N}v_{i}m_{i,j}v_{j}
	\ge\sum_{i,j=1}^{N}u_{i}m_{i,j}u_{j}.
	\label{pf}
\end{equation}
Since \( \mu_{0} \) is the lowest eigenvalue, the right-hand side 
must also be equal to \( \mu_{0} \), which means 
that \( \bf u \) is an eigenvector of \( M \).
Then the above 1) implies \( u_{i}>0 \), and hence \( v_{i}\ne0 \) for 
all \( i \).
Recalling the connectivity ii), the latter property implies we can 
find \( i \) and \( j \) such that 
\( v_{i}m_{i,j}v_{j}>0 \).
Then we have 
\( u_{i}m_{i,j}u_{j}<0 \) for the same \( i \) and \( j \), which 
means (\ref{pf}) is valid with \( \ge \) replaced by \( > \).
This contradicts with the assumption that \( \mu_{0} \) is the lowest 
eigenvalue.

3)~Finally suppose that the lowest eigenvalue of \( M \) is 
degenerate.
Then we can find two mutually orthogonal eigenvectors 
\( \bf u \) and \( \bf v \).
But from 2), we must have \( {\bf u}\cdot{\bf v}\ne0 \), which is a 
contradiction.\qed

We wish to apply this theorem to the present problem of 
the \( U=\infty \) Hubbard model with a single hole.
As for the matrix \( M \), we take the matrix representation
(\ref{ytHxs}) of the Hamiltonian for the basis 
states (\ref{Nbasis1}) with a fixed 
\( \Sztot=\sum_{y\in\La\bs x}\sigma_{y} \).
Because of (\ref{ytHxs}) and the assumption \( t_{x,y}\ge0 \),
the condition~i) of Theorem~\ref{t:PF} is satisfied.
The condition~ii), on the other hand, is not always valid.
This motivates us to consider the following {\em connectivity 
condition}.

\begin{definition}[Connectivity condition]
	\label{d:con}
	A Hubbard model with \( U=\infty \) and \( \Ne=\Ns-1 \)
	(or more precisely the hopping matrix \( \mT \) )
	is said to satisfy the {\em connectivity condition} if all the basis 
	states \( \Phi_{x,\tsigma} \) with common 
	\( \Sztot=\sum_{y\in\La\bs x}\sigma_{y} \) 
	are connected with each other through nonvanishing matrix elements of
	 \( H \).
\end{definition}

As we shall see in the next section, the connectivity condition can be 
easily verified in the Hubbard model with nearest neighbor hoppings
on most standard lattices 
including triangular, square, 
 simple cubic,  fcc, and  bcc.

Consider a model which satisfies the connectivity condition.
Then we can readily apply the Perron-Frobenius theorem to see that 
the ground state in each subspace with a fixed \( \Sztot \) is unique.
Then Theorem~\ref{t:N1} implies that this ground state must be  
ferromagnetic.
So we have proved the following, which is the full (and most 
generalized) version of Nagaoka's famous theorem.

\begin{theorem}[Generalized Nagaoka theorem]
	\label{t:N2}
	Consider an arbitrary Hubbard model with 
	\( t_{x,y}\ge0 \) for any \( x,y\in\La \),
	\( \Ne=\Ns-1 \), and \( U=\infty \),
	and further assume that the model satisfies the connectivity condition.
	Then the ground states have total spin 
	\( \Stot=\Smax(=\Ne/2) \), 
	and are non-degenerate apart from the trivial 
	\( (2\Smax+1) \)-fold degeneracy.
\end{theorem}

As stressed earlier, the knowledge that all the ground states 
are ferromagnetic is of fundamental importance.
The theorem is  of course no longer valid for the trivial model with 
\( t_{x,y}=0 \) for all \( x \) and \( y \).
Since the theorem asserts the nondegeneracy of the ferromagnetic ground 
state for \( U=\infty \), the continuity of energy eigenvalues in \( U 
\) implies that the statement of the theorem is valid also for 
sufficiently large but finite \( U \).
However, we have no meaningful estimates of how large \( U \) should 
be.

\subsection{Connectivity condition}
\label{s:con}
We still have to verify the connectivity condition for some systems 
to make Theorem~\ref{t:N2} meaningful.
It seems, however, that to write down a simple necessary and 
sufficient condition for the connectivity condition is a  
nontrivial problem.
We here follow Nagaoka's original spirit, and provide a 
constructive criterion (i.e., a sufficient condition) for the 
connectivity condition.

\begin{figure}
\centerline{\epsfig{file=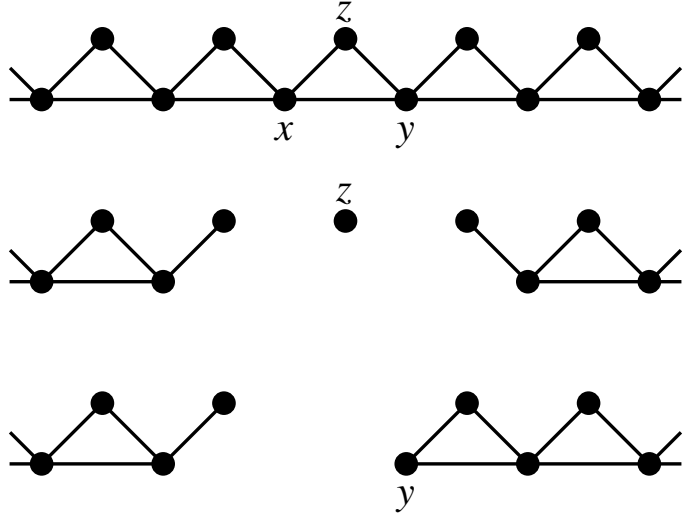,width=9cm}}
\caption[dummy]{
{\em Exchange bonds} in the ``delta-chain.''
The horizontal bond \( \{x,y\} \) is not an exchange bond since 
the site \( z \) is disconnected from the rest when \( x \) and \( y \) 
are removed.
The bond \( \{x,z\} \) is an exchange bond since the lattice 
remains connected after the removal of \( x \) and \( y \) 
provided that we use 
periodic boundary conditions.
}
\label{f:exb}
\end{figure}

Let us introduce some terminology.
By a {\em loop} of length \( m \), we mean an ordered set 
\( (x_{1},\ldots,x_{m}) \) of sites such that
\( t_{x_{i},x_{i+1}}\ne0 \) for all \( i=1,\ldots,m-1 \),
and \( t_{x_{m},x_{1}}\ne0 \).
We say that a pair \( \{x,y\} \) of lattice sites is an 
{\em exchange bond} if they belong to a loop of length three or four, 
and the whole lattice remains connected via nonvanishing 
\( t_{x,y} \) when 
the sites \( x \) and \( y \) are removed.
(See Figure~\ref{f:exb}.)

Then we have the following sufficient condition for the connectivity.

\begin{lemma}[A sufficient condition for the connectivity]
	\label{l:con}
	If the whole lattice is connected by  exchange bonds, then the 
	model satisfies the connectivity condition.
\end{lemma}

This sufficient condition can be easily verified in various systems.
In models defined on regular lattices like  triangular,  square, 
 simple cubic,  fcc, or  bcc lattices with nonvanishing 
hopping amplitudes between nearest neighbor sites, 
it is obvious that all the nearest 
neighbor bonds are exchange bonds.
Thus they trivially satisfy the condition.
A less trivial example is the delta chain with periodic boundary 
conditions in Figure~\ref{f:exb}.
Although the horizontal bonds are not exchange bonds, the whole 
lattice is connected via non-horizontal bonds, which are exchange 
bonds.

\proofof{Proof of Lemma~\ref{l:con}}
Suppose that we are given an arbitrary configuration of 
\( \Ne=\Ns-1 \)
electrons on \( \La \).
Our goal is to show that we can get an arbitrary configuration with 
the same \( \Sztot \) by moving the single hole along nonvanishing 
\( t_{x,y} \).

Let \( \{x,y\} \) be an exchange bond.
We show below that we can exchange the spins at sites \( x \) 
and \( y \) without changing the configuration outside \( \{x,y\} \).
Since the whole lattice is connected via exchange bonds, 
this means we can 
generate any permutation of spin configurations by successive 
exchanges on the exchange bonds.
This proves the connectivity condition.

\begin{figure}
\centerline{\epsfig{file=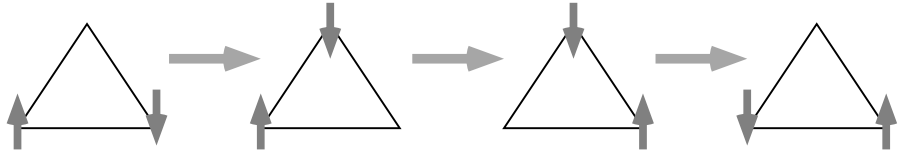,width=12cm}}
\caption[dummy]{
The two spins are exchanged when the hole hops around the loop once.
Such an ``exchange process along a triangle'' appears repeatedly in 
various examples of ferromagnetism, and is regarded as a universal and 
fundamental mechanism leading to ferromagnetism \cite{97d}.
}
\label{f:tri}
\end{figure}

\begin{figure}
\centerline{\epsfig{file=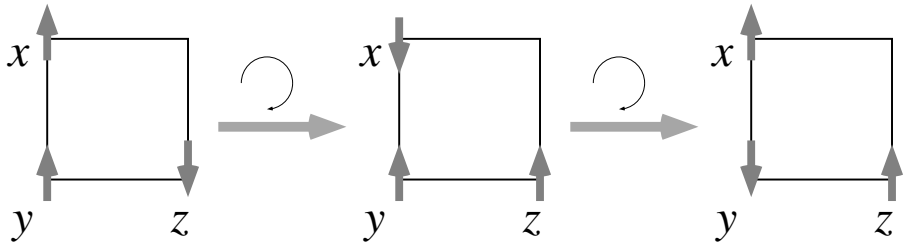,width=12cm}}
\caption[dummy]{
The spins at \( x \) and \( z \)  are exchanged when the hole hops 
around the loop once.
The spins at \( y \) and \( z \)  are exchanged when the hole hops 
around the loop twice (or once in the opposite orientation).
}
\label{f:squ}
\end{figure}

We now prove the desired property of exchange bonds.
Let \( \{x,y\} \) be an exchange bond, and assume that 
\( x \) and \( y \) are 
occupied by electrons with opposite spins.
We first bring the hole (by successive hops outside \( \{x,y\} \)) to 
a site other than \( x \) or \( y \) on the loop 
(of length three or 
four) that contains \( \{x,y\} \).
Next we let the hole move along the loop until the spins at 
\( x \) and \( y \) are exchanged.
In a loop of length three, this is realized after the hole goes 
around the loop once, as in Figure~\ref{f:tri}.
In a loop of length four\footnote{
The exchange on a length four loop is possible because electronic 
spins take (only) two values.
}, we have to move the hole along the loop
once or twice, depending on the spin configuration, as in 
Fig~\ref{f:squ}.
Finally, we bring the hole back to the original location.
By following the same path as in the first step backwards, we recover 
exactly the same configuration, except on sites 
\( x \) and \( y \).\qed

\subsection{Stability of Nagaoka's ferromagnetism}
\label{s:stabN}
It is desirable to extend Nagaoka's ferromagnetism to systems with   
finite  $U$ and with finite densities of holes.
Unfortunately, the Perron-Frobenius argument 
which works for the one-hole 
case fails even for models with two holes.
There is a considerable number of rigorous works
(including that in Nagaoka's original work \cite{Nagaoka66}) which 
establish that saturated ferromagnetism {\em does not} take place in 
certain situations.
See, for example, 
\cite{DoucotWen89,Shastry90,Toth91,Suto91b,%
HanischMullerHartmann93,HanischUhrigMuellerHartmann97}.
Most of these works are essentially based on variational arguments 
where one constructs sophisticated variational states which have lower 
energies than the ferromagnetic state.
(We have seen the most elementary versions of the argument in 
Section~\ref{s:noferro}.)
As far as we know, there are no rigorous results about the 
{\em stability} of Nagaoka's ferromagnetism\footnote{
\label{fn:stab}
Interesting lower bounds for the ground state energy of the 
\( U=\infty \) model with multiple holes are proved in
\cite{Trugman90,Tian91}.
We do not, however, interpret these bounds as ``proofs of stability of 
Nagaoka's ferromagnetism'' since the bounds 
do not rule out paramagnetism (e.g., in the 
trivial model with \( t_{x,y}=0 \) for any \( x,y \)).
See also footnote~\ref{fn:gs}.
} in the Hubbard model.

There is also a considerable number of numerical and theoretical works 
which indicate stability or instability of Nagaoka's ferromagnetism 
in various situations
\cite{BarbieriRieraYoung90,Putikka93,LiangPang94,KusakabeAoki94c}.

In spite of all these efforts, it is still not known if 
straightforward extensions of Nagaoka's ferromagnetism to finite \( U \) 
and finite hole density are possible.
It seems that a recent dominant opinion is that the possibility of 
extension depends strongly on lattice structures, where the 
triangular lattice is regarded as one of the most promising candidates.
The importance of lattice structures is closely related to the approach 
from flat-band models that we discuss in Section~\ref{s:flat}

\section{Ferromagnetism in the Hubbard model with long-range hopping}
\label{s:lrh}
\subsection{Main statement and preliminary results}
\label{s:pre}
There is at least one situation in which the extension of Nagaoka's 
ferromagnetism to finite \( U \) is possible.
This is an artificial model with long range hopping that we shall now 
define.

\begin{figure}
\centerline{\epsfig{file=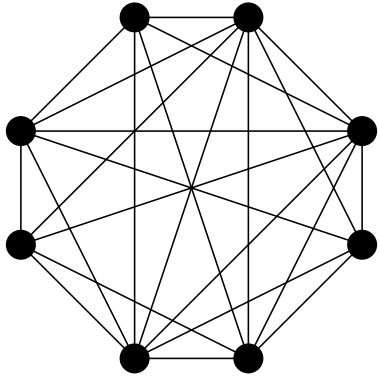,width=5cm}}
\caption[dummy]{
In the long-range hopping model, hopping between two arbitrary sites 
in the lattice is possible.
The model serves as an intermediate step in our attempt to relate 
Nagaoka's ferromagnetism and flat-band ferromagnetism.
}
\label{f:long}
\end{figure}

We associate with each site \( x \) a constant \( \lambda_{x}>0 \),
and define the hopping matrix 
\( \mT=(t_{x,y})_{x,y\in\Lambda} \) with 
long range hopping amplitudes by\footnote{
In coordinate-free notation, this becomes 
\( \mT=t\vlambda\otimes\vlambda \), where \( \otimes \) is understood 
as a Kronecker product, and 
\( \vlambda=(\lambda_{x})_{\xiL} \).
As for this construction, the Dirac notation
\( \mT=t|\vlambda\rangle\langle\vlambda| \) 
may be more informative.
}
\begin{equation}
	t_{x,y}=t\lambda_{x}\lambda_{y},
	\label{t=ll}
\end{equation}
where \( t>0 \) is a constant (which does not play any
essential role).
Note that an electron can hop from any site in the lattice to any other 
site (Figure~\ref{f:long}).
Then we have the following extension of Nagaoka's theorem.

\begin{theorem}[Ferromagnetism in the long-range hopping model]
	\label{t:cg}
	Consider the Hubbard model with the hopping (\ref{t=ll}) with 
	electron number \( \Ne=\Ns-1 \).
	For any \( U>0 \),  
	the ground states have total spin 
	\( \Stot=\Smax(=\Ne/2) \), 
	and are non-degenerate apart from the trivial 
	\( (2\Smax+1) \)-fold degeneracy.
\end{theorem}

At first glance, the theorem looks surprising and attractive since the 
precise statement of Nagaoka's theorem is extended to an arbitrary 
positive value of \( U \).
As we shall see below, this is indeed a consequence of a very special 
property of the present model.

With the hopping (\ref{t=ll}), 
the action of \( \mT \) onto any \( \vphi\in\hilb \) becomes
\( \mT\vphi=t\vlambda\bkt{\vlambda,\vphi} \),
where \( \vlambda=(\lambda_{x})_{\xiL}\in\hilb \).
Thus we find 
\( \bkt{\vphi,\mT\vphi}=t\bkt{\vphi,\vlambda}\bkt{\vlambda,\vphi}
=|\bkt{\vlambda,\vphi}|^2\ge0 \), which means that 
\( \mT\ge0 \).
On the other hand, the single-electron \Sch{} equation 
(\ref{sSch2}) becomes
\begin{equation}
	\ep\vphi=t\vlambda\bkt{\vlambda,\vphi}.
	\label{sSch3}
\end{equation}
This means that any \( \vphi \) orthogonal to \( \vlambda \) is an 
eigenstate of (\ref{sSch3}) with \( \ep=0 \), which is the minimum 
possible eigenvalue.
Clearly, there are \( (\Ns-1) \)-fold degenerate eigenstates with 
\( \ep=0 \).
The remaining eigenstate is \( \vlambda \) itself, and it has the 
eigenvalue \( \ep=t\sum_{\xiL}(\lambda_{x})^{2}>0 \).

It is crucial for the electron number \( \Ne \) to be set equal to the 
dimension \( \Ns-1 \) of the above degeneracy.
We see from (\ref{Eferro}) that the lowest energy among ferromagnetic 
states is 
\( E_{\rm ferro}=\sum_{j=1}^{\Ns-1}\ep_{j}=0 \).

On the other hand, it is clear from (\ref{Hhope})
that the lowest eigenvalue of \( \Hhop \) is 0 if 
\( \Ne\le2(\Ns-1) \).
Hence we can write \( \Hhop\ge0 \) for \( \Ne=\Ns-1 \).
(See Definition~\ref{d:pos} and Lemma~\ref{l:pos}.)
We also have \( \Hint\ge0 \) in general.
Thus (by using Lemma~\ref{l:A+B>0}) we get
\( H=\Hhop+\Hint\ge0 \).
This, along with the fact that there is a 
ferromagnetic state with energy zero, implies that the ground state 
energy of \( H \) is 0.
We have thus 
found that there are ferromagnetic states among the ground
states.

Let \( U>0 \), and let \( \GS \) be an arbitrary ground state.
Then (from Lemma~\ref{l:Ai})
\( \Hhop\ge0 \), \( \Hint\ge0 \), and 
\( H\GS=(\Hhop+\Hint)\GS=0 \) imply
\begin{equation}
	\Hhop\GS=0,\quad
	\Hint\GS=0.
	\label{HGS=0}
\end{equation}
In other words, the ground state happens to be a simultaneous ground 
state\footnote{
Of course the existence of one simultaneous eigenstate does not imply 
\( \Hhop \) and \( \Hint \) are simultaneously diagonalizable.
Recall that quantum mechanical angular momenta have simultaneous 
eigenstate with vanishing eigenvalue.
} of \( \Hhop \) and \( \Hint \).
This is of course a very special property of the present model with 
singular degeneracy in the single-electron eigenstates.
Since \( U\ne0 \), the second relation in (\ref{HGS=0}) implies
\begin{equation}
	\sum_{\xiL}\nxu\nxd=0.
	\label{nnGS=0}
\end{equation}

\subsection{First proof}
\label{s:first}
We give a proof of Theorem~\ref{t:cg} which makes explicit use 
of Nagaoka's theorem.
Since we have seen that there are ferromagnetic states among 
the ground states for \( U\ge0 \), we only have to show 
that they are the 
only ground states for \( U>0 \).

The ground states for any \( U>0 \) are fully characterized by 
\( \Hhop\GS=0 \) and (\ref{nnGS=0}).
Noting that (\ref{nnGS=0}) implies \( \Hint\GS=0 \) for any \( U>0 \),
we see that the ground states of a given \( U>0 \)
remain as ground states for 
any values of \( U>0 \), and hence in the limit \( U\to\infty \).
Since the ground states for \( U\to\infty \) are completely characterized 
by Nagaoka's theorem (Theorem~\ref{t:N2}), we have proved 
Theorem~\ref{t:cg}.

\subsection{Second proof}
\label{s:second}
We describe another proof of Theorem~\ref{t:cg}.
This proof is certainly more involved than the first one, but provides 
us with an entirely different physical picture of the problem.
In the new picture,  ferromagnetism in this model is generated by 
an ``exchange interaction'' among the spins of electrons which are 
``frozen'' in certain single-electron states.
We recall that the first proof is based on Nagaoka's theorem, which 
suggests a dynamical picture that ferromagnetism is generated 
by the motion of the hole.
It is interesting that the two totally different pictures 
apply equally to the present model.
As we shall investigate in the next sections, 
the new picture will lead us to a new class of ferromagnetism 
now known as `flat-band ferromagnetism.'

We start by constructing a (non-orthonormal) basis for the space of 
degenerate single-electron ground states.
Fix an arbitrary site \( x_{0}\in\La \), and denote by 
\( \La'=\La\bs x_{0} \) the lattice obtained by removing \( x_{0} \) 
from \( \La \).
For each \( y\in\La' \), we define a single-electron state
\( \vphi^{(y)}=(\phi^{(y)}_{x})_{\xiL}\in\hilb \)
by 
\begin{equation}
	\phi^{(y)}_{x}=\cases{
	\lambda_{x_{0}}&if \( x=y \);\cr
	-\lambda_{y}&if \( x=x_{0} \);\cr
	0&otherwise.	
	}
	\label{phiyxLR}
\end{equation}
This state satisfies 
\( \bkt{\vlambda,\vphi^{(y)}}=0 \), and hence it is a 
single-electron ground state.
Since the states 
\( \vphi^{(y)} \) with \( y\in\La' \) are linearly independent, 
they span the entire space of the single-electron ground states.
We define the corresponding fermion operators by
\( \bys=\Cds(\vphi^{(y)}) \) for \( y\in\La' \) and 
\( \sigma=\up,\dn \), which satisfies
\( [\Hhop,\bys]=0 \) because of the general commutation rule
(\ref{cAcC}) and the definition (\ref{Hhop}) of \( \Hhop \).

Let us characterize an arbitrary ground state \( \GS \) for 
\( U>0 \) in a constructive manner.
Because \( \Hhop\GS=0 \), the ground state \( \GS \) should consist 
only of the (single-electron) zero energy states.
Therefore it is written as\footnote{
In a more careful proof, one 
 uses Lemma~\ref{l:basis} with \( \Ns \) linearly independent states
\( \vphi^{(y)} \) (\( y\in\La' \)) and \( \vlambda \) 
to construct a basis 
of the entire Hilbert space \( \calH_{\Ne} \), 
formally writes \( \GS \) as a linear combination of all the basis 
states, and finally uses the 
condition \( \Hhop\GS=0 \) to see that the terms including 
\( \Cds(\vlambda) \) must be vanishing.
}
\begin{equation}
	\GS
	=
	\sumtwo{\Lup,\Ldn\subset\La'}{{\rm s.t.}\,|\Lup|+|\Ldn|=\Ne}
	f(\Lup,\Ldn)
	\rbk{\prod_{y\in\Lup}\byu}\rbk{\prod_{y\in\Ldn}\byd}
	\vac,
	\label{GSexp1}
\end{equation}
where the \( f(\Lup,\Ldn) \) are coefficients, and
\( \Lup \), \( \Ldn \) are summed over all the subsets of \( \La' \)
such that 
\( |\Lup|+|\Ldn|=\Ne=\Ns-1 \).
Each state in the sum of (\ref{GSexp1}) is nonvanishing because of
Lemma~\ref{l:CCC}.

Since \( \nxu\nxd\ge0 \), the condition (\ref{nnGS=0}) 
(and Lemma~\ref{l:Ai}) indeed implies 
a stronger statement that \( \nxu\nxd\GS=0 \) for any \( x\in\La \).
By further rewriting this relation as
\newline
\( (\axd\axu)^{\dagger}(\axd\axu)\GS=0 \) (and using
Lemma~\ref{l:BBP=0}), we find that the ground 
state must satisfy
\begin{equation}
	\axd\axu\GS=0,
	\label{ccGS=0}
\end{equation}
for each \( \xiL \).

We take \( x\in\La' \) and examine what (\ref{ccGS=0}) 
implies about (\ref{GSexp1}).
Using the anticommutation relation
\( \{\axs,b^{\dagger}_{y,\sigma'}\}=
\lambda_{x_{0}}\delta_{x,y}\delta_{\sigma,\sigma'} \),
we obtain
\begin{eqnarray}
	\axd\axu\GS
	&=&
 \sum_{\Lup,\Ldn}
	\chi[x\in\Lup\cap\Ldn]\,f(\Lup,\Ldn)\,
	{\rm sgn}[x,\Lup,\Ldn]
\times
\ret&&
	\times\rbk{\prod_{y\in\Lup\bs x}\byu}
\rbk{\prod_{y\in\Ldn\bs x}\byd}
	\vac,
	\label{ccGSa}
\end{eqnarray}
where \( \chi[{\rm true}]=1 \), \( \chi[{\rm false}]=0 \), and 
the factor \( {\rm sgn}[x,\Lup,\Ldn]=\pm1 \) comes from the exchange 
of fermion operators.
Since all the terms in the sum of (\ref{ccGSa}) are linearly 
independent, the condition (\ref{ccGS=0}) implies 
\( f(\Lup,\Ldn)=0 \) for any \( \Lup,\Ldn \) such that 
\( x\in\Lup\cap\Ldn \).
Using this for all \( x\in\La' \), we finally see that
\( f(\Lup,\Ldn)=0 \) unless \( \Lup\cap\Ldn=\emptyset \).
Since \( |\Lup|+|\Ldn|=|\La'| \), 
the condition \( \Lup\cap\Ldn=\emptyset \) implies
 \( \Lup\cup\Ldn=\La' \).
This means that we can reorganize the sum (\ref{GSexp1}) as
\begin{equation}
	\GS=
	\sum_{\tsigma\in\calS_{\La'}}g[\tsigma]
	\rbk{\prod_{y\in\La'}b^{\dagger}_{y,\sigma_{y}}}\vac,
	\label{GSexp2}
\end{equation}
where the sum is over all the spin configurations
\( \tsigma=(\sigma_{y})_{y\in\La} \) on \( \La' \),
and \( g[\tsigma] \) is a new coefficient.
The representation (\ref{GSexp2}) suggests a new physical picture that 
the motion of the electrons is ``frozen'', and only spin degrees of 
freedom are left.
Note, however, that this interpretation is somewhat arbitrary since 
the site \( x_{0} \) is chosen arbitrarily.

Now we consider the condition (\ref{ccGS=0}) with \( x=x_{0} \).
By using (\ref{GSexp2}) and the anticommutation relation
\( \{c_{x_{0},\sigma},b^{\dagger}_{y,\sigma'}\}=
-\lambda_{y}\delta_{\sigma,\sigma'} \) for any \( y\in\La' \), we get
\begin{eqnarray}
	&&
	c_{x_{0},\dn}c_{x_{0},\up}\GS
	\ret
	&&
	=
	\sumtwo{x,y\in\La'}{{\rm s.t.\ }x>y}\ 
	\sumtwo{\tsigma\in\calS_{\La'}}
	{{\rm s.t.\,}\sigma_{x}=\up,\,\sigma_{y}=\dn}
	{\rm sgn}[x,y]\,t\lambda_{x}\lambda_{y}\,
	(g[\tsigma]-g[\tsigma_{x\leftrightarrow y}])
	\rbk{\prod_{z\in\La'\bs\{x,y\}}b^{\dagger}_{z,\sigma_{z}}}
	\vac,
	\label{ccGS2}
\end{eqnarray}
where we have introduced arbitrary ordering in \( \La' \), and 
\( {\rm sgn}[x,y]=\pm1 \) is again the fermion sign\footnote{
\( \La\bs\{y,z\} \) is the set obtained by removing \( y \) and 
\( z \) from
\( \La \).
}.
The spin configuration \( \tsigma_{x\leftrightarrow y} \) is obtained 
by switching \( \sigma_{x} \) and \( \sigma_{y} \) in the original 
configuration \( \tsigma \).

Since all the states in the sum (\ref{ccGS2}) are linearly 
independent, the condition (\ref{ccGS=0}) is satisfied only when
\begin{equation}
	g[\tsigma]=g[\tsigma_{x\leftrightarrow y}]
	\label{exch}
\end{equation}
holds for all \( \tsigma \) and all \( x,y\in\La' \).
Note that we have deduced a kind of ``exchange interaction'' 
(\ref{exch}) among 
``localized spins'' from the no-double-occupancy condition 
(\ref{ccGS=0}).

The exchange relation (\ref{exch}) implies that all the 
\( g[\tsigma] \) with common 
\( \Sztot=\sum_{y\in\La'}\sigma_{y} \) assume exactly the same 
values\footnote{
Although it is not necessary in the present proof, one can 
show directly that such a state has \( \Stot=\Smax \).
See proof of Theorem~\ref{t:t-J}.
}.
This means that the lowest energy state in the space with 
a fixed \( \Sztot \) is unique.
Since we already know that there are ferromagnetic ground states, this 
completes the second proof of Theorem~\ref{t:cg}.
 
\section{Flat-band ferromagnetism}
\label{s:flat}
\subsection{Definition}
\label{s:model}
We have seen that the Hubbard model with artificial long range hopping 
exhibits ferromagnetism for arbitrary \( U>0 \).
As we shall see in the present section, it is possible to assemble 
many (identical) copies of the long-range model to get  
Hubbard models with only short range hoppings which still exhibit 
ferromagnetism \cite{92e,93d}.
Unlike Nagaoka's ferromagnetism, the electron density is away from 
half-filling, and \( U \) need only to be nonvanishing.
On the other hand, these models have very singular band structure which 
is expressed by the name ``flat-band models.''

Let us define our model by means of a ``cell construction\footnote{
A class of exactly solvable Hubbard models with very similar cell 
structures (with paramagnetic ground states) 
was discovered by Brandt and Giesekus 
\cite{BrandtGiesekus92} before the present model for 
flat-band ferromagnetism was introduced.
See also \cite{94a,YamanakaHonjouHatsugaiKohmoto96} 
and references therein for further results concerning 
this model.
}.''
Our lattice \( \La \) can be written as
\begin{equation}
	\La=C_{1}\cup\cdots\cup C_{M},
	\label{Lam=CC}
\end{equation}
where each \( C_{j} \) is called a cell.
Each cell consists of a single {\em internal site} and \( n \) {\em 
external sites}, where \( n\ge2 \) is a constant.
The simplest cell is a triangle with \( n=2 \).
The models to be constructed will again involve ``exchange processes 
along a triangle.''
This seems to be a universal and fundamental 
mechanism of ferromagnetism \cite{97d}.
See Figure~\ref{f:tri}.

\begin{figure}
\centerline{\epsfig{file=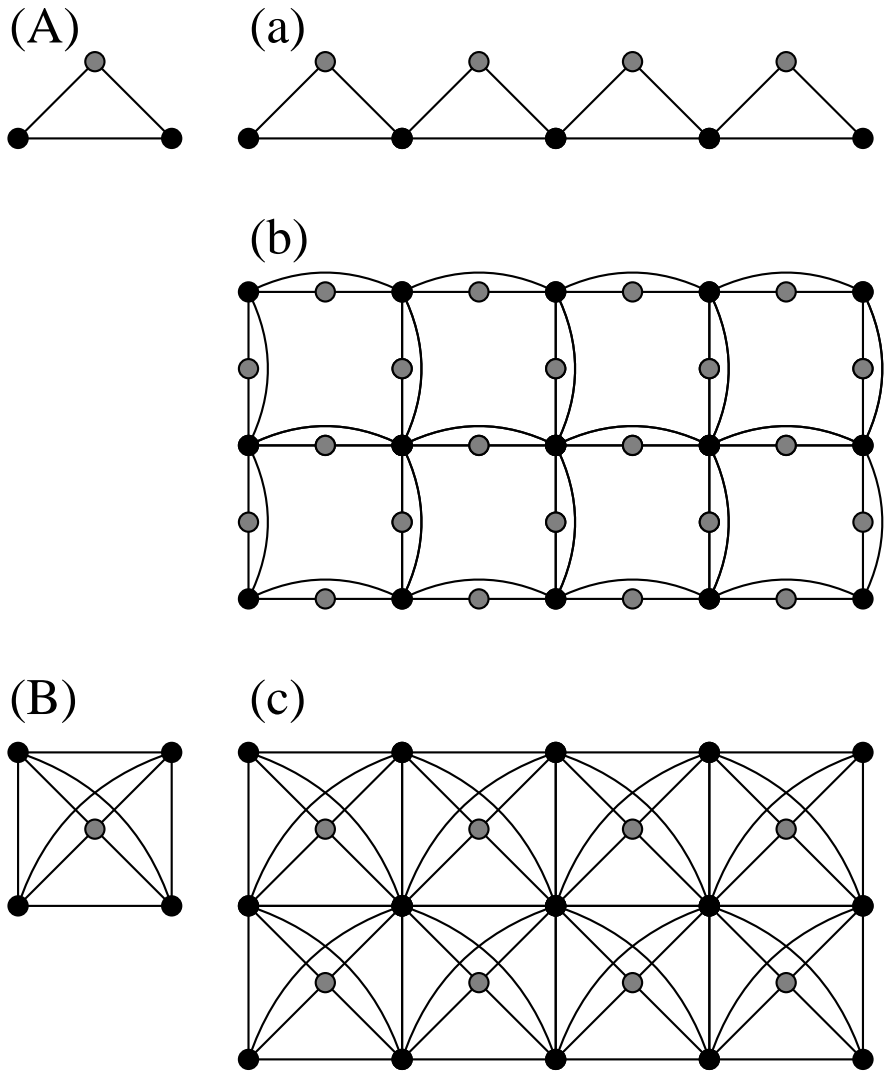,width=12cm}}
\caption[dummy]{
Examples of cells and lattices.
Gray dots represent internal sites.
From the triangular
cell with three sites (A), one can form (a)~the delta-chain by 
identifying two external sites, or (b)~a decorated square lattice by 
identifying four.
From the cell with five sites (B), one can form (c)~another 
decorated square lattice.
There are many similar examples in higher dimensions.
The flat-band Hubbard model on these lattices exhibits ferromagnetism 
for any \( U>0 \) when the filling factor \( \nu=\Ne/(2\Ns) \) is 
equal to (a)~\( 1/4 \), (b)~\( 1/6 \), or (c)~\( 1/4 \).
}
\label{f:flat}
\end{figure}

When assembling \( M \) cells to form \( \La \) in (\ref{Lam=CC}),
we identify some external sites from different cells\footnote{
There can be external sites left unidentified.
} to regard them as a single site in \( \La \).
See Figure~\ref{f:flat}.
The lattice \( \La \) is naturally decomposed as
\( \La=\calI\cup\calE \),
where \( \calI \) is the set of internal sites, and
\( \calE \) is the set of external sites.

We define the hopping amplitudes for the \( j \)-th cell by
\begin{equation}
	t^{(j)}_{x,y}=t\lambda_{x}^{(j)}\lambda_{y}^{(j)},
	\label{tj}
\end{equation}
where
\begin{equation}
	\lambda_{x}^{(j)}=\cases{
	\lambda&if \( x \) is the internal site of \( C_{j} \);\cr
	1&if \( x \) is one of the external sites of \( C_{j} \);\cr
	0&otherwise.\cr
	}
	\label{lambdaxj}
\end{equation}
Here \( t>0 \) and \( \lambda>0 \) are constants.
Note that this is the same as the hopping amplitudes
(\ref{t=ll}) of the long range model.
An essential difference, which makes the present model less 
artificial,  is that we have this type of hopping only within each 
cell.
The total hopping is defined as
\begin{equation}
	t_{x,y}=\sum_{j=1}^{M}t^{(j)}_{x,y}.
	\label{txyf}
\end{equation}
We finally assume that the whole lattice is connected via nonvanishing 
\( t_{x,y} \).
We then define the Hubbard model on \( \La \) with
the hopping matrix given by (\ref{txyf}).

\begin{theorem}[Flat-band ferromagnetism]
	\label{t:flat}
	Consider the above Hubbard model with the electron number
	\( \Ne=|\calE| \).
	For any \( U>0 \),  
	the ground states have total spin 
	\( \Stot=\Smax(=\Ne/2) \), 
	and are non-degenerate apart from the trivial 
	\( (2\Smax+1) \)-fold degeneracy.
\end{theorem}

Note that the existence of ferromagnetism has been established in 
models with finite \( U \), finite-range hoppings, and electron 
densities away from half-filling.
As one might guess, 
the straightforward (and general) methods using the Perron-Frobenius 
theorem (as in the proof of Nagaoka's theorem) do not work for this 
problem.
The proof will make use of very special properties of the model.
This ferromagnetism is closely related to Nagaoka's ferromagnetism 
(at least) via the ferromagnetism in the long-range hopping model, 
but it certainly belongs to a new class of ferromagnetism.
This new class is now known as ``flat-band ferromagnetism'' for a 
reason which we shall explain in Section~\ref{s:single}.
\subsection{Examples}
\label{s:examples}
Clearly the cell construction leads to a wide variety of models\footnote{
It is not necessary to have the 
same \( n \) or \( \lambda \) for all the 
cells.
This further extends the possibility of models.
}.
Let us discuss a class of examples which is constructed from the most 
elementary triangular cell with \( n=2 \).

We construct the \( d \)-dimensional \( L\times\dots\times L \) 
hypercubic lattice with periodic boundary conditions
by assembling together the triangular cell (A) 
of Figure~\ref{f:flat} regarding it as a basic bond (with length 1).
The resulting lattice 
\( \Lambda \) is the decorated hypercubic lattice with 
\( (d+1)L^d \) sites.
The set \( \calE \) of external sites can 
be identified with the \( d \)-dimensional \( L\times\dots\times L \) 
hypercubic lattice with \( L^d \) sites.
The internal sites in \( \calI \) are the decorating sites at the 
center of every bond.
The lattices (a) and (b) in Figure~\ref{f:flat} represent 
parts of \( \Lambda \) 
for \( d=1 \) and 2, respectively.

Then the hopping amplitude of (\ref{txyf}) becomes
\begin{equation}
	t_{x,y}=\cases{
	\lambda t&
	if \( |x-y|=1/2 \);\cr
	t&
	if \( x,y\in\calE \) and \( |x-y|=1 \);\cr
	\lambda^2t&
	if \( x=y\in\calI \);\cr
	2dt&
	if \( x=y\in\calE \);\cr
	0&
	otherwise.
	}
	\label{txyf2}
\end{equation}
Note that there are next-nearest-neighbor hoppings between nearby 
external sites.

The electron number specified in Theorem~\ref{t:flat} is
\( \Ne=L^d \).
In terms of the filling factor, this corresponds to 
\( \nu=\Ne/(2\Ns)=\{2(d+1)\}^{-1} \).

\subsection{Single-electron problem}
\label{s:single}
The present models have peculiar degeneracy in their 
single-electron ground states which are quite similar to that in the 
long range hopping model of the previous section.

We first note that the hopping matrix \( \mT \) defined by 
(\ref{txyf}) satisfies \( \mT\ge0 \).
This is because each \( (t_{x,y}^{(j)})_{x,y\in\Lambda} \) determined 
by (\ref{tj}) is positive semidefinite (which fact is clear from the 
discussion in Section~\ref{s:pre}), and the sum of positive semidefinite 
matrices 
is positive semidefinite (as in Lemma~\ref{l:A+B>0}).

\begin{figure}
\centerline{\epsfig{file=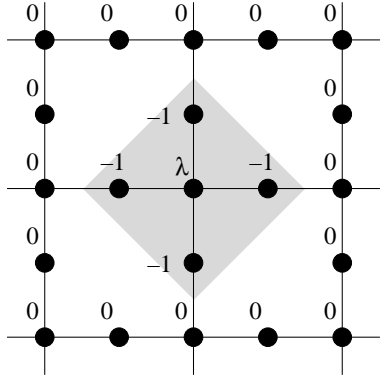,width=5cm}}
\caption[dummy]{
The components of the single-electron ground state 
\( \vphi^{(y)} \) in the model b) of Figure~\ref{f:flat}.
The state is localized around the site \( y \), which is at the 
center of the figure.
}
\label{f:locbasis}
\end{figure}

We define for each \( y\in\calE \) the 
single-electron state \( \vphi^{(y)}=(\phi^{(y)}_{x})_{\xiL} \)
by
\begin{equation}
	\phi^{(y)}_{x}=\cases{
	\lambda&if \( x=y \);\cr
	-1&if \( x \) is the internal site of one of the cells 
	containing \( y \);\cr
	0&otherwise.
	}
	\label{phiyx}
\end{equation}
See Figure~\ref{f:locbasis}.
Observe that we have
\( \bkt{\vlambda^{(j)},\vphi^{(y)}}=0 \) for any 
\( j=1,\ldots,M \) and \( y\in\calE \), where 
\( \vlambda^{(j)}=(\lambda^{(j)}_{x})_{\xiL}\in\hilb \), and hence
\begin{equation}
	\mT\vphi^{(y)}
	=t\sum_{j=1}^{M}
	\vlambda^{(j)}\bkt{\vlambda^{(j)},\vphi^{(y)}}
	=0.
	\label{Tphi=0}
\end{equation}
Since \( \vphi^{(y)} \) with \( y\in\calE \) are linearly independent, 
the single-electron \Sch{} equation (\ref{sSch2}) has 
\( |\calE| \)-fold degenerate ground states\footnote{
It is not hard to prove that \( \vlambda^{(j)} \) with
\( j=1,\ldots,M \) span the remaining space with \( \ep>0 \).
} with \( \ep=0 \).
It is needless to say that such a huge degeneracy is  
accidental, and makes the present models very special.
We also stress that such a degeneracy is lifted by an arbitrarily small 
generic perturbation applied to the hopping matrix.

In models with translational invariance, the degeneracy in 
the single-electron ground states corresponds to the lowest band 
being completely dispersionless, or ``flat.''
For example, in the models on the \( d \)-dimensional hypercubic 
lattice of Section~\ref{s:examples}, we find that there are \( d+1 \) 
bands with dispersion relations
\begin{equation}
	\ep_{j}(k)=\cases{
	0&if \( j=1 \);\cr
	\lambda^2t&if \( j=2,\ldots,d \);\cr
	(\lambda^2+2d)t+t\sum_{\mu=1}^d\cos k_{\mu}&
	if \( j=d+1 \).\cr
	}
	\label{epjkf}
\end{equation}
Note that the lowest band and the \( (d-1) \) middle bands are 
completely flat, while the upper band is dispersive.
For definitions of bands and dispersion relations as well as an 
explicit calculation for the \( d=1 \) case, see Appendix~\ref{s:band}.
\subsection{Proof}
\label{s:flatP}
We now prove Theorem~\ref{t:flat}.
Interestingly, the proof is completely analogous to and not more 
difficult than the second proof of the (less attractive) 
Theorem~\ref{t:cg} for the long range hopping model.

Since the single-electron ground states of the model are 
\( |\calE| \)-fold degenerate, and the electron number is set to 
\( \Ne=|\calE| \), all the preliminary considerations in 
Sections~\ref{s:pre} and \ref{s:second} apply to the present model as 
well.
Any ground state \( \GS \) satisfies 
the conditions (\ref{HGS=0}), and (\ref{ccGS=0})
for any \( x\in\La \).

Again introducing
\( \bys=\Cds(\vphi^{(y)}) \) for \( y\in\calE \), and using 
\( \Hhop\GS=0 \), we can represent any ground state as
\begin{equation}
	\GS
	=
	\sumtwo{\Lup,\Ldn\subset\calE}{{\rm s.t.}\,|\Lup|+|\Ldn|=\Ne}
	f(\Lup,\Ldn)
	\rbk{\prod_{y\in\Lup}\byu}\rbk{\prod_{y\in\Ldn}\byd}
	\vac,
	\label{GSexp3}
\end{equation}
where \( f(\Lup,\Ldn) \) are coefficients, and
\( \Lup \), \( \Ldn \) are summed over all the subsets of \( \calE \)
such that 
\( |\Lup|+|\Ldn|=\Ne=|\calE| \).

Exactly as before, the condition
(\ref{ccGS=0}) for \( x\in\calE \) shows that 
there can be no ``double occupancies'' in \( b^{\dagger} \), 
i.e., \( f(\Lup,\Ldn)=0 \) whenever
\( \Lup\cup\Ldn\ne\emptyset \).
We again reorganize the sum to get a spin system representation
\begin{equation}
	\GS=
	\sum_{\tsigma\in\calS_{\calE}}g[\tsigma]
	\rbk{\prod_{y\in\calE}b^{\dagger}_{y,\sigma(y)}}\vac,
	\label{GSexp4}
\end{equation}
where the sum is over all the spin configuration 
\( \tsigma=(\sigma_{y})_{y\in\calE} \) on \( \calE \),
and \( g[\tsigma] \) is a new coefficient.

Next we examine the implication of the condition (\ref{ccGS=0}) for 
\( x\in\calI \) on the ground state (\ref{GSexp4}).
Unlike in the long range hopping model, there are still many relations 
to use.
Let \( \calE_{x} \) be the set of external sites contained in the cell 
which contains an internal site \( x \).
By using (\ref{GSexp4}) and the anticommutation relation
\( \{c_{x,\sigma},b^{\dagger}_{y,\sigma'}\}=
-\delta_{\sigma,\sigma'}\chi[y\in\calE_{x}] \) 
for any \( x\in\calI \) and \( y\in\calE \), we get
\begin{equation}
	c_{x_,\dn}c_{x,\up}\GS
	=
	\sumtwo{y,z\in\calE_{x}}{{\rm s.t.\ }y>z}\ 
	\sumtwo{\tsigma\in\calS_{\calE}}
	{{\rm s.t.\,}\sigma_{y}=\up,\,\sigma_{z}=\dn}
	{\rm sgn}[x,y]
	(g[\tsigma]-g[\tsigma_{y\leftrightarrow z}])
	\rbk{\prod_{u\in\calE\bs\{y,z\}}b^{\dagger}_{u,\sigma_{u}}}
	\vac.
	\label{ccGS3}
\end{equation}
Since this quantity vanishes for all \( x\in\calI \), we finally find 
that
\begin{equation}
	g[\tsigma]=g[\tsigma_{y\leftrightarrow z}]
	\label{exch2}
\end{equation}
for any \( y,z\in\calE \) which belong to a common cell.
Note that the ``exchange interaction'' is short range because we are 
treating models with short range hoppings.
Since the entire lattice is connected, (\ref{exch2}) ensures that the 
lowest energy state is unique in each sector with a fixed
\( \Sztot \).
This completes the proof of Theorem~\ref{t:flat}.

\subsection{Mielke's flat-band ferromagnetism}
\label{s:mielke}
A slightly different class of Hubbard models which have highly
degenerate 
single-electron ground states and exhibit ferromagnetism was 
discovered by Mielke \cite{Mielke91b,Mielke92} 
{\em before} the above models were 
introduced.
Here we briefly summarize Mielke's beautiful construction based on  
graph theoretic notion.
For a general theory of flat-band ferromagnetism
obtained by Mielke, see \cite{Mielke93}.

We start from abstract notation.
Let $G=(V,E)$ be an abstract graph,
where $V$ is the set of vertices (sites)
 $\alpha,\beta,\ldots\in V$, and
$E$ is the set of edges (bonds) which are nothing but pairs of
vertices like $\{\alpha,\beta\}$.
Given a graph $G$, one can construct the
corresponding line graph $L(G)=(V_{\rm L},E_{\rm L})$
by the following procedure.
The set of vertices (sites) $V_{\rm L}$ (whose elements are denoted as
$x,y,\ldots\in V_{\rm L}$) is taken to be identical to the set $E$.
This means that we identify edges in $G$ with the vertices (sites) in
$L(G)$ as, for example, $x=\{\alpha,\beta\}$, $y=\{\alpha,\gamma\}$,
etc.
Next we declare that two vertices $x,y\in V_{\rm L}$ are adjacent to
each other if the corresponding two edges in $E$ share a common vertex
in $V$.
The vertices
$x$ and $y$ in the above example are adjacent to each other since
the corresponding edges in $E$ have a common vertex $\alpha$.
$E_{\rm L}$ is the set of edges (bonds) in $L(G)$, which
consists of all the adjacent pairs (like $\{x,y\}$)
of vertices (sites) in $V_{\rm L}$.
Finally we set $M(G)=|E|-|V|+1$ if $G$ is bipartite\footnote{
$G$ is bipartite if it can be decomposed into two disjoint sublattices
as $G=A\cup B$ with the property that any edge in $E$ joins a vertex
in $A$ with a vertex in $B$.
}, and $M(G)=|E|-|V|$ if $G$ is non-bipartite.

We define the Hubbard model on the line graph $L(G)$.
With each site $x\in V_{\rm L}$, we associate the fermion operator
$c_{x,\sigma}$, and consider the Hamiltonian
\begin{equation}
	H = t \sumtwo{\{x,y\}\in E_{\rm L}}{\sigma=\uparrow,\downarrow}
	(c^{\dagger}_{x,\sigma}c_{y,\sigma}
	+c^{\dagger}_{y,\sigma}c_{x,\sigma})
	+
	U\sum_{x\in V_{\rm L}}n_{x,\uparrow}n_{x,\downarrow},
	\label{Ham2}
\end{equation}
where \( t>0 \) is a constant.
Then the main result of \cite{Mielke91b} is the following.

\begin{theorem}
[Mielke's flat-band ferromagnetism]
\label{t:mielke}
Suppose that the graph $G$ is twofold connected\footnote{
A graph is twofold connected if and only if one cannot make it
disconnected by the removal of a single vertex.
}.
Consider the above Hubbard model with \( \Ne=M(G) \).
Then for any \( U>0 \), the ground states of the model 
have total spin \( \Stot=\Smax(=\Ne/2) \), and are nondegenerate
apart from the trivial \( (2\Smax+1) \)-fold degeneracy.
\end{theorem}

\begin{figure}
\centerline{\epsfig{file=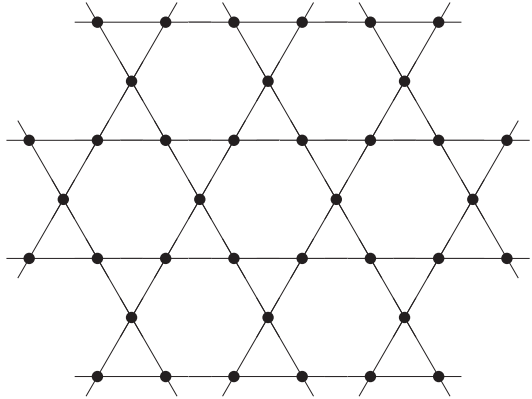,width=7cm}}
\caption[dummy]{
The kagom\'{e} lattice is the line graph of the hexagonal lattice.
Mielke showed that the Hubbard model on the kagom\'{e} lattice 
exhibits ferromagnetism when the filling factor is \( \nu=1/6 \)
for any \( U>0 \).
This 
is the most beautiful example of flat-band ferromagnetism.
}
\label{f:kagome}
\end{figure}

This theorem applies to the Hubbard model defined on various line
graphs, a typical one being the kagom\'{e} lattice of 
Figure~\ref{f:kagome}.

\subsection{Beyond flat-band ferromagnetism}
\label{s:beyond}
Let us once again take a look at the mechanism by which ferromagnetism 
is generated in a flat-band model.
Reflecting the bulk degeneracy in the single-electron ground states, 
the ground states of the non-interacting model with \( H=\Hhop \) are 
highly degenerate.
There are ground states for any possible values of total spin
\( \Stot \), including the smallest \( \Stot=0 \) or \( 1/2 \) and the 
maximum \( \Stot=\Smax(=\Ne/2) \).
When we introduce the Hubbard interaction \( \Hint \), the energy of 
the ferromagnetic ground states (of \( \Hhop \)) do not change since 
ferromagnetic states {\em do not feel} on-site repulsion.
These observations are indeed trivial.
A truly nontrivial point (which is proved by imposing some conditions 
on models) is that all the other ground states of \( \Hhop \) receive 
extra energy from \( \Hint \), and the ferromagnetic ground states 
become the only ground states of 
\( H=\Hhop+\Hint \).

In this sense, the flat-band ferromagnetism certainly takes into 
account nontrivial interplay of \( \Hhop \) and 
\( \Hint \).
However, there is no true ``competition'' between \( \Hhop \) and
\( \Hint \).
The ferromagnetic ground states are already present in the \( U=0 \) 
model among highly degenerate ground states.
The only role of the interaction is to lift the degeneracy, and 
``select'' the ferromagnetic ground states as the only ground states.
This is why flat-band ferromagnetism takes place for any \( U>0 \).

As we have seen in Theorem~\ref{t:var1}, we never realize saturated 
ferromagnetism for values of \( U \) which are too small 
in a system without
bulk degeneracy in the single-electron ground states.
Of course, the
bulk degeneracy in a flat-band model is far from being robust, 
and it is easily destroyed by adding to the hopping matrix \( \mT \) an 
arbitrarily small generic perturbation.
Then, an essential question is whether ferromagnetism found in a 
flat-band model remains stable after adding a small perturbation to 
\( \mT \) which makes the lowest band ``nearly flat.''
If the ferromagnetism 
were unstable against perturbations, we would have to conclude 
 that 
flat-band ferromagnetism is a mere mathematical game.

It was conjectured \cite{92e,93d} that the flat-band ferromagnetism 
is stable against small perturbations to the hopping Hamiltonian.
Among the main arguments used in that discussion
was that an approximate 
low-energy effective Hamiltonian of the flat-band model 
has the precise 
form of the ferromagnetic Heisenberg spin system.
Kusakabe and Aoki \cite{KusakabeAoki94a,KusakabeAoki94b} 
made the first systematic study about 
stability of flat-band ferromagnetism.
They argued that the flat-band models possess spin-wave excitations 
which have healthy dispersions, and this fact guarantees the 
robustness of the flat-band ferromagnetism.
They also found numerical evidence that ferromagnetism 
remains stable if we add sufficiently small 
 perturbations to the hopping 
matrix, thus making the lowest flat band ``nearly flat.''

As for rigorous results, stability of ferromagnetism under a single-spin 
flip is proved in \cite{94c,95b} for the model obtained by adding an
{\em arbitrary} small translation invariant perturbation to the
hopping matrix of the translation invariant flat-band Hubbard model 
of Section~\ref{s:model}.
Although this only establishes the local stability of flat-band 
ferromagnetism, it is a remarkable nonperturbative 
result which applies to a robust class of models.
Given the global stability of ferromagnetism 
in the original flat-band model,
the local stability of ferromagnetism in a nearly-flat-band model 
provides very strong evidence that the nearly-flat-band model indeed 
exhibits globally stable ferromagnetism.

Finally, 
the problem of stability of ferromagnetism 
was completely solved in \cite{95c,97e} for a special class of models.
Let us explain the results briefly.

Take the same lattice \( \Lambda \) as in Section~\ref{s:examples},
i.e., the \( d \)-dimensional decorated hypercubic lattice.
We define special perturbations to the hopping amplitudes as follows.
For each \( z\in\calE \), we let
\begin{equation}
	s_{x,y}^{(z)}=-s\mu_{x}^{(z)}\mu_{y}^{(z)},
	\label{sxyz}
\end{equation}
where 
\begin{equation}
	\mu_{x}^{(z)}=\cases{
	\lambda&if \( x=z \);\cr
	1&if \( x \) is an internal site adjacent to \( z \);\cr
	0&otherwise.\cr
	}
	\label{muxz}
\end{equation}
Here \( s>0 \) is a new constant, and \( \lambda>0 \) is the same as 
in Section~\ref{s:model}.

\begin{figure}
\centerline{\epsfig{file=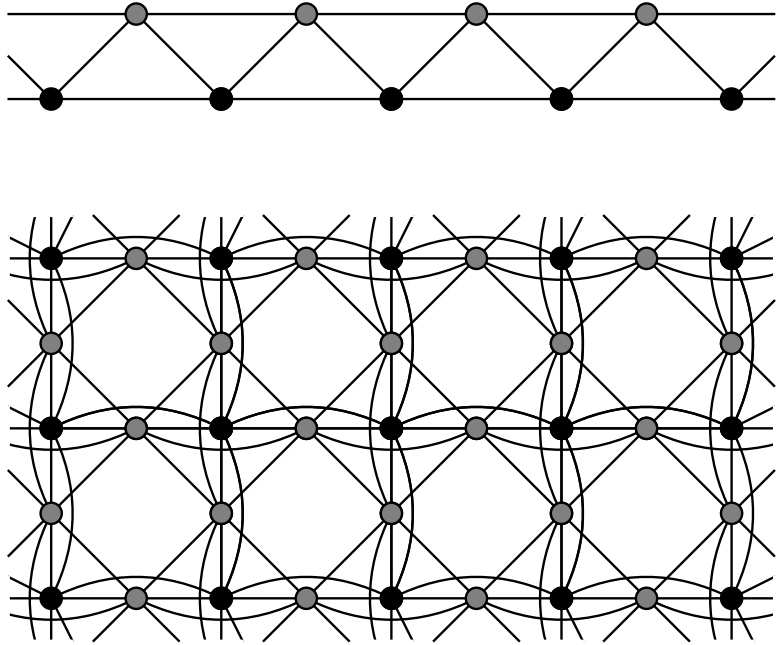,width=12cm}}
\caption[dummy]{
The \( d \)-dimensional decorated hypercubic lattice with the 
nearest neighbor and the next nearest neighbor hoppings
for \( d=1 \) and 2.
These models are obtained by adding special perturbations to the 
models (a) and (b) in Figure~\ref{f:flat}, respectively.
In a range of parameters, we can prove that the corresponding Hubbard 
model with sufficiently large but finite \( U \) exhibits 
ferromagnetism even when the lowest band is not flat.
}
\label{f:nonflat}
\end{figure}

Then the hopping matrix \( \mT=(t_{x,y})_{x,y\in\Lambda} \) of 
the perturbed models is defined by\footnote{
The term with the Kronecker delta is included only to make the 
expression (\ref{txyp2}) simpler and is not essential.
}
\begin{equation}
	t_{x,y}=
	\sum_{j=1}^{M}t_{x,y}^{(j)}
	+\sum_{z\in\calE}s_{x,y}^{(z)}
	+(\lambda^2s-2dt)\delta_{x,y},
	\label{txyp1}
\end{equation}
where \( t_{x,y}^{(j)} \) is the same as in (\ref{tj}).
This \( t_{x,y} \) can be written more explicitly as
\begin{equation}
	t_{x,y}=\cases{
	\lambda(t+s)&
	if \( |x-y|=1/2 \);\cr
	t&
	if \( x,y\in\calE \) and \( |x-y|=1 \);\cr
	-s&
	if \( x,y\in\calI \) and \( |x-y|=1 \) or \( 1/\sqrt{2} \);\cr
	(\lambda^2-2d)t+(\lambda^2-2)s&
	if \( x=y\in\calI \);\cr
	0&
	otherwise.
	}
	\label{txyp2}
\end{equation}
As can be seen also from Figure~\ref{f:nonflat}, this model contains 
nearest-neighbor hoppings and
some next-nearest-neighbor hoppings (but not more than 
that).
The amplitudes for the different hoppings and the on-site potential 
must satisfy special relations because there are only three 
controllable parameters, \( t>0 \), \( s>0 \) and \( \lambda>0 \).

The single-electron energy bands of the model can be easily obtained.
There are \( (d+1) \) bands with dispersion relations
\begin{equation}
	\ep_{j}(k)=\cases{
	-2d(t+s)-2s\sum_{\mu=1}^{d}\cos k_{\mu}&if \( j=1 \);\cr
	\lambda^{2}(t+s)-2dt&if \( j=2,\cdots,d \);\cr
	\lambda^{2}(t+s)+2t\sum_{\mu=1}^{d}\cos k_{\mu}&if \( j=d+1 \).
	}
	\label{ejk}
\end{equation}
If \( d>1 \), the middle bands with \( j=2,\cdots,d \) have  
constant values (i.e., are flat and degenerate), reflecting the 
geometry of the decorated hypercubic lattice\footnote{
It is possible to design short range perturbations to \( \mT \) 
which make these bands non-flat and non-degenerate, while allowing 
Theorem~\ref{t:nonflat} to hold.
}.
For us it is crucial that the most important lowest band has a  
healthy non-constant dispersion.

We consider the Hubbard model on \( \Lambda \) with the above 
\( t_{x,y} \) and  electron number \( \Ne=|\calE|=L^d \).
If \( s=0 \), the model exhibits the flat-band ferromagnetism of 
Theorem~\ref{t:flat} for any \( U>0 \).
When \( s>0 \) and the lowest band is no longer flat, the model 
exhibits Pauli paramagnetism for \( U=0 \).
Moreover, from the elementary variational estimate of 
Theorem~\ref{t:var1}, we find that 
the model does not exhibit ferromagnetism when \( U<4ds \).
We can say that there is a true ``competition'' between \( \Hhop \)
and \( \Hint \) in this model.
If the model exhibits ferromagnetism, it must be in a non-perturbative
region with sufficiently large \( U \).
The following theorem establishes such a non-perturbative statement 
in the case that the lowest band is sufficiently flat.

\begin{theorem}[Ferromagnetism in non-singular Hubbard models]
\label{t:nonflat}
	Consider the above Hubbard model with  electron number
	\( \Ne=|\calE|=L^d \), and let\footnote{
	In \cite{95c}, we presented a complete proof for \( d=1 \) which, 
	however, assumed \( \lambda>\lambda_{\rm c}>0 \).
	The improved proof in \cite{97e} applies to general 
	\( d \) and only requires \( \lambda>0 \).
	} \( \lambda>0 \).
	If \( s/t \) is sufficiently small, and \( U/t \) is sufficiently 
	large, the ground states have  total spin 
	\( \Stot=\Smax(=\Ne/2) \), 
	and are non-degenerate apart from the trivial 
	\( (2\Smax+1) \)-fold degeneracy.
\end{theorem}

For given \( d\ge1 \) and \( \lambda>0 \), the range of 
\( s/t \) and \( U/t \) in which the theorem holds can be determined 
by finite calculations which can be (in principle) executed by a 
computer.
This means that one can construct a computer-aided proof if 
one desires.
We have only performed very elementary calculations for \( d=1 \) 
with a personal computer.
When we set \( \lambda=\sqrt{2} \) (in which case \( t_{x,x}=0 \) for 
all \( x \)), the theorem is valid, for example\footnote{
For sufficiently large \( U \), the theorem is valid for \( t=1.6s \),
in which case the lower band
occupies more than \( 1/4 \) of the whole range of the 
single-electron energy spectrum (including the gap).
In this case the lower band may not even be ``nearly-flat''.
}, if \( s/t\le0.4 \) and \( U/t\ge40 \).

Although the model is rather artificial, this is the first rigorous 
example of saturated ferromagnetism in a non-singular \Hub\ in which we 
must deal with the competition between $\Hint$ and $\Hhop$.
If we further have \( s/t\ll1 \), \( U/t\gg1 \) and \( \lambda\gg1 \), 
it has also been proved that low-lying excitation 
above the ground state has a normal dispersion relation of 
spin-wave excitations \cite{95c,95b}.
Starting from a \Hub\ of itinerant electrons, the existence of a 
``healthy'' ferromagnetism has been established rigorously.

Recently, Penc, Shiba, Mila, and Tsukagoshi 
\cite{PencShibaMilaTsukagoshi96} made a systematic study of related 
one-dimensional Hubbard models and found various pieces of evidence
suggesting that the 
ferromagnetism in these models are indeed robust.

\section{Possible experimental realizations of (nearly-)flat-band 
ferromagnetism}
\label{s:exp}
\subsection{Ferromagnetism in La$_{4}$Ba$_{2}$Cu$_{2}$O$_{10}$}
\label{s:LBCO}
In 1990 (when the only example of saturated ferromagnetism in the 
Hubbard model was that of Nagaoka), Mizuno, Masuda, Hirabayashi, 
Tanaka, Hasegawa and Mizutani 
\cite{Mizunoea90,Mizunoea91}
reported that the tetragonal cuprate 
La$_{4}$Ba$_{2}$Cu$_{2}$O$_{10}$ 
(also called La$_{2}$BaCuO$_{5}$),
which we shall abbreviate as \LBCO, exhibits  
ferromagnetism\footnote{
Ferromagnetism was discovered while investigating the 
superconductor LaBa$_{2}$Cu$_{3}$O$_{7-y}$.
}.
A series of experiments
\cite{Mizunoea90,Mizunoea91,MizunoMasudaHirabayashi93}
has revealed that \LBCO\ is an ideal 
insulating ferromagnet with Curie temperature 5.2~K, where most of 
the magnetic moment comes from the spin 1/2 moments of the 
Cu\( ^{2+} \) ions.

\begin{figure}
\centerline{\epsfig{file=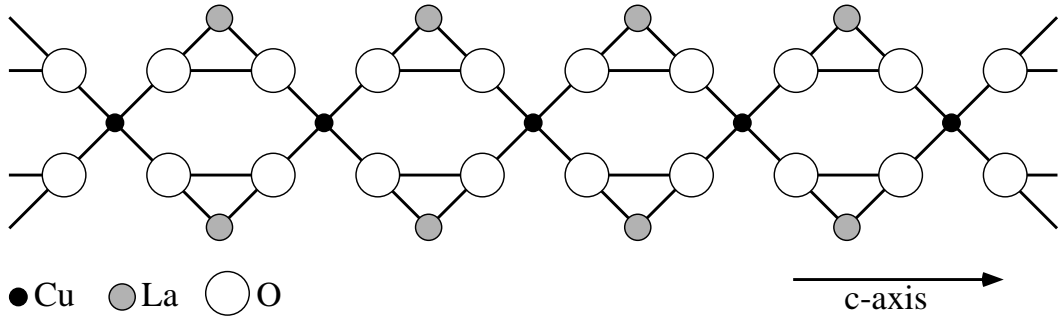,width=14cm}}
\caption[dummy]{
A chain of Cu, O, and La formed along the c-axis of \LBCO.
One finds a triangular structure formed by La and O, which reminds us of 
the (nearly-)flat-band models of Section~\ref{s:flat}.
}
\label{f:LBCO}
\end{figure}

The origin of the ferromagnetic coupling between Cu-ions within the 
ab-plane of \LBCO\ is expected to be described by the GKA rules 
\cite{Mizunoea91,Feldkemperea95}, which involve exchange
interactions  between excited orbital states and ground
orbital states. 
Ferromagnetic coupling along the
c-axis seems to require more careful  treatment.
Along the c-axis, \LBCO\ forms a characteristic chain structure 
consisting of Cu, O, and La.
This is shown in Figure~\ref{f:LBCO}.
Interestingly enough, we see La and O forming a triangular structure 
similar to those in the (nearly-)flat-band models of
Section~\ref{s:flat}. We stress that there are experimental results
which strongly indicate  that the existence of La is essential for
the occurrence of  ferromagnetism.
One such result 
is the observation that the cuprate Nd$_{2}$BaCuO$_{5}$, which 
has the same crystal structure as \LBCO, exhibits antiferromagnetic 
ordering \cite{PaukovPopovaMill91,GolosovskyBoniFischer93}.
It is pointed out that the essential difference is that Nd has a 
magnetic moment, while La does not.
This point is supported by the result  \cite{SalinasSaez93} that
ferromagnetism remains stable upon 
partial replacement of La in \LBCO\ by 
Eu, which has no magnetic moment.
Another such result is from NMR measurement in \LBCO\ that there is a 
large hyperfine field at La sites \cite{Pieperea93}.
This suggests that the magnetic moments are not sharply localized at 
Cu-sites, but partially exist on La-sites as well (exactly as in the 
(nearly-)flat-band models).

On the other hand,
 a band calculation based on the Local Density Approximation 
(LDA) \cite{EyertHockRiseborough95} revealed that two bands of 
\LBCO\ near the fermi level are almost flat\footnote{
In LDA calculations, the Coulomb interaction between electrons 
is supposed 
to be taken into account in a certain self-consistent manner.
Therefore the flatness of two bands in the LDA calculation does not 
directly indicate that the corresponding 
tight-binding model has a flat band.
}.

These observations about the lattice structure and the band structure
 may be regarded as clues that the ferromagnetism in 
\LBCO\ is related to the ferromagnetism in the Hubbard model with 
(nearly-)flat-band.
As far as the present author knows, this possible relation was first 
realized by Hirabayashi \cite{Hirabayashi97}.

In fact it 
is not hard to construct a flat-band Hubbard model on the lattice of 
Figure~\ref{f:LBCO} where hopping amplitudes are nonvanishing only on the 
bonds in the figure.
We can prove that the model exhibits ferromagnetism when the number of 
electrons is identical to the number of Cu-sites in the lattice.

This theoretical result, however, seems to be almost irrelevant to 
the ferromagnetism in \LBCO.
Analysis of the orbital structures of \LBCO\ \cite{Feldkemperea95} 
suggests that a simplified\footnote{
This model is a slight simplification of the tight-binding model 
discussed in \cite{Feldkemperea95}.
} but reasonable tight-binding description of the system
is given by the lattice 
structure of Figure~\ref{f:LBCO} with one non-degenerate orbit per each 
site, and with the electron number equal to the number of  
Cu-sites plus the twice the number of  O-sites.
(In other words, originally Cu orbits are singly occupied, O orbits 
are doubly occupied, and La orbits are empty.)
Then the resulting ferromagnetism is expected to be a ``partial 
ferromagnetism'' in which only spins from Cu-sites align 
ferromagnetically.
Unfortunately, we still do not know how to treat such partial 
ferromagnetism theoretically.

Nevertheless we believe that there is a deep relation between the 
ferromagnetism in \LBCO{} and that in Hubbard 
models with (nearly-)flat-bands.
It is a challenging problem to further investigate this relation both from 
the above described tight-binding model\footnote{
Apart from numerical experiments (which are always possible), 
systematic 
perturbative calculations of ``effective Hamiltonian'' along 
the line of \cite{PencShibaMilaTsukagoshi96,Feldkemperea95} 
are also possible and may be enlightening.
} and from a continuum model.

\begin{figure}
\centerline{\epsfig{file=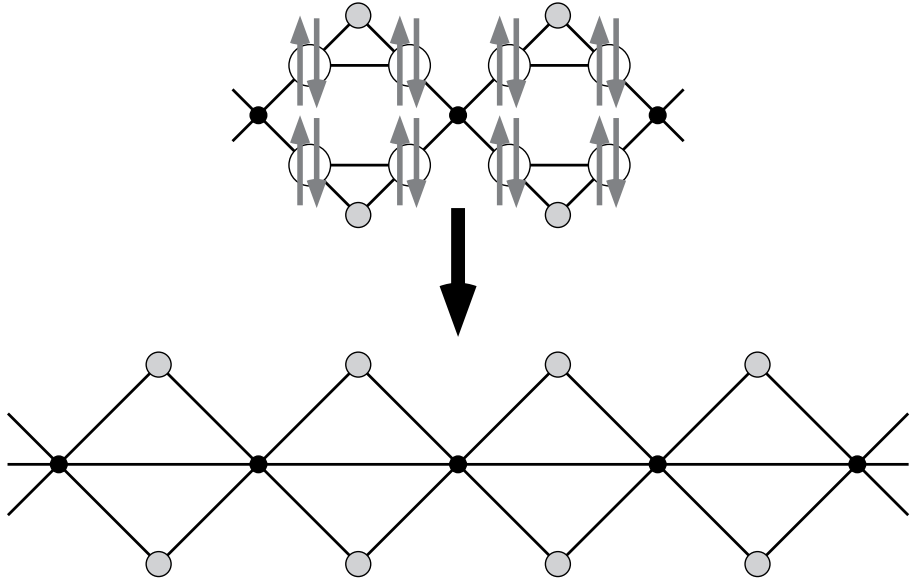,width=12cm}}
\caption[dummy]{
In the second order perturbation in \( 1/\epsilon_{\rm O} \), the 
model of Figure~\ref{f:LBCO} reduces to a Hubbard model on the Cu-La 
chain in the figure with the electron number equal to the number of 
the Cu-sites.
The new model falls into the class of (nearly-)flat-band ferromagnetism
for suitable hopping amplitudes.
This observation provides a crude link between the ferromagnetism in
\LBCO{} and the (nearly-)flat-band ferromagnetism.
}
\label{f:CuLa}
\end{figure}

If one wishes to see a more direct connection between the 
ferromagnetism in \LBCO\ and the (nearly-)flat-band ferromagnetism,
one may try reducing the above Hubbard model to a simpler one by 
relying on 
some approximate argument.
For example, one can consider the limit in which the on-site 
potential \( \epsilon_{\rm O} \) for O-sites (which is \( t_{x,x} \),
with \( x \) being an O-site) is negative, and its absolute value is 
much larger than other \( |t_{x,y}| \).
In this limit, all the O-sites are doubly occupied, and electrons 
(including those on Cu- and La-sites) cannot move on the lattice.
If we consider a perturbation expansion in \( 1/\epsilon_{\rm O} \) up 
to  second order, there arise effective hoppings between Cu-sites 
and La-sites, and between nearby Cu-sites.
The model then reduces to an effective model with Cu-sites and La-sites, as 
in Figure~\ref{f:CuLa}, with the electron number equal to the number of the 
Cu-sites.
For suitable values of hopping amplitudes, the model falls into the 
class of (nearly-)flat-band ferromagnetism of Section~\ref{s:flat}.
This observation provides a crude link between the ferromagnetism in
\LBCO{} and the (nearly-)flat-band ferromagnetism.

\subsection{Possible ferromagnetism in quantum wires}
\label{s:QW}
\begin{figure}
\centerline{\epsfig{file=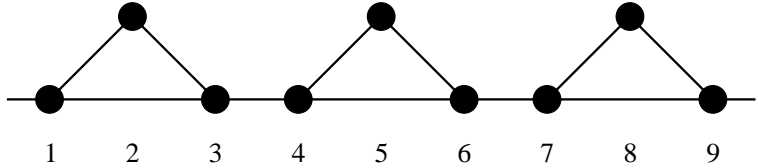,width=10cm}}
\caption[dummy]{
The effective tight-binding model for the ``quantum wire.''
It is possible to design Hubbard model with the same structure 
which shows flat-band ferromagnetism \cite{Ichimuraea98}.
}
\label{f:QW}
\end{figure}

Recent developments in nanotechnology are making it possible to design 
various atomic scale structures which should show nontrivial quantum 
behavior.
In \cite{Watanabeea97}, 
Watanabe, Ichimura, Onogi, Ono, Hashizume, and Wada 
made a theoretical study of the electronic properties of 
Ga adsorbates around dangling-bond wires on an 
H-terminated Si surface, and 
pointed out the 
possibility of ferromagnetism.
The effective tight-binding model (which is not quite faithful to 
the actual lattice structure) used in \cite{Watanabeea97} has the 
lattice and hopping structure shown in Figure~\ref{f:QW} and has  
electron number\footnote{
To be precise, 5/6-filling with \( \Ne=5\Ns/3 \) was considered in 
\cite{Watanabeea97}.
This electron number is changed into \( \Ne=\Ns/3 \) by using the 
hole-particle transformation.
(See Appendix~\ref{s:HP}.)
One should note that the hole-particle transformation does not only 
change the signs of the hopping amplitudes, but it also introduces a
difference in on-site potentials when \( U\ne U' \) in the Hamiltonian
(\ref{HQW}).
} \( \Ne=\Ns/3 \).

It is possible to construct Hubbard models with exactly the same hopping 
structure so that the models have flat lowest bands
\cite{Ichimuraea98}.
Let the lattice be
\( \Lambda=\{1,2,\ldots,\Ns\} \), where \( \Ns \) is a multiple of 
three, and impose periodic boundary conditions.
The Hamiltonian is then
\begin{eqnarray}
	H
	&=&
	t\sum_{j=1}^{\Ns/3}\sum_{\sigma=\up,\dn}
	(\lambda c_{3j-1,\sigma}+c_{3j-2,\sigma}+c_{3j,\sigma})^\dagger
	(\lambda c_{3j-1,\sigma}+c_{3j-2,\sigma}+c_{3j,\sigma})
	\ret
	&&
	+s\sum_{j=1}^{\Ns/3}\sum_{\sigma=\up,\dn}
	(c_{3j,\sigma}+c_{3j+1,\sigma})^\dagger
	(c_{3j,\sigma}+c_{3j+1,\sigma})
	\ret
	&&
	+U\sum_{j=1}^{\Ns/3}(n_{3j-2,\up}n_{3j-2,\dn}+n_{3j,\up}n_{3j,\dn})
	+U'\sum_{j=1}^{\Ns/3}n_{3j-1,\up}n_{3j-1,\dn},
	\label{HQW}
\end{eqnarray}
with \( t,s,U,U'>0 \) and \( \lambda\ne0 \).
This model has a flat lowest band with energy 0.
For electron number \( \Ne=\Ns/3 \), one can prove
that the model exhibits ferromagnetism \cite{Ichimuraea98}.
Since the discussion about the robustness of the flat-band 
ferromagnetism applies to the present models as well, 
the present ferromagnetism is also expected to be stable against
perturbations.

\section{Towards metallic ferromagnetism}
\label{s:metallic}
\subsection{Conjectures and some evidence}
\label{s:conj}
Metallic ferromagnetism is a fascinating phenomenon in which  
electrons exhibiting ferromagnetism contribute to 
electric conductivity as well.
Whether such simple models as the single-orbital Hubbard model can 
describe metallic ferromagnetism is an unsolved and intriguing problem.

Nagaoka's ferromagnetism is certainly motivated by metallic 
ferromagnetism, and we believe it reveals some important aspects of 
a possible mechanism of metallic ferromagnetism.
But for the models where the theorem is proved, the only dynamical 
freedom comes form the motion of the single ``hole.''
We cannot expect the single hole to contribute to appreciable 
electric current in a bulk system.

In the flat-band models of Section~\ref{s:model} and 
the related nearly-flat band models of Section~\ref{s:beyond},
the existence of ferromagnetism is proved for special electron 
numbers.
These electron numbers
correspond to the half-filling of the lowest bands, but since 
the ground states are ferromagnetic, the lowest bands become effectively 
fully filled.
Then the systems should behave as (Mott) insulators.

In the flat-band models of Section~\ref{s:examples} 
defined in dimensions greater than one, 
the existence of ferromagnetism is proved 
\cite{92e,93d} for (not 
too small) electron numbers less than \( |\calE| \).
There is a similar rigorous result for Mielke's model 
in two dimensions\cite{Mielke92}.
We expect the models to
describe a kind of metallic ferromagnetism, but 
the situation is not clear because of pathological degeneracy in the 
(many-body) ground states.
In any case, the flat-band models, which are quite useful in describing the 
origin of exchange interaction in certain systems, are too singular to 
discuss electric conductivity.

Promising candidates of simple models exhibiting metallic ferromagnetism 
are the nearly-flat-band models (obtained by adding perturbations to the 
flat-band models) at filling factor
\( \nu=\Ne/(2\Ns) \) different from 
\( \nu_{0}=|\calE|/(2\Ns) \).
For the case \( \nu<\nu_{0} \),
the approximate projection method (similar to those
described in \cite{93d,95b}, but
based on a local Wannier basis)
indicates that the low energy effective theory of these models is
represented by 
the ferromagnetic \( t \)-\( J \) models in any dimension \cite{97e}.
(The definition of the ferromagnetic \( t \)-\( J \) model
can be found in Section~\ref{s:t-J}.)
This observation leads us to the conjecture that these models exhibit 
metallic ferromagnetism at least 
when \( \nu<\nu_{0} \) is not too small 
and the model is sufficiently close to a flat-band model.

Penc, Shiba, Mila, and Tsukagoshi 
\cite{PencShibaMilaTsukagoshi96} made a detailed 
study  of related problems in one dimension, and found  
both theoretical and numerical evidences that the model exhibits 
metallic ferromagnetism for arbitrary filling factor within the 
range \( 0<\nu<\nu_{0}=1/4 \).
That the ferromagnetism is stable for an arbitrarily small density 
reflects the special character of one-dimensional systems as noted 
in Section~\ref{s:noferro}.
Sakamoto and Kubo \cite{SakamotoKubo96} also found strong numerical 
evidence that related one-dimensional models exhibit metallic 
ferromagnetism for \( 0<\nu<\nu_{0} \).

Watanabe and Miyashita 
\cite{WatanabeMiyashita97a,WatanabeMiyashita97b}
treated 
the one-dimensional flat-band Hubbard model
of Section~\ref{s:examples} for a filling factor
\( \nu \) in the range 
\( 1/2>\nu>\nu_{0}=1/4 \).
Since the lower band is totally filled for such electron densities, 
the flatness of the lower band is regarded as irrelevant.
Note that the case with \( \Ne=\Ns-1 \) is nothing but Nagaoka's 
ferromagnetism;
we have come back to Nagaoka's ferromagnetism following a 
different path!
Watanabe and Miyashita found numerical evidence that the model 
exhibits metallic ferromagnetism for all densities they considered.

A different (but in some sense related) candidate for 
a model with metallic 
ferromagnetism is the so-called \( t \)-\( t' \) model.
It is defined on the one-dimensional lattice 
\( \La=\{1,\ldots,\Ns\} \), 
and its hopping amplitudes are defined by
\( t_{x,x+1}=t_{x+1,x}=-t \), \( t_{x,x+2}=t_{x+2,x}=t' \)
for all \( x\in\La \) and \( t_{x,y}=0 \) otherwise.
From a first order perturbation theory, it was suggested
\cite{LongCastletonHayward94,UedaNishinoTsunetsugu94}
that the model with \( U=\infty \) exhibits ferromagnetism if
\( t'>0 \).
By considering a continuum limit theory, M\"{u}ller-Hartmann
\cite{MullerHartmann95} argued that the model exhibits metallic 
ferromagnetism when \( 4t'>|t|>0 \), \( U=\infty \), and the electron 
density is sufficiently low.
There is  numerical evidence that the \( t \)-\( t' \) model
exhibits ferromagnetism \cite{Pieri97,DaulNoack97}.

Kohno \cite{Kohno97} discussed 
the possibility of metallic ferromagnetism in the 
Hubbard model on a ladder.
Both 
perturbation theory and numerical calculations suggest that 
metallic ferromagnetism appears for electron numbers
satisfying \( \Ns/2<\Ne<\Ns \).
It is interesting that, in this model, itinerant electrons in upper 
bands play essential roles in generating ferromagnetism.

Unfortunately, none of the above conjectures have yet been confirmed 
rigorously\footnote{
In a fermion system on a finite lattice, a formal perturbation series 
alway converges because the operators are finite dimensional.
Then one might think that the  
first order perturbation theories of 
\cite{LongCastletonHayward94,UedaNishinoTsunetsugu94,Kohno97}
imply weak rigorous results that a finite model exhibits 
ferromagnetism for sufficiently small \( t'/t \)
(or \( t_{\parallel}/t_{\perp} \)).
However, this is not the case since there are no estimates of the 
energy difference between the ground state and the first excited state, 
and there remains a possibility that the higher order perturbations 
change the nature of the ground state for any small values of the 
expansion parameter.
The situation is different from that of Nagaoka's ferromagnetism, 
where the result for \( U=\infty \) automatically implies the same 
result for sufficiently large \( U \) (in a finite system).
For an example of rigorous (and general) perturbation theory for 
quantum many-body systems, see \cite{92b}.
}.

\subsection{Ferromagnetic $t$-$J$ model}
\label{s:t-J}
In the present and following sections, we discuss some 
rigorous results concerning 
(metallic) ferromagnetism in one-dimensional
models with \( U\to\infty \) related to the Hubbard model.
Thanks to the special nature of one-dimensional models,
all the results can be obtained by straightforward applications of 
the Perron-Frobenius theorem.
As far as we know, such an application of 
the Perron-Frobenius theorem to 
metallic ferromagnetism in one dimension was developed in the 
pioneering work of Kubo \cite{Kubo82}, where he studied the double 
exchange model.
Although the results we shall discuss 
appear to be somewhat similar to the 
conjectures in the previous section,  
we still have no idea if they shed light 
on the truly fascinating problem of metallic ferromagnetism in the 
Hubbard model.

We first discuss the ferromagnetic \( t \)-\( J \) model.
This is not an interesting model (in one dimension), but 
helps us in illustrating the basic structure of the proof that we 
use for other models.
Unlike the Hubbard model, this model contains spin-spin interactions 
which explicitly favor ferromagnetism.
The problem is whether ferromagnetism is realized in the presence of 
hopping.
As we shall see (and is well-known), the problem is trivial in 
one dimension.

Consider the one-dimensional lattice
\( \La=\{1,\ldots,\Ns\} \).
The Hamiltonian of the ferromagnetic \( t \)-\( J \) model is 
\begin{equation}
	H=
	-t\sum_{x=1}^{\Ns}\sum_{\sigma=\up,\dn}
	(\cxs c_{x+1,\sigma}+\mbox{h.c.})
	-J\sum_{x=1}^{\Ns}
	(\Sop_{x}\cdot\Sop_{x+1}-\frac{n_{x}n_{x+1}}{4})
	+U\sum_{x=1}^{\Ns}\nxu\nxd,
	\label{HtJ}
\end{equation}
with \( t>0 \), \( J>0 \).
Here h.c. represents the hermitian conjugate.
We let \( U\to\infty \) to inhibit 
double occupancies.
We use periodic boundary conditions and identify 
the site \( \Ns+1\) with 1.
The spin operator at site \( x \) is defined as
\( \Sop_{x}=\sum_{\sigma,\tau}\cxs(p^{(\alpha)})_{\sigma,\tau}\,\axs \).
(See (\ref{Stot}).)

\begin{theorem}[Ferromagnetism in the one-dimensional ferromagnetic
\( t \)-\( J \) model]
	\label{t:t-J}
	Assume that the electron number \( \Ne \) satisfies \( \Ne\le\Ns \) 
	and is odd\footnote{
	We can allow any \( \Ne\le\Ns \) if we use open boundary conditions.
	Peculiar dependence of the nature of the ground states on the parity 
	of the electron number and boundary conditions is related to the 
	appearance of spiral states.
	See \cite{Kubo82,KusakabeAoki95,PencShibaMilaTsukagoshi96,%
WatanabeMiyashita97b}.
	}.
	Then the ground states of the present 
	\( t \)-\( J \) model have  total spin 
	\( \Stot=\Smax(=\Ne/2) \), 
	and are non-degenerate apart from the trivial 
	\( (2\Smax+1) \)-fold degeneracy.
\end{theorem}

\proof
The proof is based on an elementary observation which one may call a 
rigorous ``spin-charge separation'' argument.
(See \cite{PencShibaMilaTsukagoshi96} and references therein.)
To take into account the \( U\to\infty \) limit
(with the help of Lemma~\ref{l:U}), we use as our basis
the collection of the states (\ref{basis1}) with 
\( x_{1}<x_{2}<\cdots<x_{\Ne} \) and \( \sigma_{j}=\up,\dn \)
such that 
\( \Sztot=\sum_{j=1}^{\Ne}\sigma_{j}=1/2 \).
The theory of angular momenta ensures that any eigenstate of \( H \) 
has its copy in the sector with the lowest \( |\Sztot| \).
This
sector is spanned by this basis.
We first claim that all the matrix elements of \( H \) in the present 
basis are nonpositive.
As for the hopping term, this is trivial since there is no exchange 
in electron ordering and one does not have to worry about fermion 
signs\footnote{
Hops between sites 1 and \( \Ne \) are exceptional, but these do not 
produce any signs if \( \Ne \) is odd.
When \( \Ne \) is even, such hops generate a ``wrong'' sign for  
the Perron-Frobenius theorem.
This ``frustration'' can be regarded as the origin of the spiral 
states \cite{Kubo82,KusakabeAoki95,PencShibaMilaTsukagoshi96,%
WatanabeMiyashita97b}.
}.
As for the exchange term, this is easily verified by using the identity
\( \Sop_{x}\cdot\Sop_{x+1}=
(\hat{S}^+_{x}\hat{S}^-_{x+1}+\hat{S}^-_{x}\hat{S}^+_{x+1})/2
+\Sopc{3}{x}\Sopc{3}{x+1} \).
It is also verified that all the basis states are connected via 
nonvanishing matrix elements of \( H \).
Note that the existence of the 
exchange term is essential for this property.
Therefore we can readily apply the Perron-Frobenius theorem 
(Theorem~\ref{t:PF}) to conclude 
that the ground state \( \GS \) is unique (in this sector), and it is a 
linear combination of all the basis states with positive coefficients.

This fact indeed implies that for \( \GS \), \( \Stot=\Smax \).
To see this, we first note that uniqueness of the ground state implies 
that \( \GS \) is an eigenstate of \( (\Sopt)^2 \).
We then take a reference state
\( \Phi=\rbk{\hat{S}^{-}_{\rm tot}}^{(\Ne-1)/2}
\rbk{\prod_{x=1}^{\Ne}\cxu} \vac \), which obviously 
has \( \Sztot=1/2 \)
and \( \Stot=\Smax \).
Clearly \( \Phi \) is written as a linear combination of basis states
(\ref{basis1}) with nonnegative coefficients.
Thus we see \( \bkt{\Phi,\GS}\ne0 \), which implies that
 \( \Stot=\Smax \) also holds for \( \GS \).\qed

Once knowing that the ground states have \( \Stot=\Smax \), we can 
look at the ground state with \( \Sztot=\Smax \).
Then it is apparent that the term
\( (\Sop_{x}\cdot\Sop_{x+1}-n_{x}n_{x+1}/4) \)
always vanishes, and hence the ground state is identical to that of 
\( \Hhop \).
Then we have the exact expression for the ground state
\begin{equation}
	\GS=
	\rbk{\prod_{k\ {\rm s.t.\ }|k|\le k_{\rm F}}
	C^{\dagger}_{\up}(\veta^{(k)})}
	\vac,
	\label{GStJ}
\end{equation}
where \( \veta^{(k)} \) is the plane wave state (\ref{veta}), and 
\( k_{\rm F}=\pi(\Ne-1)/(2L) \) is the fermi momentum.
The ground state allows gapless (charge) excitations, 
and hence describes a 
metallic system.

Unfortunately, the above theorem and proof do not shed {\em any} light on 
the corresponding problems in higher dimensions, which are much more 
important and interesting.
The proof for the one-dimensional case relies heavily on the fact that 
essentially no nontrivial spin exchanges take place when \( U=\infty \)
and \( J=0 \).
The system suddenly becomes ferromagnetic when we turn on 
arbitrarily small \( J>0 \).

In the corresponding problem in higher dimensional lattices (or even 
on slightly more complicated one-dimensional lattices, like a ladder), 
there are highly nontrivial exchange processes even for \( J=0 \).
There is no guarantee that one gets ferromagnetism when introducing 
an explicit ferromagnetic interaction, \( J>0 \).
We of course believe that the ferromagnetic \( t \)-\( J \) models in 
higher dimensions also exhibit ferromagnetism for sufficiently large
\( J/t \) and \( \rho=\Ne/\Ns \),
but we have no idea how one can  prove such a statement.
Although ``proving ferromagnetism in the ferromagnetic \( t \)-\( J \)
models'' might sound like a tautology at first glance, it is indeed a 
very difficult and deep problem, whose solution should shed light on 
various aspects of strongly interacting itinerant electron systems.

\subsection{Hubbard model with correlated hopping}
\label{s:two}
In this section, we discuss an artificial model which (like 
the Hubbard model, and unlike the \( t \)-\( J \) model) have no 
interactions explicitly favoring ferromagnetism, but
in which we can still prove 
the appearance of (probably metallic) ferromagnetism in 
a wide range of parameters and electron density.

Take the one-dimensional lattice
\( \La=\{1,\ldots,\Ns\} \)
with even \( \Ns \), and impose periodic boundary conditions by 
identifying \( \Ns+x \) with \( x \).
We consider the Hamiltonian
\begin{equation}
	H=
	-t\sum_{x=1}^{\Ns}\sum_{\sigma=\up,\dn}
	(\cxs c_{x+1,\sigma}+\mbox{h.c.})
	+t'\sum_{j=1}^{\Ns/2}\sum_{\sigma=\up,\dn}
	(c^\dagger_{2j,\sigma}c_{2j+2,\sigma}n_{2j+1}+\mbox{h.c.})
	+U\sum_{x=1}^{\Ns}\nxu\nxd,
	\label{Hcor}
\end{equation}
where we set \( t>0 \), \( t'>0 \).
We will take the \( U\to\infty \) limit.

This resembles the Hubbard model with next nearest neighbor hopping, 
but the hopping term with \( t' \) has an extra number operator 
\( n_{2j+1}=n_{2j+1,\up}+n_{2j+1,\dn} \).
Thus a hop between \( 2j \) and \( 2j+2 \) is possible only when 
\( 2j+1 \) is occupied by an electron.
Note, however, that such density-dependent hopping terms themselves 
are completely spin-independent, and do not prefer ferromagnetism.
Nevertheless we have the following.

\begin{theorem}[Ferromagnetism in the model with correlated hopping]
	\label{t:corhop}
	Consider the model (\ref{Hcor}) in the limit \( U\to\infty \)
	with any odd electron number \( \Ne \).
	Then the ground states have  total spin 
	\( \Stot=\Smax(=\Ne/2) \), 
	and are non-degenerate apart from the trivial 
	\( (2\Smax+1) \)-fold degeneracy.
\end{theorem}

\proof
Exactly as in the proof of Theorem~\ref{t:t-J},
 we take  account of the \( U\to\infty \) limit by using the basis 
states (\ref{basis1}) with 
\( x_{1}<x_{2}<\cdots<x_{\Ne} \) and \( \sigma_{j}=\up,\dn \)
such that 
\( \Sztot=\sum_{j=1}^{\Ne}\sigma_{j}=1/2 \).
It is again trivial to check that the nearest-neighbor hopping terms 
in (\ref{Hcor}) have nonpositive matrix elements.
By definition, an electron hops 
between even sites \( x \) and \( x+2 \) only when \( x+1 \) is 
occupied.
Then, such a hop is always associated with a 
change of ordering in electrons.
(See Figure~\ref{f:chop}.)
This yields a minus sign, resulting again in nonpositive matrix 
elements.

It remains to prove that all the basis states (with \( \Sztot=1/2 \))
are connected with each other by nonvanishing matrix elements.
However, this is already proved in the one hole case in 
Section~\ref{s:con}, and extensions to the multiple holes are trivial.
Thus we can apply the 
Perron-Frobenius theorem exactly as in the proof of 
Theorem~\ref{t:t-J}.\qed

We believe that the ferromagnetic ground states of the present model
have a metallic character.
But this is not as obvious as in the case of the \( t \)-\( J \) model.
The model remains strongly interacting 
in the sector with \( \Sztot=\Smax \), and it is not easy to 
investigate the properties of 
the ground states.

It is clear that the above theorem extends to much more general models 
with additional density-density interactions and on-site potential provided that there are nearest neighbor and 
next nearest neighbor (possibly site-dependent) hoppings with the
correct signs.
An interesting extension is obtained by replacing the second sum in
(\ref{Hcor}) over even sites with a sum over all the sites as
\begin{equation}
	H_{1}=
	-t\sum_{x=1}^{\Ns}\sum_{\sigma=\up,\dn}
	(\cxs c_{x+1,\sigma}+\mbox{h.c.})
	+t'\sum_{x=1}^{\Ns}\sum_{\sigma=\up,\dn}
	(c^\dagger_{x,\sigma}c_{x+2,\sigma}n_{x+1}+\mbox{h.c.})
	+U\sum_{x=1}^{\Ns}\nxu\nxd.
	\label{Hcor2}
\end{equation}
This defines a translation invariant model which resembles the 
\( t \)-\( t' \) model.
Of course the same proof as above 
shows that the model exhibits ferromagnetism 
for \( U=\infty \).

To see the relation between the model (\ref{Hcor2}) and the 
\( t \)-\( t' \) model, consider the Hamiltonian
\begin{equation}
	H_{2}=H_{1}+t''\sum_{x=1}^{\Ns}\sum_{\sigma=\up,\dn}
	\{c^\dagger_{x,\sigma}c_{x+2,\sigma}(1-n_{x+1})+\mbox{h.c.}\}.
	\label{Ht''}
\end{equation}
When \( t''=t' \), this becomes the Hamiltonian of the 
\( t \)-\( t' \) model, while it defines a controllable model
(\ref{Hcor2}) when \( t''=0 \).
One finds that the \( t'' \)-term in (\ref{Ht''}) (along with other 
terms) generates an antiferromagnetic interaction.
Therefore in order to control the possible ferromagnetism in the 
\( t \)-\( t' \) model, one has to deal with the competition between 
the \( t' \)-term and the \( t'' \)-term in (\ref{Ht''}), which seems 
to be a nontrivial task\footnote{
One does not face this competition in first order 
perturbation theory
\cite{LongCastletonHayward94,UedaNishinoTsunetsugu94}.
}.

\subsection{Limiting $U$-$V$ model}
\label{s:three}
We consider another toy model (which perhaps is slightly less 
artificial than the one in the previous section)
obtained by adding a strong nearest-neighbor repulsion
to the Hubbard model.
Again we can prove 
the existence of (metallic) ferromagnetism easily.

Take the one-dimensional lattice
\( \La=\{1,\ldots,\Ns\} \)
with even \( \Ns \), and impose periodic boundary conditions.
We consider 
a model with electron number \( \Ne \) in the range
\( \Ns>\Ne>\Ns/2 \), and with the Hamiltonian
\begin{eqnarray}
	H&=&
	-t\sum_{x=1}^{\Ns}\sum_{\sigma=\up,\dn}
	(\cxs c_{x+1,\sigma}+\mbox{h.c.})
	+t'\sum_{j=1}^{\Ns/2}\sum_{\sigma=\up,\dn}
	(c^\dagger_{2j,\sigma}c_{2j+2,\sigma}+\mbox{h.c.})
\ret&&
	+U\sum_{x=1}^{\Ns}\nxu\nxd
	+V\cbk{
	\rbk{\sum_{x=1}^{\Ns}n_{x}n_{x+1}}-(2\Ne-\Ns)
	},
	\label{HV}
\end{eqnarray}
where we require \( t>0 \) and \( t'>0 \).
We will consider the 
\( U,V\to\infty \) limits.
The new nearest neighbor repulsion term is normalized so 
that the minimum energy for this term is zero.
Note that the hopping amplitudes in the above model can be 
gauge transformed\footnote{
Take \( \La' \) in Appendix~\ref{s:gauge} as the set of even sites.
} to satisfy the condition for Nagaoka's theorem.

As in the Hubbard model, none of the terms in (\ref{HV}) explicitly 
favor ferromagnetism.
Nevertheless we can prove that they together generate ferromagnetism.

\begin{theorem}[Ferromagnetism in the limiting \( U \)-\( V \) model]
	\label{t:Vmodel}
	Consider the model (\ref{HV}) in the limits \( U\to\infty \)
	and \( V\to\infty \) with odd \( \Ne \) such that
	\( \Ns>\Ne>\Ns/2 \).
	Then the ground states have  total spin 
	\( \Stot=\Smax(=\Ne/2) \), 
	and are non-degenerate apart from the trivial 
	\( (2\Smax+1) \)-fold degeneracy.
\end{theorem}

\begin{figure}
\centerline{\epsfig{file=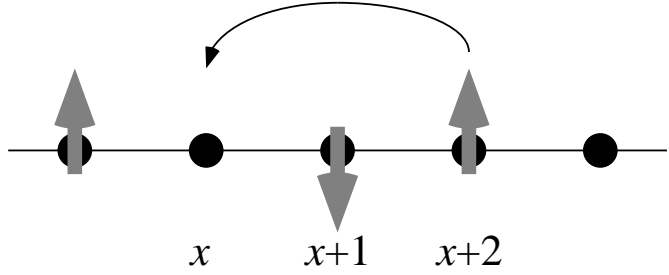,width=9cm}}
\caption[dummy]{
The essential electron exchange process in the correlated hopping 
model and the limiting $U$-$V$ model.
In the limiting $U$-$V$ model with \( \Ne>\Ns/2 \),
sites \( x-1 \) and \( x+1 \) must be occupied when \( x \) is 
empty.
Then the ordering of electrons always changes when an electron hops 
from \( x+2 \) to \( x \).
In the model with correlated hopping, the same fact is guaranteed by 
the definition.
}
\label{f:chop}
\end{figure}

\proof
We take into account the \( V\to\infty \) limit exactly as in
Lemma~\ref{l:U}.
Now our basis states are  (\ref{basis1}) with 
\( x_{1}<x_{2}<\cdots<x_{\Ne} \) 
and \( \sigma_{j}=\up,\dn \),
where \( x_{1},\ldots,x_{\Ne} \) must further satisfy the condition 
that the nearest neighbor repulsion term in (\ref{HV}) is minimized.
When \( \Ne>\Ns/2 \), this condition implies that there can be no 
neighboring holes in the relevant basis states.
This guarantees that an electron hops 
between even sites \( x \) and \( x+2 \) only when \( x+1 \) is 
occupied.
(See Figure~\ref{f:chop}.)
The remainder of the proof is the same as that of 
Theorem~\ref{t:corhop}.\qed

Again, this result can be extended to various more complicated
models.

Note that the nature of the model is quite different for  electron 
numbers \( \Ne\le\Ns/2 \) (in which case we drop 
\( -(2\Ne-\Ns) \) from the Hamiltonian (\ref{HV})).
In the limits \( U\to\infty \) and \( V\to\infty \),
configurations with neighboring electrons are inhibited, and hence 
exchange processes never take place.
In this case the model should exhibit paramagnetism.
This is consistent with a
numerical work \cite{AritaShimoiKurokiAoki97} where 
a shift in the ferromagnetic region as a result of 
finite \( V \) is reported for \( \Ne\le\Ns/2 \).
We expect that a finite value of \( V \) enlarges the region of ferromagnetism 
for \( \Ns/2<\Ne<\Ns \).

\appendix
\section{Gauge transformation}
\label{s:gauge}
The signs of the hopping amplitude
\( t_{x,y} \) can be partially changed by means of gauge 
transformations.
Let \( \La' \) be an arbitrary subset of \( \La \).
We define the new operators \( \tc_{x,\sigma} \) by
\( \tc_{x,\sigma}=-\axs \) if \( x\in\La' \) and
\( \tc_{x,\sigma}=\axs \) if \( x\not\in\La' \).
Since \( \tc_{x,\sigma} \) also satisfy the 
canonical anticommutation relations (\ref{ac1})
and  (\ref{ac2}),
we can use these operators to describe the system.
Then the hopping Hamiltonian (\ref{Hhop}) is written as
\begin{equation}
	\Hhop
	=
	\sumtwo{x,y\in\Lambda}{\sigma=\up,\dn}
	t'_{x,y}\tc^{\dagger}_{x,\sigma}\tc_{y,\sigma},
	\label{Hhop'}
\end{equation}
with \( t'_{x,y}=t_{x,y} \) if both \( x \) and \( y \) are in 
\( \La' \) or 
if neither \( x \) nor \( y \) is in \( \La' \), and
\( t'_{x,y}=-t_{x,y} \) if 
exactly one of \( x \) and \( y \) is in \( \La' \).

The model is said to be bipartite when there is a subset \( \La' \) with 
the property that \( t_{x,y}=0 \) 
if both \( x \) and \( y \) are in \( \La' \) or 
if neither \( x \) nor \( y \) is in \( \La' \).
In a bipartite system, we can use the above gauge transformation to 
change the signs of all the hopping amplitudes.
A typical example is the model on the simple cubic lattice with only 
nearest neighbor hoppings.
\section{Hole-particle transformation}
\label{s:HP}
Let us discuss a simple and standard transformation which maps a 
Hubbard model onto a different Hubbard model with (usually) different
electron number.

We define the new operators \( \tc_{x,\sigma} \) by
\( \tc_{x,\sigma}=\cxs \).
This simply means that we switch the creation and the annihilation 
operators.
 Since the \( \tc_{x,\sigma} \) also satisfy the 
canonical anticommutation relations (\ref{ac1}) and (\ref{ac2}),
we can use these operators to describe the system.
From (\ref{ac1}), the new number operator 
\( \tilde{n}_{x,\sigma}=\tc^{\dagger}_{x,\sigma}\tc_{x,\sigma} \)
is related to the original number operator by
\( \tilde{n}_{x,\sigma}=1-\nxs \).
Therefore the new total electron number \( \tilde{N}_{\rm e} \)
(which is an eigenvalue of 
\( \sum_{\xiL,\sigma=\up,\dn}\tilde{n}_{x,\sigma} \))
is related to the original electron number by
\( \tilde{N}_{\rm e}=2\Ns-\Ne \).

The hopping Hamiltonian (\ref{Hhop}) is written as
\begin{equation}
	\Hhop
	=
	\sumtwo{x,y\in\Lambda}{\sigma=\up,\dn}
	(-t_{x,y})\tc^{\dagger}_{x,\sigma}\tc_{y,\sigma}
	+2\sum_{x\in\Lambda}t_{x,x},
	\label{Hhop''}
\end{equation}
and the interaction Hamiltonian (\ref{Hint}) as
\begin{eqnarray}
	\Hint
	&=&
	U\sum_{\xiL}(1-\tilde{n}_{x,\up})(1-\tilde{n}_{x,\dn})
	\ret
	&=&
	U\sum_{\xiL}\tilde{n}_{x,\up}\tilde{n}_{x,\dn}
	-U\sumtwo{\xiL}{\sigma=\up,\dn}\tilde{n}_{x,\sigma}
	+\Ns U.
	\label{Hint''}
\end{eqnarray}
Note that (\ref{Hint''}) has exactly the same form as the original 
interaction Hamiltonian (\ref{Hint}) apart from the constant term 
and the term 
proportional to the total electron number, which simply shift the 
total energy.

\section{Positive semidefinite operators}
\label{s:positivity}
We summarize 
the definition and elementary properties of positive semidefinite
operators (or matrices) which we used in the main body of the paper.
They should be well-known to readers with a mathematical background.

Let \( \calH \) be a finite dimensional Hilbert space with inner 
product \( \bkt{\cdot,\cdot} \).
In the paper, \( \calH \) may be \( \calH_{\Ne} \) or \( \hilb \).

\begin{definition}[Positive semidefiniteness]
	\label{d:pos}
	For an operator (or matrix) on \( \calH \), we write 
	\( A\ge0 \) and say \( A \) is positive semidefinite
	if \( A \) is self-adjoint (or hermitian) and we have
	\( \bkt{\Phi,A\Phi}\ge0 \) for any \( \Phi\in\calH \).
	For two self-adjoint operators \( A \) and \( B \), we write
	\( A\ge B \) if \( A-B\ge0 \).
\end{definition}

The following statement is easily proved by diagonalizing \( A \).

\begin{lemma}[Positive semidefiniteness and eigenvalues]
	\label{l:pos}
	A self-adjoint operator (or a hermitian matrix) \( A \) is positive 
	semidefinite if and only if all the eigenvalues of \( A \) are 
	nonnegative.
\end{lemma}

The following lemma provides a standard way of constructing a 
positive semidefinite operator.

\begin{lemma}
	\label{l:BB>0}
	Let \( B \) be an arbitrary operator (or matrix) on \( \calH \).
	Then \( A=B^\dagger B \) is positive semidefinite.
\end{lemma}

\proof
Observe that for any \( \Phi \), we have
\( \bkt{\Phi,B^\dagger B\Phi}=\bkt{B\Phi,B\Phi}\ge0 \).\qed

Conversely, any \( A\ge0 \) can be expressed as \( A=B^{2} \)
with some \( B\ge0 \).
This is most easily verified by diagonalizing \( A \).

The following lemma is used repeatedly in the main body of the paper.

\begin{lemma}
	[The sum of positive semidefinite operators]
	\label{l:A+B>0}
	If \( A\ge0 \) and \( B\ge0 \), we have \( A+B\ge0 \).
\end{lemma}
\proof
Assume that \( A+B \) has an eigenstate \( \Phi \) with a 
negative eigenvalue.
Then we get
\( 0>\bkt{\Phi,(A+B)\Phi}
=\bkt{\Phi,A\Phi}+\bkt{\Phi,B\Phi}\ge0 \),
which is a contradiction.\qed

The following is trivial if we expand \( \Phi \) into eigenstates 
of \( A \).

\begin{lemma}
	\label{l:postrivial}
	Let \( A\ge0 \).
	Then \( \bkt{\Phi,A\Phi}=0 \) is equivalent to \( A\Phi=0 \).
\end{lemma}

The following lemma is also useful.

\begin{lemma}
	\label{l:Ai}
	Let \( A_{i}\ge0 \) for \( i=1,\ldots,n \).
	Then \( \sum_{i=1}^n A_{i}\Phi=0 \) implies
	\( A_{i}\Phi=0 \) for each \( i=1,\ldots,n \).
\end{lemma}
\proof
Since \( \sum_{i=1}^n A_{i}\Phi=0 \), we have
\( 0=\bkt{\Phi,\sum_{i=1}^n A_{i}\Phi}=
\sum_{i=1}^n \bkt{\Phi,A_{i}\Phi} \).
By noting that \( \bkt{\Phi,A_{i}\Phi}\ge0 \),
this means \( \bkt{\Phi,A_{i}\Phi}=0 \) 
for each \( i=1,\ldots,n \).
We then use Lemma~\ref{l:postrivial}.\qed

We finally state the following lemma, which sometimes provides us with 
a powerful information.

\begin{lemma}
	\label{l:BBP=0}
	Assume that \( A\ge0 \) is expressed as \( A=B^\dagger B \)
	where \( B \) is not necessarily self-adjoint.
	Then \( A\Phi=0 \) implies \( B\Phi=0 \).
\end{lemma}
\proof
Since \( B^\dagger B\Phi=0 \), we have
\( 0=\bkt{\Phi,B^\dagger B\Phi}=\bkt{B\Phi,B\Phi} \)
which means \( B\Phi=0 \).\qed

\section{Explicit construction of the Hilbert space and fermion 
operators}
\label{s:Hilb}
In section~\ref{s:fo}, we first introduced the algebra 
of fermion operators, and then defined the Hilbert space by operating 
(a representation of) the algebra onto a single state \( \vac \).
There is no problem in making the discussion mathematically 
rigorous  since both the algebra and the Hilbert space are finite 
dimensional.
However,  readers familiar with fields like functional analysis might 
feel more comfortable if the Hilbert space is 
first defined explicitly and then the operators are defined.
Here we explicitly define the Hilbert space and the fermion 
operators following the standard approach\footnote{
For more details, see, for example, Section 5.2 of 
\cite{BratteliRobinson81}.
}.

Let \( \tLa=\Lambda\times\{\up,\dn\} \) be the configuration space for 
a single electron.
We denote its elements as\footnote{
Of course \( u=(x,\sigma) \) with \( x\in\Lambda \) and 
\( \sigma=\up,\dn \).
} \( u,u_{1},u_{2},\cdots\in\tLa \).
The Hilbert space for a single 
electron is \( \ell^2(\tLa;\comp) \).
For \( n=0,1,\ldots,2\Ns \), we define the \( n \)-electron 
Hilbert space \( \calH_{n} \) by \( \calH_{0}=\comp \), and
\begin{equation}
	\calH_{n}=
	\Pas\underbrace{
	\ell^2(\tLa;\comp)\otimes\cdots\otimes\ell^2(\tLa;\comp)
	}_{n},
	\label{Hilbn}
\end{equation}
where \( \Pas \) is the projection operator onto functions 
which are antisymmetric under 
exchanges of any two variables.
More precisely, it is defined as
\begin{equation}
	\Pas\psi(u_{1},\ldots,u_{n})
	=
	\frac{1}{n!}
	\sum_{p:(1,\ldots,n)\to(p(1),\ldots,p(n))}
	(-1)^{p}\psi(u_{p(1)},\ldots,u_{p(n)}),
	\label{Pas}
\end{equation}
where \( p \) is summed over all permutations of \( (1,\ldots,n) \), and 
\( (-1)^{p} \) denotes the parity of \( p \).

We now define the fermion operators. 
For \( \psi\in\calH_{0} \), we let \( c_{u}\psi=0 \).
For \( \psi\in\calH_{n} \) with \( n\ge1 \), we let
\begin{equation}
	(c_{u}\psi)(u_{1},\ldots,u_{n-1})=
	\sqrt{n}\,\psi(u,u_{1},\ldots,u_{n-1}),
	\label{cudef}
\end{equation}
where \( c_{u}\psi\in\calH_{n-1} \).
On \( \psi\in\calH_{n} \) with \( n<2\Ns \), 
the adjoint operator \( c^\dagger_{u} \) acts as
\begin{equation}
	(c^\dagger_{u}\psi)(u_{1},\ldots,u_{n+1})=
	\frac{1}{\sqrt{n+1}}
	\sum_{j=1}^{n+1}(-1)^{j+1}\delta_{u,u_{j}}
	\psi(u_{1},\ldots,u_{j-1},u_{j+1},\ldots,u_{n+1}),
	\label{cdudef}
\end{equation}
where \( c^\dagger_{u}\psi\in\calH_{n+1} \).
For \( \psi\in\calH_{2\Ns} \), we let \( c^\dagger_{u}\psi=0 \).

It is natural to regard the operators
\( c_{u}:\calH_{n}\to\calH_{n-1} \) and
\( c^\dagger_{u}:\calH_{n}\to\calH_{n+1} \) 
as acting on the Fock space
\begin{equation}
	\calF=\bigoplus_{n=0}^{2\Ns}\calH_{n}.
	\label{Fock}
\end{equation}

Finally, we take a basis state (say 1) of the Hilbert space 
\( \calH_{0}=\comp \), and identify it with \( \vac \).
Then it is not difficult to check that 
 states of the form (\ref{basis1}) 
form a basis of the Hilbert space \( \calH_{\Ne} \).

As examples, let us calculate 
\( f=(\cd_{v}\vac)\in\calH_{1} \) and 
\( g=(\cd_{w}f)=(\cd_{w}\cd_{v}\vac)\in\calH_{2} \),
where we understand that \( \vac=1 \).
From (\ref{cdudef}), we have
\( f(u_{1})=(\cd_{v}1)(u_{1})=\delta_{v,u_{1}} \).
Similarly we get
\begin{eqnarray}
	g(u_{1},u_{2})
	&=& 
	(\cd_{w}f)(u_{1},u_{2})
	\ret
	&=&
	\frac{1}{\sqrt{2}}
	\cbk{\delta_{w,u_{1}}f(u_{2})-\delta_{w,u_{2}}f(u_{1})}
	\ret
	&=&
	\frac{1}{\sqrt{2}}
	\rbk{\delta_{w,u_{1}}\delta_{v,u_{2}}
	-\delta_{w,u_{2}}\delta_{v,u_{1}}}.
	\label{guu}
\end{eqnarray}
Note that \( g \) is antisymmetric under the exchange of the names 
of the two particles.
This is a special case of more general states
\begin{equation}
	(\cd_{v_{1}}\cdots\cd_{v_{n}}\vac)(u_{1},\ldots,u_{n})
	=
	\frac{1}{\sqrt{n!}}
	\abs{\matrix{
	\delta_{v_{1},u_{1}}&\delta_{v_{2},u_{1}}&
	\cdots&\delta_{v_{n},u_{1}}\cr
	\delta_{v_{1},u_{2}}&\delta_{v_{2},u_{2}}&
	\cdots&\delta_{v_{n},u_{2}}\cr
	\vdots&\vdots&\ddots&\vdots\cr
	\delta_{v_{1},u_{n}}&\delta_{v_{2},u_{n}}&
	\cdots&\delta_{v_{n},u_{n}}\cr
	}},
	\label{Slater2}
\end{equation}
which are known as the Slater determinant states.

\section{Band structures in single-electron energy spectra}
\label{s:band}
Here we explain (for readers without background in condensed matter 
physics) the notion of band structure in a single-electron 
problem.
The readers will find the notion quite elementary\footnote{
In the band theory of electrons in solids, one usually talks about 
band structures in an ``effective single-electron problem'' (defined 
through complicated self-consistency arguments) in interacting 
many-electron problems.
To explain this theory is beyond the scope of the present article (and 
the ability of the author).
},
especially in a tight-binding model.

We realize the lattice \( \Lambda \) as a subset of the 
\( d \)-dimensional \( L\times\cdots\times L \) torus 
\begin{equation}
	T_{L}
	=\set{(x_{1},\ldots,x_{d})}{0\le x_{j}< L}
	=\real^d/{\sim},
	\label{TL}
\end{equation}
where the identification \( \sim \) is defined by
\( x\sim y \Longleftrightarrow x-y\in L\inte^d \).
Here \( L \) is a positive integer.
Let the unit cell \( \calU\subset[0,1)^{d} \) be a set of finite points 
(sites).
The number of sites \( |\calU| \) in the unit cell will determine the 
number of bands.
Assume that the lattice \( \Lambda \) can be written as
\begin{equation}
	\Lambda=\set{x+z}{x\in\calU, z\in\{0,1,\ldots,L-1\}^d}.
	\label{Lambda}
\end{equation}
We then assume that the hopping matrix 
\( \mT=(t_{x,y})_{x,y\in\Lambda} \) has  translational
invariance\footnote{
We of course identify \( x+z \) and \( y+z \) as elements in 
\( T_{L} \) by using 
the identification \( \sim \) if necessary.
}
\begin{equation}
	t_{x,y}=t_{x+z,y+z},
	\label{trans}
\end{equation}
for any \( x,y\in\Lambda \) and \( z\in\inte^d \).
We note that perfect translational invariance is never possible in 
real physical systems which always have boundaries\footnote{
The presence of boundaries does not change the band structure
drastically, but introduces extra eigenstates which mainly live on 
the boundaries.
Analysis of single-particle eigenstates may not be easy.
}.
We often take artificial periodic boundary conditions because we want 
to concentrate on universal behavior taking place in 
the bulk of the system.

We are interested in the eigenvalue problem 
\( \ep\vphi=\mT\vphi \) or
\begin{equation}
	\ep\phi_{x}=\sum_{y\in\Lambda}t_{x,y}\,\phi_{y},
	\label{SchA}
\end{equation}
for a single electron.
The translational invariance (\ref{trans}) suggests 
that we look for eigenstates 
in the form 
\begin{equation}
	\phi_{x}=e^{ik\cdot x}v_{u(x)},
	\label{phik}
\end{equation}
where \( k\in\calK \) is a wave vector, 
\( u(x) \) is the unique element in \( \calU \) such that
\( x-u(x)\in\inte^d \),
and \( {\bf v}=(v_{u})_{u\in\calU} \) is a \( |\calU| \)-dimensional 
vector.
The set of wave vectors \( \calK \) consists of
\( k=((2\pi/L)n_{1},\ldots,(2\pi/L)n_{d}) \)
with
\( n_{j}=0,\pm1,\ldots,\pm\{(L/2)-1\},L/2 \)
(where we assume \( L \) to be even).
By substituting (\ref{phik}) into (\ref{SchA}), we get
\begin{equation}
	\ep\,v_{u}=\sum_{u'\in\calU}\tau_{u,u'}^{(k)}\,v_{u'},
	\label{SchB}
\end{equation}
with
\begin{equation}
	\tau_{u,u'}^{(k)}
	=\sum_{y\in\Lambda\,{\rm s.t.}\,u(y)=u'}
	t_{u,y}\,e^{ik\cdot(y-u)}.
	\label{tauuu}
\end{equation}
For each \( k\in\calK \), (\ref{SchB}) has \( |\calU| \) eigenvalues 
that we denote as \( \ep_{j}(k) \) with \( j=1,\ldots,|\calU| \).
We can choose \( \ep_{j}(k) \) so that
\( \ep_{j}(k) \) is an analytic function of 
\( k\in(-\pi,\pi]^d \) for each \( j \).
Consequently, the spectrum of the hopping matrix \( \mT \) is 
decomposed as
\begin{equation}
	\sigma(\mT)=\bigcup_{j=1}^{|\calU|}\sigma_{j}(\mT),
	\label{sigmaT}
\end{equation}
with
\( \sigma_{j}(\mT)=\set{\ep_{j}(k)}{k\in\calK} \).
Each \( \sigma_{j}(\mT) \) is called an {\em energy band}.
The function \( \ep_{j}(k) \) is called the {\em dispersion relation} 
of the \( j \)-th band.

\begin{figure}
\centerline{\epsfig{file=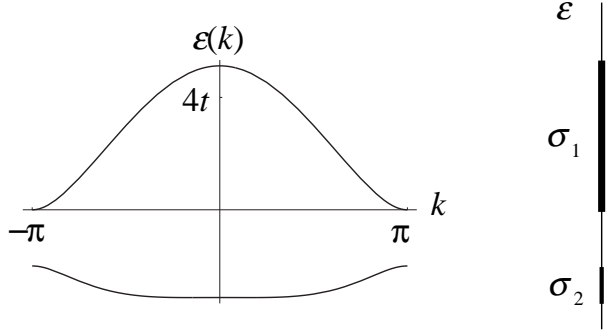,width=8cm}}
\caption[dummy]{
The dispersion relations and the corresponding band structure of the 
simple one-dimensional model with next-nearest-neighbor hoppings
with \( t=t'/2>0 \) and \( V=0 \).
The spectrum consists of the two bands separated by a finite band gap.
}
\label{f:band}
\end{figure}

Let us discuss a simple example with two bands.
Let \( d=1 \), and take \( \calU=\{0,1/2\} \).
The resulting lattice is 
\( \Lambda=\{0,1/2,1,3/2,\ldots,L-(1/2)\} \).
We define the hopping matrix 
\( \mT=(t_{x,y})_{x,y\in\Lambda} \) by
\( t_{x,y}=t' \) if \( |x-y|=1/2 \),
\( t_{x,y}=t \) if \( x,y\in\inte \) and \( |x-y|=1 \),
\( t_{x,x}=V \) if \( x+1/2\in\inte \),
and 
\( t_{x,y}=0 \) otherwise.
Then the eigenvalue equation (\ref{SchB}) becomes
\begin{equation}
	\ep\rbk{\matrix{
	v_{0}\cr v_{1/2}
	}}
	=
	\rbk{\matrix{
	2t\cos k&2t'\cos(k/2)\cr
	2t'\cos(k/2)&V
	}}
	\rbk{\matrix{
	v_{0}\cr v_{1/2}
	}}.
	\label{SchC}
\end{equation}
The two eigenvalues of (\ref{SchC}) define two dispersion relations
\begin{equation}
	\ep_{1,2}(k)
	=
	\frac{1}{2}
	\cbk{
	V+2t\cos k\mp\sqrt{(V-2t\cos k)^2+(4t'\cos(k/2))^2}
	}.
	\label{ep12}
\end{equation}
See Figure~\ref{f:band}.

Consider special cases where the parameters \( t' \) and \( V \) can 
be written as
\( t'=\lambda t \) and \( V=(\lambda^2-2)t \) with the parameter 
\( \lambda>0 \).
Then (\ref{ep12}) becomes
\( \ep_{1}(k)=-2t \)
and
\( \ep_{2}(k)=\lambda^2t+2t\cos k \).
The model becomes the flat-band model\footnote{
We recover the dispersion relation at the end of 
Section~\ref{s:single} if we shift 
the energy by \( 2t \).
} discussed in 
Section~\ref{s:single}.

\section{Proof of Theorem~\protect\ref{t:var2}}
\label{s:detail}
We describe estimates required for the proof of Theorem~\ref{t:var2}.
Although the calculations are straightforward, they illustrate some 
standard techniques we encounter in many-fermion problems and may serve 
as a good exercise for beginners.

First note that the sate \( \Psi \) defined in (\ref{var1}) satisfies
\( n_{x,\dn}n_{y,\dn}\Psi=0 \) for any \( x\ne y \) since there is 
only one \( \dn \) electron.
Then the definition (\ref{var2}) of the variational state \( \tPsi \) 
simplifies as
\begin{equation}
	\tPsi=P_{0}\Psi=(1-\hnu)\Psi,
	\label{var3}
\end{equation}
with
\( \hnu=\sum_{x\in\Lambda}n_{x,\up}n_{x,\dn} \).
To evaluate the energy expectation value of (\ref{varesti}),
we observe that
\begin{equation}
	\bkt{\tPsi,\tPsi}=
	\bkt{P_{0}\Psi,P_{0}\Psi}=
	\bkt{\Psi,P_{0}\Psi}=
	1-\bkt{\Psi,\hnu\Psi},
	\label{PP}
\end{equation}
and
\begin{eqnarray}
	\bkt{\tPsi,\Hhop\tPsi} 
	& = &
	\bkt{\Psi,(1-\hnu)\Hhop(1-\hnu)\Psi} 
	\ret
	 & = & 
	\bkt{\Psi,\Hhop\Psi}
	-\bkt{\Psi,\hnu\Hhop\Psi}
	-\bkt{\Psi,\Hhop\hnu\Psi}
	+\bkt{\Psi,\hnu\Hhop\hnu\Psi}
	\ret
	&=&
	E_{\Psi}(1-2\bkt{\Psi,\hnu\Psi})
	+\bkt{\Psi,\hnu\Hhop\hnu\Psi},
	\label{PHP}
\end{eqnarray}
where we used
\( \Hhop\Psi=E_{\Psi}\Psi \)
and
\( \bkt{\Psi,\Psi}=1 \).
Thus we need to estimate the quantities
\( \bkt{\Psi,\hnu\Psi} \) and
\( \bkt{\Psi,\hnu\Hhop\hnu\Psi} \).

The best way to evaluate these quantities is to go into a Fourier 
representation and express everything 
in terms of the \( a_{j,\sigma} \) operators of (\ref{aj}).
By using the completeness relation
\( \sum_{j=1}^{\Ns}(\psi_{x'}^{(j)})^*\psi_{x}^{(j)}
=\delta_{x,x'} \),
we find the inverse of (\ref{aj}) is
\begin{equation}
	c_{x,\sigma}=\sum_{j=1}^{\Ns}\psi_{x}^{(j)}a_{j,\sigma}.
	\label{ajinv}
\end{equation}
This leads us to the representations
\begin{equation}
	\hnu=
	\sum_{p,q,r,s=1}^{\Ns}
	V_{p,q,r,s}a^\dagger_{p,\up}a_{q,\up}a^\dagger_{r,\dn}a_{s,\dn},
	\label{rep1}
\end{equation}
with
\begin{equation}
	V_{p,q,r,s}=\sum_{x\in\Lambda}
	(\psi_{x}^{(p)})^*\psi_{x}^{(q)}(\psi_{x}^{(r)})^*\psi_{x}^{(s)},
	\label{Vpqrs}
\end{equation}
and
\begin{equation}
	\Hhop=\sum_{j=1}^{\Ns}
	\sum_{\sigma=\up,\dn}
	\ep_{j}\,a^\dagger_{j,\sigma}a_{j,\sigma}.
	\label{rep2}
\end{equation}
All we have to do now is substitute (\ref{rep1}) and (\ref{rep2}) 
into the desired quantities, and use the anticommutation relations
\( \{a_{j,\sigma},a_{k,\tau}\}=0 \),
\( \{a^\dagger_{j,\sigma},a_{k,\tau}\}
=\delta_{j,k}\delta_{\sigma,\tau} \),
the definition (\ref{var1}) of \( \Psi \),
and the fact that
\( a_{j,\sigma}\vac=0 \).
Then we obtain
\begin{eqnarray}
	\bkt{\Psi,\hnu\Psi} 
	& = & 
	\sum_{p=1}^{\Ne-1}V_{p,p,1,1}
	\ret
	&=&
	\sum_{\xiL}\rbk{
	|\psi_{x}^{(1)}|^2
	\sum_{p=1}^{\Ne-1}|\psi_{x}^{(p)}|^2
	},
	\label{PnP2}
\end{eqnarray}
\begin{eqnarray}
	\bkt{\Psi,\hnu\Hhop^\up\hnu\Psi}
	&=&
	\rbk{\sum_{j=1}^{\Ne-1}\ep_{j}}
	\bkt{\Psi,\hnu\Psi}
	-\sum_{\xiL}\rbk{
	|\psi_{x}^{(1)}|^2
	\sum_{p=1}^{\Ne-1}\ep_{p}|\psi_{x}^{(p)}|^2
	}
	\ret
	&&
	+\sum_{\xiL}\rbk{
	t_{x,x}|\psi_{x}^{(1)}|^2
	\sum_{p=1}^{\Ne-1}|\psi_{x}^{(p)}|^2
	},
	\label{PnHupP}
\end{eqnarray}
and
\begin{eqnarray}
	\bkt{\Psi,\hnu\Hhop^\dn\hnu\Psi}
	&=&
	\sum_{x,y\in\Lambda}
	\cbk{
	t_{x,y}
	\rbk{\sum_{p=1}^{\Ne-1}|\psi_{x}^{(p)}|^2}
	\rbk{\sum_{q=1}^{\Ne-1}|\psi_{y}^{(q)}|^2}
	(\psi_{x}^{(1)})^*\psi_{y}^{(1)}
	}
	\ret
	&&
	+\sum_{x,y\in\Lambda}
	\cbk{
	t_{x,y}
	\rbk{\sum_{p=1}^{\Ne-1}(\psi_{x}^{(p)})^*\psi_{y}^{(p)}}
	\rbk{\sum_{q=\Ne}^{\Ns}\psi_{x}^{(q)}(\psi_{y}^{(q)})^*}
	(\psi_{x}^{(1)})^*\psi_{y}^{(1)}
	}
	\ret
	&=&
	\sum_{\xiL}
	\cbk{
	t_{x,x}
	\rbk{\sum_{p=1}^{\Ne-1}|\psi_{x}^{(p)}|^2}
	|\psi_{x}^{(1)}|^2
	}
	+R,
	\label{PnHdnP}
\end{eqnarray}
with
\begin{eqnarray}
	R&=&
	\sumtwo{x,y\in\Lambda}{{\rm s.t.}\,x\ne y}
	\cbk{
	t_{x,y}
	\rbk{\sum_{p=1}^{\Ne-1}|\psi_{x}^{(p)}|^2}
	\rbk{\sum_{q=1}^{\Ne-1}|\psi_{y}^{(q)}|^2}
	(\psi_{x}^{(1)})^*\psi_{y}^{(1)}
	}
	\ret
	&&
	-\sumtwo{x,y\in\Lambda}{{\rm s.t.}\,x\ne y}
	\cbk{
	t_{x,y}
	\rbk{\sum_{p=1}^{\Ne-1}(\psi_{x}^{(p)})^*\psi_{y}^{(p)}}
	\rbk{\sum_{q=1}^{\Ne-1}\psi_{x}^{(q)}(\psi_{y}^{(q)})^*}
	(\psi_{x}^{(1)})^*\psi_{y}^{(1)}
	},
	\label{R}
\end{eqnarray}
where we have used the fact that
\( \sum_{p=1}^{\Ns}\ep_{p}(\psi_{x}^{(p)})^*\psi_{y}^{(p)}
=t_{x,y} \)
and
\( \sum_{p=1}^{\Ns}(\psi_{x}^{(p)})^*\psi_{y}^{(p)}
=\delta_{x,y} \).
We also made the obvious decomposition
\( \Hhop=\Hhop^{\up}+\Hhop^{\dn} \).

By collecting all the estimates, we get
\begin{eqnarray}
	\Delta E
	&=&
	\frac{\bkt{\tPsi,\Hhop\tPsi}}{\bkt{\tPsi,\tPsi}}
	-E_{\rm ferro}
	\ret
	&=&
	\ep_{1}-\ep_{\Ne}
	+\frac{2(\bar{t}-\ep_{1})\bkt{\Psi,\hnu\Psi}}
	{1-\bkt{\Psi,\hnu\Psi}}
	+\frac{R}{1-\bkt{\Psi,\hnu\Psi}},
	\label{dE<}
\end{eqnarray}
with
\begin{equation}
	\bar{t}
	=
	\frac{
	\sum_{\xiL}\rbk{
	t_{x,x}|\psi_{x}^{(1)}|^2
	\sum_{p=1}^{\Ne-1}|\psi_{x}^{(p)}|^2
	}
	}{
	\sum_{\xiL}\rbk{
	|\psi_{x}^{(1)}|^2
	\sum_{p=1}^{\Ne-1}|\psi_{x}^{(p)}|^2
	}
	}
	\le
	\max_{x\in\Lambda}t_{x,x}.
	\label{bart}
\end{equation}

To proceed further, we assume that there is a constant \( \alpha>0 \)
independent of \( x \), \( \Ns \) and \( p=1,\ldots,\Ns \), and we 
have\footnote{
This assumption rules out the possibility of localized eigenstates.
}
\begin{equation}
	|\psi_{x}^{(p)}|\le\frac{\alpha}{\sqrt{\Ns}}.
	\label{psi<a}
\end{equation}
Note that if the model possesses translational invariance as in 
Appendix~\ref{s:band}, then the bound is true with 
\( \alpha=\sqrt{|\calU|} \).
We also assume that 
\( \max_{\xiL}t_{x,x} \),
\( \max_{\xiL}\sum_{y\in\Lambda}|t_{x,y}| \),
and
\( \ep_{1} \) converge to finite quantities as \( \Ns\to\infty \),
which is again trivially valid in translation invariant models.
With the requirement (\ref{psi<a}), we obtain
\( |R|\le{\rm const.}\rho^{2} \), which means that \( R \) gives
only negligible 
contributions for small \( \rho \).
Since we also find
\( \bkt{\Psi,\hnu\Psi}\propto\rho \), the bound (\ref{dE<}) 
implies the desired estimate (\ref{varesti}).
\par\bigskip
It is a pleasure to thank Izumi Hirabayashi and Koichi Kusakabe for 
valuable discussions related to the material in 
Section~\ref{s:LBCO},
and Shun-Qing Shen for valuable discussions about the 
\( t \)-\( t' \) model.
I also thank Fumihiko Nakano and Yuuki Watanabe for 
various useful comments on 
the manuscript.
\bibliography{myWorks,CM,MP}

\begin{thebibliography}{10}

\bibitem{Heisenberg28}
W.~J. Heisenberg, Z. Phys. {\bf 49},  619  (1928).

\bibitem{Bloch29}
F.~Bloch, Z. Phys. {\bf 57},  545  (1929).

\bibitem{Kanamori63}
J.~Kanamori, Prog. Theor. Phys. {\bf 30},  275  (1963).

\bibitem{Gutzwiller63}
M.~C. Gutzwiller, Phy. Rev. Lett. {\bf 10},  159  (1963).

\bibitem{Hubbard63}
J.~Hubbard, Proc. Roy. Soc. (London) {\bf A276},  238  (1963).

\bibitem{Slater53}
J.~C. Slater, H.~Statz, and G.~F. Koster, Phy. Rev. {\bf 91},  1323  (1953).

\bibitem{92e}
H.~Tasaki, Phy. Rev. Lett. {\bf 69},  1608  (1992).

\bibitem{93d}
A.~Mielke and H.~Tasaki, Commun. Math Phys. {\bf 158},  341  (1993),
  cond-mat/9305026.

\bibitem{Thouless65}
D.~J. Thouless, Proc. Phys. Soc. London {\bf 86},  893  (1965).

\bibitem{Nagaoka66}
Y.~Nagaoka, Phy. Rev. {\bf 147},  392  (1966).

\bibitem{Lieb89}
E.~H. Lieb, Phy. Rev. Lett. {\bf 62},  1201  (1989).

\bibitem{ShenQiu94}
S.~Q. Shen, Z.~M. Qiu, and G.~S. Tian, Phy. Rev. Lett. {\bf 72},  1280  (1994).

\bibitem{Mielke91a}
A.~Mielke, J. Phys. {\bf A24},  L73  (1991).

\bibitem{Mielke91b}
A.~Mielke, J. Phys. {\bf A24},  3311  (1991).

\bibitem{Mielke92}
A.~Mielke, J. Phys. {\bf A25},  4335  (1992).

\bibitem{94c}
H.~Tasaki, Phy. Rev. Lett. {\bf 73},  1158  (1994).

\bibitem{95b}
H.~Tasaki, J. Stat. Phys. {\bf 84},  535  (1996).

\bibitem{95c}
H.~Tasaki, Phy. Rev. Lett. {\bf 75},  4678  (1995), cond-mat/9509063.

\bibitem{97e}
H.~Tasaki, in prepration  (1998).

\bibitem{Lieb95}
E.~H. Lieb,  in {\em Advances in Dynamical Systems and Quantum Physics} (World
  Scientific, 1995), cond-mat/9311033.

\bibitem{97d}
H.~Tasaki, J. Phys. Cond. Matt. to appear (1998), cond-mat/9512169.

\bibitem{Fazekas96}
P.~Fazekas, Philosophical Magazine B  to appear  (1997), cond-mat/9612090.

\bibitem{Vollhardt97}
D.~Vollhardt, N.~Bl\"umer, K.~Held, M.~Kollar, J.~Schlipf, and M.~Ulmke, Z.
  Phys. B {\bf 103},  283  (1997), cond-mat/9701150.

\bibitem{HanischUhrigMuellerHartmann97}
T.~Hanisch, G.~S. Uhrig, and E.~Mueller-Hartmann, preprint  (1997),
  cond-mat/9707286.

\bibitem{Shen97}
S.-Q. Shen, Int. J. Mod. Phys. B  to appear  (1997).

\bibitem{Pieri97}
P.~Pieri, S.~Daul, D.~Baeriswyl, M.~Dzierzawa, and P.~Fazekas, Phys. Rev. B
  {\bf 45},  9250  (1996), cond-mat/9603163.

\bibitem{Shastry90}
B.~S. Shastry, H.~R. Krishnamurthy, and P.~W. Anderson, Phy. Rev. {\bf B41},
  2375  (1990).

\bibitem{Roth67}
L.~M. Roth, J. Phys. Chem. Solids {\bf 28},  1549  (1967).

\bibitem{LiebMattis62}
E.~H. Lieb and D.~Mattis, Phy. Rev. {\bf 125},  164  (1962).

\bibitem{89c}
H.~Tasaki, Phy. Rev. {\bf B40},  9192  (1989).

\bibitem{Lieb71}
E.~H. Lieb,  in {\em Phase Transitions} (Wiley, Interscience, 1971),
  p.\ 45.

\bibitem{BrandtGiesekus92}
U.~Brandt and A.~Giesekus, Phy. Rev. Lett. {\bf 68},  2648  (1992).

\bibitem{94a}
H.~Tasaki, Phy. Rev. {\bf B49},  7763  (1994).

\bibitem{Simon93}
B.~Simon, {\em The statistical mechanics of lattice gases. Vol. 1} (Princeton
  University Press, Princeton, New Jersey, 1993).

\bibitem{LandauQM}
L.~D. Landau and E.~M. Lifshitz, {\em Quantum Mechanics (Non-relativisitic
  Theory)} (Pergamon Press, 1977).

\bibitem{DoucotWen89}
B.~Dou\c{c}ot and X.~G. Wen, Phy. Rev. {\bf B40},  2719  (1989).

\bibitem{Toth91}
B.~T\'{o}th, Lett. Math. Phys. {\bf 22},  321  (1991).

\bibitem{Suto91b}
A.~S\"{u}t{\H{o}}, Commun. Math Phys. {\bf 140},  43  (1991).

\bibitem{HanischMullerHartmann93}
T.~Hanisch and E.~M\"{u}ller-Hartmann, Ann. Physik {\bf 2},  381  (1993).

\bibitem{Trugman90}
S.~A. Trugman, Phys. Rev. B {\bf 42},  6612  (1990).

\bibitem{Tian91}
G.~S. Tian, Phys. Rev. B {\bf 44},  4444  (1991).

\bibitem{BarbieriRieraYoung90}
A.~Barbieri, J.~A. Riera, and A.~P. Young, Phys. Rev. B {\bf 41},  11697
  (1990).

\bibitem{Putikka93}
W.~O. Putikka, M.~U. Luchini, and M.~Ogata, Phy. Rev. Lett. {\bf 69},  2288
  (1992).

\bibitem{LiangPang94}
S.~Liang and H.~Pang, Europhys. Lett. {\bf 32},  173  (1995), cond-mat/9404067.

\bibitem{KusakabeAoki94c}
K.~Kusakabe and H.~Aoki, Phys. Rev. B {\bf 50},  12991  (1994),
  cond-mat/9407036.

\bibitem{YamanakaHonjouHatsugaiKohmoto96}
M.~Yamanaka, S.~Honjo, Y.~Hatsugai, and M.~Kohmoto, J. Stat. Phys. {\bf 84},
  1133  (1996), cond-mat/9512065.

\bibitem{Mielke93}
A.~Mielke, Phys. Lett. {\bf A174},  443  (1993).

\bibitem{KusakabeAoki94a}
K.~Kusakabe and H.~Aoki, Physica {\bf B194--B196},  215  (1994).

\bibitem{KusakabeAoki94b}
K.~Kusakabe and H.~Aoki, Phy. Rev. Lett. {\bf 72},  144  (1994).

\bibitem{PencShibaMilaTsukagoshi96}
K.~Penc, H.~Shiba, F.~Mila, and T.~Tsukagoshi, Phys. Rev. B {\bf 54},  4056
  (1996), cond-mat/9603042.

\bibitem{Mizunoea90}
F.~Mizuno, H.~Masuda, I.~Hirabayashi, S.~Tanaka, M.~Hasegawa, and U.~Mizutani,
  Nature {\bf 345},  788  (1990).

\bibitem{Mizunoea91}
H.~Masuda, F.~Mizunoand, I.~Hirabayashi, and S.~Tanaka, Physical Review B {\bf
  43},  7871  (1991).

\bibitem{MizunoMasudaHirabayashi93}
F.~Mizuno, H.~Masuda, and I.~Hirabayashi,  in {\em Studies of High Temperature
  Superconductors, vol. 10}, edited by A.~Narlikar (Nova Science Publisher,
  Commack, NY, 1993).

\bibitem{Feldkemperea95}
S.~Feldkemper, W.~Weber, J.~Schulenburg, and J.~Richter, Phys. Rev. B {\bf 52},
   313  (1995).

\bibitem{PaukovPopovaMill91}
I.~V. Paukov, M.~N. Popova, and B.~V. Mill', Phys. Lett. A {\bf 157},  306
  (1991).

\bibitem{GolosovskyBoniFischer93}
I.~V. Golosovsky, P.~B\"{o}ni, and P.~Fischer, Phys. Lett. A {\bf 182},  161
  (1993).

\bibitem{SalinasSaez93}
A.~Salinas-S\'{a}nchez and R.~S\'{a}ez-Puche, Solid State Ionics {\bf 63-65},
  927  (1993).

\bibitem{Pieperea93}
M.~W. Pieper, E.-G. Caspary, P.~Adelmann, and G.~Roth, J. Mag. Mag. Mat. {\bf
  153},  319  (1994).

\bibitem{EyertHockRiseborough95}
V.~Eyert, K.-H. H\"{o}ck, and P.~S. Riseborough, Europhys. Lett. {\bf 31},  385
   (1995).

\bibitem{Hirabayashi97}
I.~Hirabayashi, private communication  .

\bibitem{Watanabeea97}
S.~Watanabe, M.~Ichimura, T.~Onogi, Y.~A. Ono, T.~Hashizume, and Y.~Wada, Jpn.
  J. Appl. Phys. {\bf 36},  L929  (1997).

\bibitem{Ichimuraea98}
M.~Ichimura, K.~Kusakabe, S.~Watanabe, and T.~Onogi, in preparation  (1998).

\bibitem{SakamotoKubo96}
H.~Sakamoto and K.~Kubo, J. Phys. Soc. Jpn. {\bf 65},  3732  (1996).

\bibitem{WatanabeMiyashita97a}
Y.~Watanabe and S.~Miyashita, J. Phys. Soc. Jpn. {\bf 66},  2123  (1997).

\bibitem{WatanabeMiyashita97b}
Y.~Watanabe and S.~Miyashita, J. Phys. Soc. Jpn. {\bf 66},  to appear  (1997).

\bibitem{LongCastletonHayward94}
M.~W. Long, C.~W.~M. Castleton, and C.~A. Hayward, J. Phys. Condens. Matter
  {\bf 6},  481  (1994).

\bibitem{UedaNishinoTsunetsugu94}
K.~Ueda, T.~Nishino, and H.~Tsunetsugu, Phys. Rev. B {\bf 50},  612  (1994).

\bibitem{MullerHartmann95}
E.~M\"{u}ller-Hartmann, J. Low Temp. Phys. {\bf 99},  349  (1995),
  cond-mat/9502104.

\bibitem{DaulNoack97}
S.~Daul and R.~Noack, preprint  (1996), cond-mat/9612056.

\bibitem{Kohno97}
M.~Kohno, Phys. Rev. B  to appear  (1997), cond-mat/9709170.

\bibitem{92b}
T.~Kennedy and H.~Tasaki, Commun. Math Phys. {\bf 147},  431  (1992).

\bibitem{Kubo82}
K.~Kubo, J. Phys. Soc. Jpn. {\bf 51},  782  (1982).

\bibitem{KusakabeAoki95}
K.~Kusakabe and H.~Aoki, Phys. Rev. B {\bf 52},  R8684  (1995),
  cond-mat/9509034.

\bibitem{AritaShimoiKurokiAoki97}
R.~Arita, Y.~Shimoi, K.~Kuroki, and H.~Aoki, preprint  (1997),
cond-mat/9801242.

\bibitem{BratteliRobinson81}
O.~Bratteli and D.~W. Robinson, {\em Operator algebras and quantum statistical
  mechanics. Vol. 2} (Springer-Verlag, New York, Heidelberg, Berlin, 1997).

\end{thebibliography}
\bibliographystyle{Newprsty}
\end{document}